\documentclass[useAMS,usenatbib]{mnras}

\usepackage{graphicx}
\usepackage{times}
\usepackage[usenames, dvipsnames]{xcolor}
\usepackage{rotating}
\usepackage{txfonts}
\usepackage{longtable,lscape}
\usepackage{anysize}
\usepackage{amssymb}
\usepackage{natbib}
\usepackage{hyperref}
\usepackage{booktabs}
\usepackage{bm}
\usepackage{multirow}

\newcommand{\kms}{${\rm km~s}^{-1}$}
\newcommand{\mgc} {\emph{mGC3}}
\newcommand{\ngc} {\emph{nGC3}}
\newcommand{\gc} {\emph{GC3}}
\newcommand{\Gaia} {\emph{Gaia}}
\newcommand{\FeH} {[\mathrm{Fe}/\mathrm{H}]}

\newcommand{\gaiaerrors} {{\tt Gaia-errors}}
\newcommand{\insitu}{in situ}
\newcommand{\Msun}{{\rm M}_\odot}
\newcommand{\Lsun}{{\rm L}_\odot}
\newcommand{\tm}{t_{\rm m}}
\newcommand{\ljmu}{HYDRO-zoom}
\newcommand{\ljmus}{HYDRO-zooms}

\title[Predictions for the detection of streams with Gaia]{Predictions for the detection of Tidal Streams with Gaia using Great Circle Methods}
\author[C. Mateu et. al.]{Cecilia~Mateu$^{1,2}$\thanks{E-mail:cmateu@cida.gob.ve}, Andrew~P.~Cooper$^{3}$, Andreea~S.~Font$^{4}$, Luis~Aguilar$^{2}$, Carlos~Frenk$^{3}$,
\newauthor
Shaun~Cole$^{3}$, Wenting~Wang$^3$ and Ian~G.~McCarthy$^4$\\
$^{1}${{Centro de Investigaciones de Astronom\'{\i}a, AP 264, M\'erida 5101--A, Venezuela}}\\
$^{2}$Instituto de Astronom\'ia, Universidad Nacional Aut\'onoma de M\'exico, Apartado Postal 877, 22860 Ensenada, B.C., M\'exico \\
$^{3}${{Institute for Computational Cosmology, Department of Physics, University of Durham, South Road, Durham DH1 3LE}}\\
$^{4}${{Astrophysics Research Institute, Liverpool John Moores University, 146 Brownlow Hill, Liverpool L3 5RF, UK}}\\
}

\begin{document}

\date{Accepted. Received ; in original form }
\pagerange{\pageref{firstpage}--\pageref{lastpage}} \pubyear{2007}
\maketitle

\label{firstpage}

\begin{abstract}
The \Gaia~astrometric mission may offer an unprecedented opportunity to discover new tidal streams in the Galactic halo. To test this, we apply \ngc, a great-circle-cell count method that combines position and proper motion data to identify streams, to ten mock \Gaia~catalogues of K giants and RR Lyrae stars constructed from cosmological simulations of Milky Way analogues. We analyse two sets of simulations, one using a combination of $N$-body and semi-analytical methods which has extremely high resolution, the other using hydro-dynamical methods, which captures the dynamics of baryons, including the formation of an {\it in situ} halo. These ten realisations of plausible Galactic merger histories allow us to assess the potential for the recovery of tidal streams in different Milky Way formation scenarios. We include the \Gaia~selection function and observational errors in these mock catalogues. We find that the \ngc~method has a well-defined detection boundary in the space of stream width and projected overdensity, that can be predicted based on direct observables alone. We predict that about \emph{4--13 dwarf galaxy streams can be detected in a typical Milky Way-mass halo with \Gaia+\ngc}, with an estimated efficiency of $>$80\% inside the detection boundary. The progenitors of these streams are in the mass range of the classical dwarf galaxies and may have been accreted as early as redshift $\sim3$. Finally, we analyse how different possible extensions of the \Gaia~mission will improve the detection of tidal streams.

\end{abstract}


\section{Introduction}

The \Gaia~mission, whose first data release is now publicly available, is expected to revolutionise our knowledge of the formation of the Milky Way, by mapping, for the first time, close to a billion stars in the disc, bulge and halo with exquisite astrometric precision \citep{Perryman2001,deBruijne2012}. It is anticipated that this detailed information will enable a breakthrough in understanding the formation history of the Milky Way.
 
The stellar halo, in particular, holds a wealth of information about the merger history of the Galaxy, being a repository of most of the tidal debris from the past merger events. The number of tidal streams surviving at the present day in the halo, their morphologies, their total luminosities and their chemical abundance patterns, all encode important information from which the series of accretion events can be reconstructed \citep{Helmi1999,Bullock2005,Johnston2008,Cooper2010,Helmi2011}. Tidal streams can also be used to infer the gravitational potential of the Milky Way \citep[e.g.][]{PriceWhelan2013,Sanderson2015,Sanderson2016}. Increasing the number of stream detections can improve this measurement \citep{Deg2014}.

While it is expected that \Gaia~will uncover new tidal streams in the halo \citep{Helmi2000,Gomez2010}, quantitative theoretical predictions for the likely number of such discoveries have not been made to date, mainly because of the uncertainties in modelling the physical processes associated with baryons in the framework of hierarchical structure formation, the computational resolution and re-sampling issues associated with producing adequate simulated catalogues at the level of individual stars, and the need to develop algorithms for making such detections by mining the Gaia dataset. In this study, we aim to make progress by employing a series of state-of-the-art simulations of Milky Way-mass haloes from which we construct mock \Gaia~star catalogues, which we search for tidal streams with a robust, quantifiable method. 

We use two suites of cosmological simulations to produce the mock \Gaia~catalogues: the Aquarius simulations, a set of high resolution dark matter only simulations of Milky Way-mass haloes \citep{Springel2008a}, combined with the  GALFORM semi-analytic prescriptions \citep{Cooper2010}; and a second set, called \ljmus, which comprises several medium resolution hydro-dynamical simulations of Milky Way-mass disc galaxies (Font et al., in prep., hereafter  F17), the initial conditions of which were extracted from the EAGLE simulation \citep{Schaye2015}. 
Aquarius allows us to study tidal streams from progenitors that span a wide range of masses and orbits, and hence to test our method on a realistic set of stream luminosities and morphologies. On the other hand, the \ljmus, although of lower resolution than Aquarius, have the benefit of modelling the hydro-dynamical effects of baryons self-consistently. Baryonic effects, including modification of the density profiles of satellites by stellar feedback and interactions between satellites and the central stellar disc may alter the morphology of tidal streams, and, together with the possible presence of an in situ halo, this may change (most likely decrease) the number of streams that can be detected. The objective of this paper is not to perform a detailed comparison between these two simulation techniques, but rather to estimate the detectability of the tidal streams they predict.  

This work goes beyond earlier studies of tidal stream detection in several ways. For the first time, we make predictions based on fully cosmological simulations of Milky Way-mass galaxies that we combine with the most up-to-date \Gaia~error estimates and selection function. The simulated tidal streams evolve within a realistic gravitational potential (non-axisymmetric and changing in time). Thus, the mock \Gaia~star catalogues constructed here complement existing \Gaia~mocks which do not include substructure in the stellar halo \citep[e.g.][]{Robin2012}. Examining a number of Milky Way-mass haloes with a variety of merger histories helps to make our predictions robust against our ignorance of the details of the Galaxy's accretion history. 
This is a step forward towards comparing the models and observations on a level playing field.
Also, with the \ljmu~simulations, the effect of halo component formed {\it in situ}  \citep{Zolotov2010,Font2011a,McCarthy2012,Cooper2015,Pillepich2015} on the detectability of tidal streams can be taken into account. To our knowledge, the contaminating effect of combined {\it in situ} and accreted halo components has only been estimated for \Gaia~by \citet{Brown2005} and \citet{Mateu2011}, who embedded a set of stellar streams in a smooth Galactic background with a constrained luminosity normalization. However, these streams were evolved in a fixed axisymmetric potential and their progenitors selected \emph{ad hoc}. 

Rather than starting from the information available in the simulations, in which every star particle and hence every stream can be identified unambiguously with a specific progenitor, we first apply an observational stream finding algorithm based on the Great Circle Counts (GC3) method. This method, described in detail below, uses combined positions and proper motions to assign stars to discrete groups with common orbital poles. GC3 methods are an efficient way to search for tidal streams in the Galactic halo. They exploit the fact that streams will be approximately confined to planes in potentials that are close to spherical, by searching for overdensities of stars along great circles (as seen from the Galactic centre). The idea was initially proposed by \citet{LyndenBell1995} and \citet{Johnston1996} and later modified by \citet[][hereafter M11]{Mateu2011} to improve its efficacy by including kinematical information (mGC3), with the Gaia~mission in mind. Its main advantage is that, with the implementation proposed in M11, the GC3 family of methods works directly in observable space (positions, parallax, proper motion, radial velocity), rather than using physical parameters such as energy or angular momentum, greatly reducing the effect of the propagation of observational errors, which \citet{Brown2005} have shown can be quite substantial even for \Gaia. 

Finally, we assess the efficiency of our stream detection method by using our knowledge of the `true' population of streams in the simulations to determine which progenitors are recovered and with what `purity'. This knowledge of the method's efficiency and selection biases will be a key ingredient in the inverse process of inferring the Galactic accretion history. 

The paper is structured as follows: Section \ref{s:simulations} summarises the simulations employed in this study. Section \ref{s:mock_catalogues} describes the construction of the mock \Gaia~catalogues and the \Gaia~error simulation. Section \ref{s:gaia_can_see} presents what \Gaia~like surveys would `see' in the simulated stellar haloes based on a selection of specific stellar tracers. Section \ref{s:mgc3} describes the Great Circle method used to identify tidal streams. The appearance of observable tidal streams in the diagnostic space of the method, which we call pole count maps, is explored in detail for a fiducial halo in Section \ref{s:full_PCMs}. Section \ref{s:all_pcms} summarises the results of applying our algorithm to all the other haloes in our sample. In Section \ref{s:recovery_all}, we investigate the properties of progenitors of the streams that are detected in the mock \Gaia~surveys of our simulations. In Section \ref{s:gaia_extensions} we analyse how the detectability of tidal streams changes under various scenarios for extending the lifetime of the \Gaia~mission. Finally, in Section \ref{s:recommendations} we discuss several ways in which this stream finding method can be further improved and give a summary of our conclusion in Section \ref{s:concl}.

\section[]{Cosmological Simulations}\label{s:simulations}

\subsection{Aquarius Simulations}\label{s:aquarius}

Aquarius is a set of six collisionless cosmological `zoom' simulations of individual dark matter haloes of mass $\sim10^{12}\Msun$ \citep{Springel2008a, Springel2008b, Navarro2010}. The simulations assume a ${\Lambda}\rm{CDM}$ cosmogony with parameters determined from the WMAP 1-year results \citep{Spergel2003} and the 2dF Galaxy Redshift Survey data \citep{Colless2001}: $\Omega_{\mathrm{M}}=0.25$,  $\Omega_{\mathrm{\Lambda}}=0.75$, $n_{\mathrm{S}}=1$, $\sigma_{8} = 0.9$ and Hubble parameter $h = 0.73$. The six haloes were selected randomly from a parent sample of isolated halos of similar mass in a lower resolution $(100 \, h^{-1})^{3} \mathrm{Mpc^{3}}$ cosmological volume simulation \citep{Gao2008}. Isolation was defined by the absence of any neighbours with more than half the mass of the target halo within $1\,h^{-1}$ Mpc. A Lagrangian region several times larger than the $z=0$ virial radius of each target halo was resimulated with a much larger number of lower-mass particles, coarsely sampling the surrounding large-scale structure with a smaller number of higher mass particles, subject to exactly the same spectrum of initial density perturbations.

The Aquarius simulations are labelled Aq-A to Aq-F; we do not use Aq-F in this paper because its recent merger history makes it highly unlikely to be representative of a system like the Milky Way \citep{BoylanKolchin2010, Cooper2010}. We use the level 2 set of simulations, the highest resolution level at which all six haloes were simulated. The particle mass varies slightly between the level 2 simulations in the range $0.6 < m_{\mathrm{p}}(\times10^{4} \,\Msun) < 1.4$. The Plummer-equivalent gravitational softening length is $\epsilon\sim66$~pc.    

The Aquarius simulations use a single high-resolution particle species to model the collisionless dynamics of  both dark matter and baryons. To represent the stellar component, we use the `particle tagging' models described by \citep{Cooper2010}. This technique first uses a semi-analytic galaxy formation model to determine the star formation history of each dark matter halo in the simulation, and then applies dynamical criteria to select subsets of collisionless particles occupying regions in phase space associated with each distinct single-age stellar population at the time of its formation. The \citet{Cooper2010} technique improves on earlier tagging approaches \citep[e.g.][]{Bullock2005} in the use of a single, self-consistent cosmological simulation to treat the dynamics of the satellites and the host halo, and in the use of a galaxy formation model constrained by large cosmological datasets as well as the properties of Milky Way and M31 satellites \citep{Bower2006,Cooper2010,Font2011b}. The five Aquarius simulations we use show considerable diversity in the properties of their stellar haloes, owing to their range of virial masses and, more significantly, to the intrinsically stochastic nature of dwarf galaxy accretion and disruption in $\Lambda$CDM.  

The particle tagging technique involves a dynamical approximation with clear limitations, and unlike \citet{Bullock2005} the \citet{Cooper2010} simulations do not include the gravitational contribution of a massive stellar disc at the centre of the host potential. The presence or absence of a disk may accelerate the tidal disruption of some satellites. This is likely to affect predominantly those substructures with orbits passing through the inner $\sim$20 kpc of the galaxy after $z>2$, which nevertheless may include satellites and streams located far from the disk at $z=0$. \citet{Errani2017} find the total number of potentially luminous subhaloes disrupted in the inner region of the halo changes by a factor of $\sim$2 when an idealised disk component is added to the potential in one of the Aquarius simulations. They demonstrate that the inner slope of the satellite mass density profile (which depends on the physics of galaxy formation) has an even larger effect on the number of surviving satellites (almost an order of magnitude; the conclusions of Errani et al. relate only to whether or not an identifiable self-bound core survives, rather than to the presence of tidal streams). Likewise, \citep{GarrisonKimmel2017} find a factor 2--5 depletion of massive subhaloes in a dark matter only simulation when they introduce a growing analytic disk potential based on a hydrodynamical realization from the same initial conditions.  Our results here concern the disruption of well-resolved satellites with very high mass-to-light ratios, predominantly in the outer halo; for further discussion of related issues we refer the reader to \citet{Cooper2010,Cooper2013,Cooper2016} and \citet{LeBret2015}. The number of these more distant satellite halos surviving at $z=0$ may therefore be considered uncertain by no more than factor of $\sim2$ as the result of neglecting the (still somewhat uncertain) influence of a disk potential. As we describe in the following subsection, we also analyse a suite of lower-resolution gas-dynamical simulations that account self-consistently for the gravitational effects of baryons neglected by the particle tagging approach. This allows us to check for large differences in the number of streams from bright satellites that could be due to the presence of a disk, albeit in the context of only one hydrodynamical model and in different dark matter haloes to Aquarius. If the MW disk has significantly depleted the number of luminous satellite subhaloes surviving to $z=0$, our predictions based on Aquarius are likely to provide a lower limit to the total number of streams that Gaia will discover.

\subsection{Gas Dynamical Simulations}\label{s:gas_dynamical}

For the gas-dynamical simulations, we use a suite of `zoom' simulations of Milky Way-mass haloes using the high-resolution `Recal' model from the recent EAGLE project \citep{Schaye2015,Crain2015}.  The zoom simulations will be described in more detail in a future study (F17), so we provide only a brief description here.

We recall that the main aim of the EAGLE project was to simulate, at relatively high resolution (baryon particle mass $\approx10^6 \ {\rm M}_\odot$, softening length of 500 pc), the evolution of the main galaxy population.  The stellar and AGN feedback parameters were adjusted so as to reproduce the observed galaxy stellar mass function and the size$-$mass relation of local galaxies.  Unfortunately, the resolution of the main EAGLE box (L100N1504) is too low for our purposes, motivating our use of significantly higher-resolution zoom simulations.  Note that \citet{Schaye2015} have found that when the resolution is increased, some re-calibration of the stellar and AGN feedback is required to preserve a match to the galaxy stellar mass function.  Using this re-calibrated model (called `Recal') they have simulated a 25~Mpc volume with a factor of 8 (2) better mass (spatial) resolution (i.e., L025N0752).  This simulation volume served as the parent volume from which several haloes were selected for re-simulation.

Specifically, F17 identified a volume-limited sample of 25 haloes which fall in the mass range $7\times10^{11} < M_{200}/{\rm M}_\odot < 3\times10^{12}$ at $z=0$ ($M_{200}$ denotes the mass within the virial radius $r_{200}$).  Inspection of the visual morphologies indicates that not all of these systems have significant stellar disc components.  While such systems are interesting in their own right (and the intention is to eventually simulate all 25 haloes), priority was given to 10 systems which have the most disc-like morphology.  F17 have carried out zoom simulations with a factor of 8 (2) better mass (spatial) resolution than the parent volume (i.e., baryon particle mass of $\approx1.5\times10^{4} \ {\rm M}_\odot$, Plummer-equivalent softening length of 125 pc) using the Recal model\footnote{No additional re-calibration of the model was performed when increasing the resolution beyond that of the Recal-L025N0752 parent volume, but F17 have verified that the stellar masses of the zoomed haloes agree with those of the parent volume to typically better than 10\%.}. 
For further details of the Recal model, including a description of the employed hydrodynamic solver and subgrid prescriptions for radiative cooling, star formation, stellar and chemical evolution, and feedback we refer the reader to \citet{Schaye2015}.

In the present study we analyse a random subset of  5 of the 10 zoom simulations carried out by F17. At $z=0$, this sub-set spans virial masses $7.14\times10^{11} < M_{200}/{\rm M}_\odot < 1.93\times10^{12}$ and stellar masses $7.33\times10^{9} < M_{*}(<30~\rm{kpc})/{\rm M}_\odot < 1.99\times10^{10}$, respectively, similar to the corresponding values of the five Aquarius haloes \citep{Cooper2010}. Apart from the fact that these galaxies resemble the Milky Way in terms of total and stellar mass, the properties of their bound substructure also match the main properties of Milky Way satellites, e.g. the luminosity function and the stellar mass - metallicity relation (see also \citet{Schaye2015} for the properties of low mass galaxies in the Recal model). A more detailed investigation of the properties of these galaxies will be presented in F17. We note, however, that, due to the limited numerical resolution, these gas-dynamical zoom simulations can follow reliably only the properties of satellites in the classical dwarf galaxies regime ($M_{*}\geq 10^7 M_{\odot}$).

Following the methods described in \citet{Font2011a}, we construct simple merger histories for each of the simulated galaxies, identifying which star particles were formed `in situ' (i.e., within the main progenitor branch), which were brought in via mergers/tidal disruption of infalling satellites, and which star particles still reside in orbiting satellites at the present day.  For the star particles that were/are in satellites, we record the properties of the halo to which the particles belonged just prior to joining the main Friends-of-Friends group.

The \ljmu~simulations have the benefit of treating various baryonic physical processes, such as gas infall, star formation and stellar feedback, self-consistently. Anticipating the results, we expect that the gas-dynamical simulations will obtain a somewhat different number of tidal streams and different stream morphologies, than in the case of the particle tagging methodology. For example, the stellar feedback may change the internal spatial and kinematical distributions of stars in satellite galaxies and may transform cuspy density profiles into cored ones. This, in turn, can affect the rate at which material is tidally stripped from satellites which changes the time when tidal streams are formed and their morphological properties.  The presence of a disc may influence the spatial distribution of satellites in the inner region of the galaxy, by inducing changes in the orientation of their angular momentum and by accelerating their tidal disruption. Additionally, the hydro-dynamical gas-dynamical simulations have been shown to produce stellar haloes with dual components: accreted and in situ \citep{Zolotov2010,Font2011a,McCarthy2012,Cooper2015,Pillepich2015}. We caution that the origin of the in situ component is still debated, current gas-dynamical simulations suggesting different scenarios: stars being ejected from the disc by disc-satellites interactions, or formed in the wake of the gas stripped from infalling satellites, or formed in cold gas filaments. Understanding the origin of in situ halo stars is crucial for predicting the physical properties of this halo component and, implicitly, for modelling the environment in which tidal streams evolve. Strictly from the point of view of the detectability of tidal streams, the in situ component of the stellar halo is another source of foreground/background contamination, whose effect needs to be assessed. 

Overall, the additional effects present in gas-dynamical simulations are expected to diminish the number of tidal streams that are dynamically cold at present day and therefore, those that are most likely to be detected. We note, however, that this discussion is mainly qualitative at this point, and a more rigorous assessment of the significance of the various baryonic processes will require an in-depth quantitative investigation. This is, however, beyond the scope of this present paper since the differences in the initial conditions and numerical resolution between the two types of simulations presented here do not allow for a fair comparison. In the case of these two types of simulations, we estimate that the main differences in the number of tidal streams are most likely do to the differences in the numerical resolution. We note, however, that in the range in which the \ljmu~ simulations are able to resolve the halo substructure $-$ roughly, the domain of the classical dwarf galaxies $-$, the two types of simulations predict similar number of surviving satellites and of tidal streams.
\section{Mock Gaia catalogues}\label{s:mock_catalogues}

\subsection{Re-sampling the Simulations} 

\subsubsection{Phase space expansion of tracer particles}

We use the method described by \citet{Lowing2015} to convert the massive `star particles' in our simulations into mock catalogues of individual stars. Briefly, the steps are as follows. Star particles are partitioned into disjoint sets according to the progenitor subhalo to which they were bound at the time of infall into the Milky Way analogue halo (for this purpose, a small number of particles not bound to any halo at the time of infall and stars formed in situ are classified as a single set). The \textsc{enbid} code \citep{Sharma2006} is run separately on each of these sets to estimate the 6-dimensional phase space volume associated with every star particle (the separation into sets avoids cross-talk between different streams in this estimate). The volume identified by \textsc{enbid} is translated to an equivalent 6D Gaussian kernel. A sample of mock stars is generated from an isochrone appropriate to the stellar population represented by the parent star particle \citep[we use the \textsc{parsec} isochrones from][]{Bressan2012}  and positions and velocities assigned to each of these by randomly sampling from the kernel. The advantage of using a 6D smoothing kernel is that `thin' structures in configuration and velocity space are preserved -- mock stars are distributed preferentially `along' the streams defined by their parent particles, rather than orthogonal to them, as would be the case for an isotropic kernel. 

As described in \citet{Lowing2015}, mock catalogues of stars in the Aquarius simulations (based on a slightly updated version of the \citealt{Cooper2010} galaxy formation model) are publicly available as online databases\footnote{\url{http://virgodb.dur.ac.uk:8080/StellarHalo}}. Our Aquarius simulation catalogues were drawn from these databases according to the criteria described in the following section. Analogous catalogues for the \ljmu{} simulations were generated by applying the \citet{Lowing2015} procedure in the same way as for the particle tagging models.

\subsubsection{Stellar Tracers}\label{s:tracer_choice}

To produce \Gaia~mock catalogues we generate samples of K giants and RR Lyrae stars (RRLS), two bright stellar tracers that can be observed by \Gaia~to large distances with reasonably small proper motion errors (see Sec. \ref{s:gerrs}). 
Both tracers have been widely used in Galactic Halo surveys \citep[e.g.][and references therein]{Morrison2000,Starkenburg2009,Xue2014,Vivas2004,Sesar2009,Sesar2013}. K giants are found in any stellar population older than a few giga years, of any metallicity, and they are bright ($1<M_r<-3$) and relatively numerous \citep[e.g.]{Xue2014}. RRLS are pulsating Horizontal Branch (HB) giants that trace old ($>10$ Gyr) and metal-poor populations \citep[$\FeH<-0.5$, e.g.][]{Smith1995}; although they are sparser and not as luminous ($M_V\sim0.55$) as the brightest K giants, RRLS are well known for being excellent standard candles. 

We select K giants using the colour and $M_g$ cuts described in 
\citet[][Sec. 3.4.3]{Lowing2015}, defined by \citet{Xue2014}. These cuts select all simulated K giant stars brighter than the HB and filter out any Red Clump or red dwarf contaminants. To select RRLS we use the effective temperature and surface gravity cuts suggested by \citet[][i.e. $6100 < T_{\rm eff}\rm{(K)} < 7400$, $2.5 < \log g < 3.0$]{Baker2015}. 

When dealing with real data, the actual samples of K giants and RRLS will be prone to some degree of contamination. For K giant samples contamination can come from foreground Main Sequence dwarfs, which should be effectively filtered out as these will be nearby stars with very precise \Gaia~parallaxes (see Sec. \ref{s:gerrs}). RRLS can be very reliably identified based on their photometric variability and well known light curve shapes, so little contamination from other types of stars is expected \citep[see e.g.][]{Vivas2004,Mateu2012}. Therefore, in both cases, we expect the effect of contamination to be small and manageable.  

\subsection{The Gaia errors simulation}\label{s:gerrs}

We simulate observational errors using the \gaiaerrors~ software from \citet{RomeroGomez2015}\footnote{The code is publicly available at \url{https://github.com/mromerog/Gaia-errors}}, which implements the latest post-launch end-of-mission prescriptions provided by the Data Processing and Analysis Consortium (DPAC, \citealt{Mignard08}) and described in \cite{Rygl2014}. 

The \gaiaerrors~ code simulates Gaussian errors for the positions, parallaxes, proper motions and radial velocities, with a standard deviation that depends on the apparent magnitude and colour of each star, accounting for the ecliptic latitude dependence introduced by the \Gaia~scanning law \citep{deBruijne2012}. Reddening is simulated based on the 3D extinction maps from \citet{Drimmel2003} while the \Gaia~selection function is assumed to have 100\% completeness down to $G=20$ and $G=16$ respectively for the astrometric observables (position, parallax and proper motion) and for radial velocities  \citep{deBruijne2014}. We simulate end-of-mission combined errors for the nominal lifetime of 5 yr for the \Gaia~mission \citep{deBruijne2014}, although the \gaiaerrors~code allows for the simulation of errors at an arbitrary mission operation time. In Sec. \ref{s:gaia_extensions} we discuss the effect of possible extensions for the mission lifetime. 

In Fig.~\ref{f:Rhel_Mv_gaia} we illustrate the volume that can be probed with \Gaia~at a fixed relative precision, for stars of different brightness. The plot shows proper motion, radial velocity and parallax relative error horizons in the heliocentric distance $R_{\rm hel}$ versus absolute magnitude $M_V$ plane, and the colour scale is proportional to the apparent $G$ magnitude. Dashed, dashed-dotted and dotted lines respectively represent loci of 10, 30 and 50\% relative errors in parallax (white), proper motion (black) and radial velocity (grey). To be able to plot relative rather than absolute errors in proper motion and radial velocity, we have assumed the stars are moving at half the escape velocity $v_e$ for the corresponding distance\footnote{We assume the radial velocity is on average $v_{r}\sim v/\sqrt{3}$ and the total proper motion $\mu \sim \sqrt{2/3}v$, where we assume $v\sim v_e/2$ and approximate the escape velocity as $v_e=v_c \sqrt{2(1-\ln(R_{\rm gal}/r_t))}$, with $v_c=200$ km s$^{-1}$, $r_t=200$ kpc and $R_{\rm gal}$ the galactocentric distance.}. Therefore, for the proper motion and radial velocity, the different lines represent the \emph{best} relative precision achievable at a given distance.

For bright stars ($M_V<0$), Fig.~\ref{f:Rhel_Mv_gaia} shows how \Gaia~can achieve remarkable relative proper motion precision of $\sim30$\% for HB stars at large distances as $\sim60$ kpc, and better than $30$\% for brighter K giants ($-1.5\lesssim M_V$) beyond 100 kpc. For Main Sequence Turn-Off stars (MSTO) the relative proper motion precision will be $\lesssim10$\%, but these stars are only bright enough to be observable up to $\sim 20$ kpc. 

Radial velocities will be available for the brightest stars with $G\leqslant16$ (light to dark blue areas in the figure), all of which will have radial velocity relative errors smaller than $30$\%. The maximum distance for HB and Red Giant Branch stars with radial velocities will be $\sim15$ kpc and $\sim30-60$ kpc respectively. The limiting magnitude of $G=16$ for \Gaia~radial velocities is $\sim1$ mag brighter than what was originally expected for the mission, due to the increased background caused by stray light \citep{deBruijne2014}, which is also already taken into account in the errors shown in Fig.~\ref{f:Rhel_Mv_gaia}. 
 
\Gaia~parallaxes will be useful only up to $\sim10$ kpc for most stars, and even for the very brightest giants ($M_V\gtrsim-2$) reasonable errors ($<50$\%) will be achieved only up to $\sim20$ kpc. Hence, photometric distance measurements for standard candle tracers will be crucial to probe the outer Halo.

\subsubsection{Photometric Distances}\label{s:photdists}

\begin{figure}
\begin{center}
 \includegraphics[width=1.05\columnwidth]{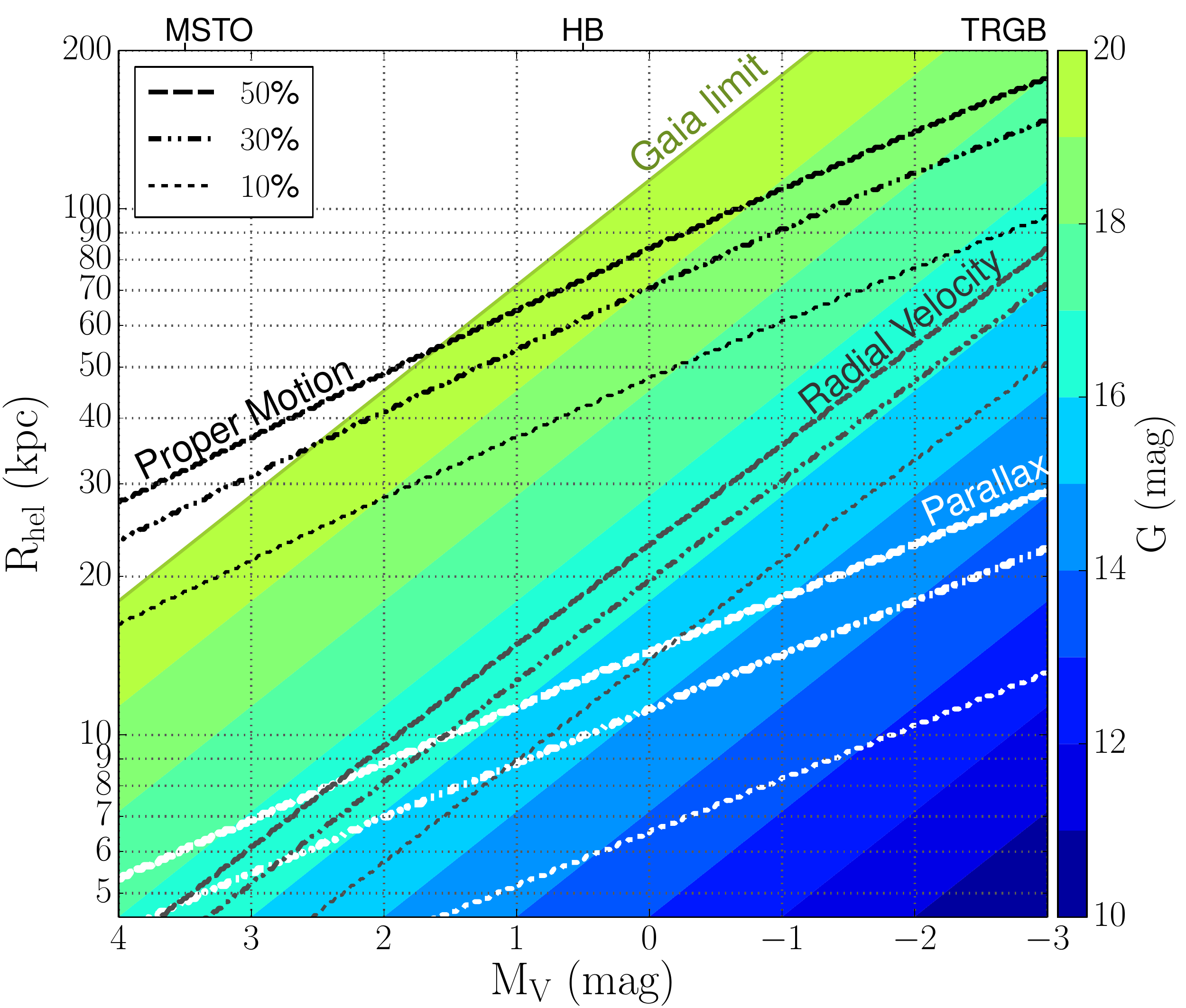} 
 \caption{\Gaia~observational error horizons in the heliocentric distance $R_{\rm hel}$ versus absolute magnitude $M_V$ plane. The colour scale is proportional to the apparent $G$ magnitude and goes up to the \Gaia~magnitude limit ($G\leqslant20$), assuming a fixed $V-I=1$ colour and $A_V=0$. Only stars brighter than $G=16$ will have radial velocity measurements (light to dark blue areas). Dashed, dashed-dotted and dotted lines respectively represent loci of 10, 30 and 50\% relative errors in parallax (white), radial velocity (grey) and proper motion (black). The absolute magnitudes of Main Sequence Turn-Off (MSTO), Horizontal Branch (HB) and Tip of the Red Giant Branch (TRGB) stars are shown for reference on the top axis. The error estimates, as well as the limiting magnitudes cited, already take into account the stray light effect \citep{deBruijne2014}.}
\label{f:Rhel_Mv_gaia}
\end{center}
\end{figure}

Fig.~\ref{f:Rhel_Mv_gaia} illustrates that \Gaia~parallax errors will be prohibitively large beyond a few kpc. However, photometric distances can be estimated for the stellar tracers we have selected, with much better precision. 

RRLS are well known standard candles, for which relative errors in distance are as low as $\sim7$\%, or even  $5$\%, if there is a relatively small (0.3 dex) uncertainty in metallicity \citep{Vivas2006,Mateu2012,Sesar2013}. 
For K giants, the dependence of $M_V$ on colour and metallicity makes photometric distance determinations more challenging. \citet{Xue2014} find that these can be estimated with a 16\% median error, based on $gr$ photometry and spectroscopic metallicities from SEGUE, using their probabilistic algorithm. \citet{Liu2014} use 2MASS photometry and LAMOST spectra and get a mean distance error of $\sim30$\%, which they attribute to the shallower photometry used in their procedure.

In what follows we simulate photometric distances in our mock catalogues, with a constant (Gaussian) error of 7\% for RRLS, and an intermediate value of 20\% for K giants.

\section{What Gaia can see}\label{s:gaia_can_see}

What \Gaia~will be able to `see' will ultimately be determined by the combination of different factors: the selected tracer,
the effect of extinction which will depend on the line of sight and, for a chosen proper motion precision, the actual
velocity distribution of the stars in the different tidal streams. Hence, the previous section and Fig.~\ref{f:Rhel_Mv_gaia} provide a simplified description.

\begin{figure}
\begin{center}
 \includegraphics[width=1.0\columnwidth]{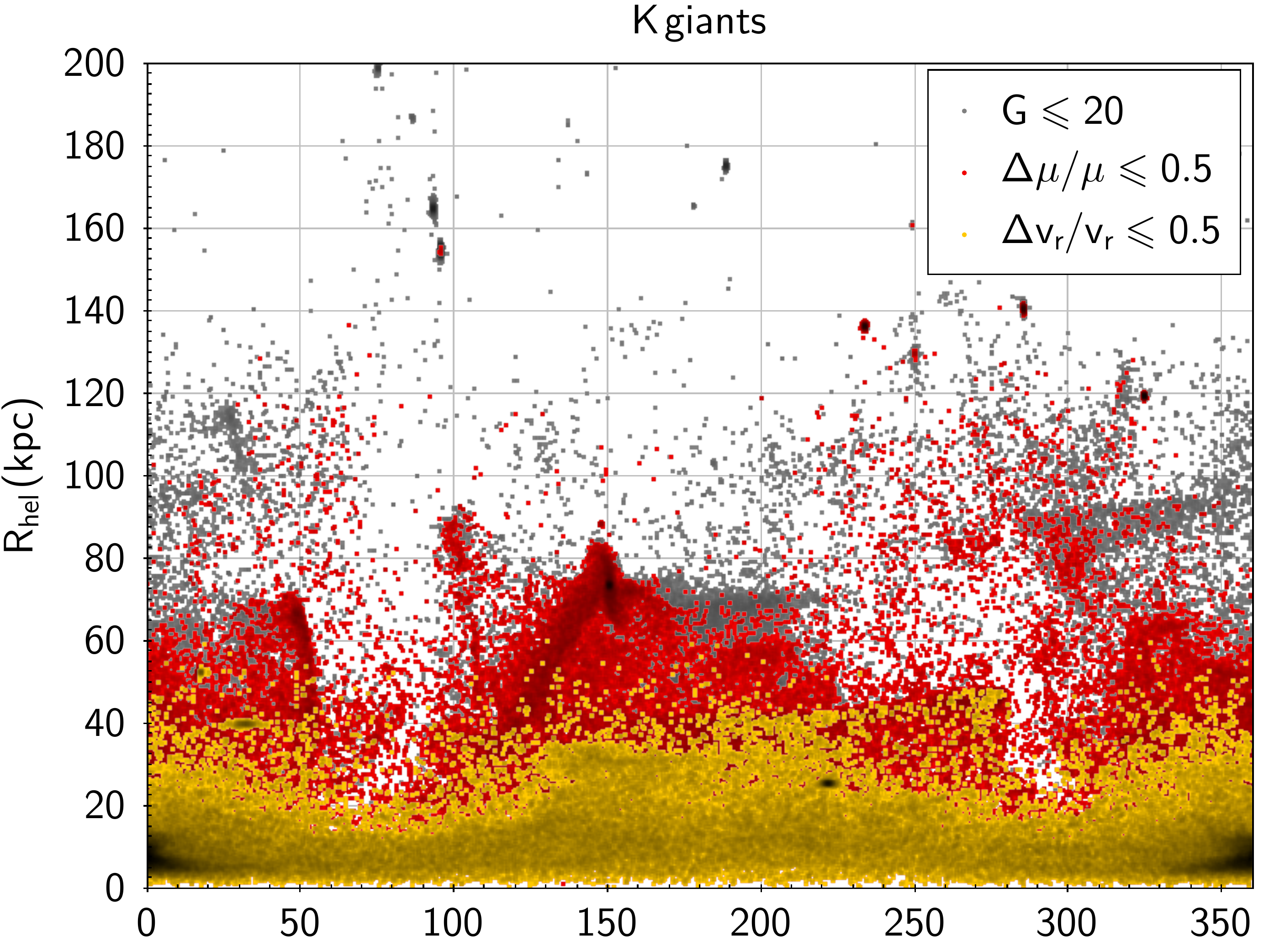} 
 \includegraphics[width=1.0\columnwidth]{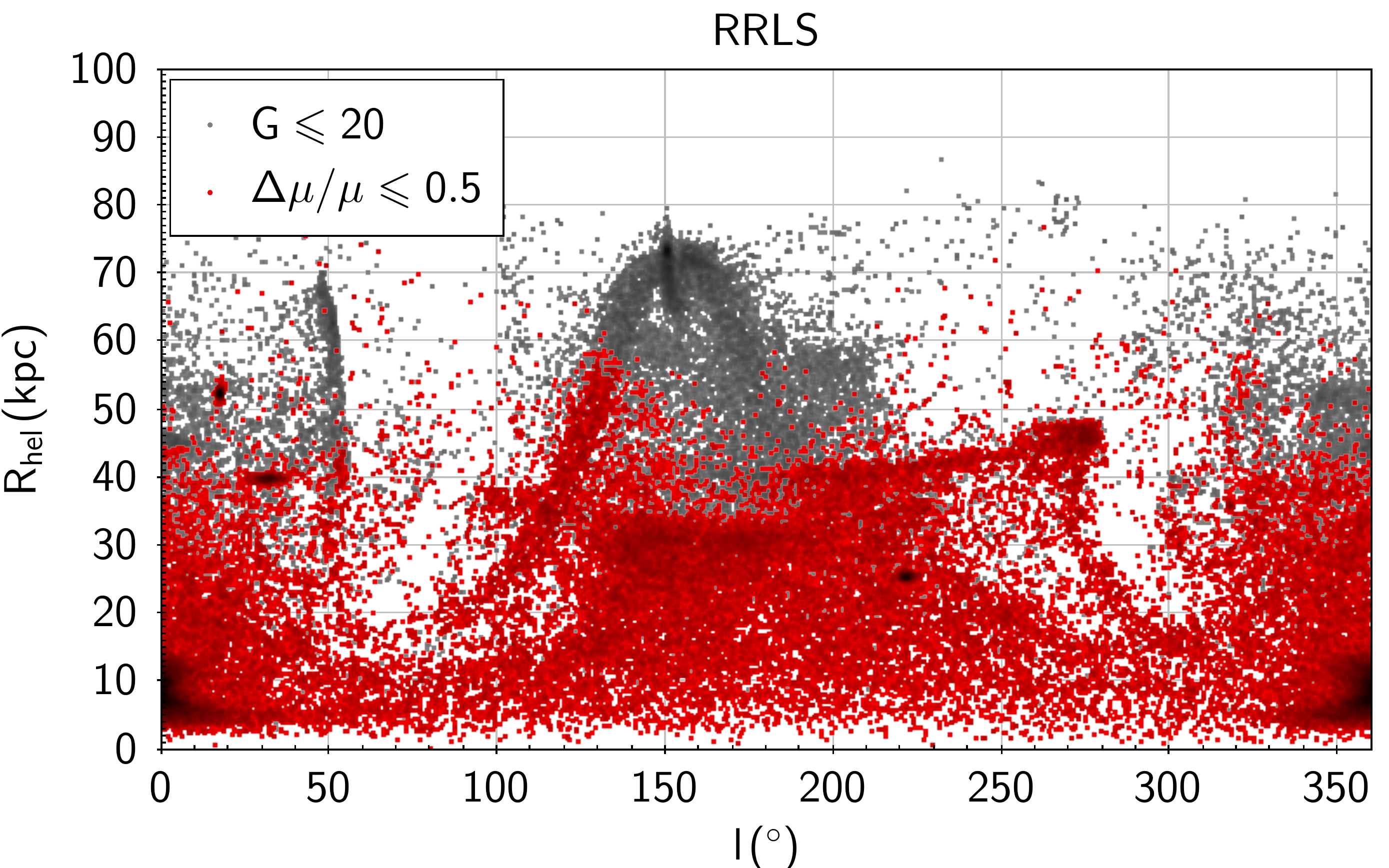} 
 \caption{Heliocentric distance $R_{\rm hel}$ versus galactic longitude $l$ for K giants (\emph{top}) and RRLS (\emph{bottom}) in the Aquarius A2 mock catalogue. \Gaia~observable stars are shown in grey, and stars with relative proper motion and radial velocity errors $<50$\% in red and ochre respectively.}
\label{f:lRhel}
\end{center}
\end{figure}

To illustrate this in a more realistic case, Fig.~\ref{f:lRhel} shows two mock \Gaia~catalogues of the Aquarius A2 Halo produced as we have described in Sec. \ref{s:gerrs}: the upper panel for K giant stars, the bottom panel for RRLSs. The plot shows heliocentric distance $R_{\rm hel}$ versus galactic longitude $l$ for:  all stars observable by \Gaia~(grey) and stars for which \Gaia~proper motions (red) and radial velocities (ochre) have relative errors better than 50\%. In this case we have used each star's own individual proper motion and radial velocity to compute the relative errors. 

In the upper panel we can see that most structures up to about 100 kpc are traced by K giants with good proper motions. Beyond this, there are a few K giant stars with good proper motions observable even as far as $\sim$150 kpc, in the denser structures which are more likely to host more of the intrinsically brightest K giants. On the other hand, note also that although all structures are very well traced by K giants below 80 kpc, there is a severe lack of observable stars with good proper motions in the tidal arm at $(l,R_{\rm hel})\sim(200 \degr, 70\, \mathrm{kpc})$. This is a case of a stream that happens to have most of its velocity along the line of sight (not shown), and so for the typical proper motion precision attainable at this distance, the fractional proper motion error is larger than the imposed cut of 50\%. The volume that can be probed with K giants including radial velocities reaches out to $\sim40$ kpc on average.

The volume that can be probed with RRLSs with good proper motions reaches out to $\sim40$ kpc. Note that, incidentally, this is roughly the same volume inside which K giants will have full 6D information with good precision. So structures in this volume can in principle be tracked down with both tracers. We do not show the analogous coverage for RRLS radial velocities as the \Gaia~standard errors are end-of-mission error prescriptions for the combined spectra, and so do not apply for pulsating stars. 
 For RRLS and other pulsating variables single epoch spectra must be used in order to account for the pulsation component in the radial velocity, however no error prescriptions are yet provided by DPAC for radial velocities from single epoch spectra. 

\section{Great Circle Methods}\label{s:mgc3}

The Great Circle Cell Counts method (GC3) was proposed by \citet{Johnston1996} to search for tidal streams in the Galactic Halo, by using the fact that stars that belong to a tidal stream produced in a spherical potential orbit in a fixed plane as the total angular momentum is conserved. This means that a tidal stream will lie approximately in a great circle band, which is the projection of the orbital plane onto the galactocentric celestial sphere. In fact, the idea of searching for great circle alignments had already been introduced by \citet{LyndenBell1995}, who looked at the intersection of great circles of orbital poles to search for alignments of dwarf galaxy satellites and globular clusters along great circles. These authors even proposed a way to include kinematic information, by assuming energy and angular momentum conservation.

In light of the (then) upcoming \Gaia~astrometric mission, starting with M11 we extended GC3 into a family of great-circle-cell methods which includes kinematic information: adding a total velocity criterion in M11 (the \mgc~method) and a proper-motion-only version in \citet{Abedi2014} (the \ngc~method).

\subsection{\gc,~\mgc~ and \ngc}\label{s:nmgc3}

In general, the GC3 family of great-circle-cell methods is defined by the use of a geometric selection criterion to choose stars along a great circle band in the sky, orthogonal to a particular direction marked with what is called its \emph{pole} vector $\bm{\hat{L}}$. The number of stars that satisfy this criterion are counted and assigned to this particular pole vector. The pole vector is then changed in direction following the nodes of a spherical coordinate mesh in the sky, and the whole operation is repeated at each node. This produces the so called \emph{pole-count maps} (hereafter PCMs) where the number of stars at each mesh node is indicated. Maxima in this map indicate the presence of substructure. To decide whether a star is associated to a given pole $\bm{\hat{L}}$ we use the following position and velocity criteria (Eq. 6 in M11):

\begin{equation}\label{e:rcriterion}
| \bm{\hat{L}} \cdot \bm{r'}_\mathrm{gal}| \leq \Vert \bm{r'}_\mathrm{gal} \Vert\delta_c
\end{equation}
\begin{equation}\label{e:vcriterion}
|\bm{\hat{L}} \cdot \bm{v'}_\mathrm{gal}| \leq  \Vert \bm{v'}_\mathrm{gal} \Vert \delta_c 
\end{equation}

\noindent where $\delta_c=\sin\delta\theta$ is sine of the tolerance that allows for the width $\delta\theta$ of each great circle band and $\bm{r'}_\mathrm{gal}$ and $\bm{v'}_\mathrm{gal}$ are simply the galactocentric position and velocity vectors $\bm{r}_\mathrm{gal}$ and $\bm{v}_\mathrm{gal}$, multiplied by the parallax\footnote{This is done in order to avoid using the reciprocal of the parallax which would introduce a bias. See M11.}, which in terms of the heliocentric observables $(l,b,\varpi,\mu_l,\mu_b,v_{r})$  are given by

\begin{equation}\label{e:rgal_vgal_prime}
\begin{array}{l}
 \bm{r'}_\mathrm{gal} = \varpi\bm{r}_\odot + A_p(\cos{l}\cos{b}) \bm{\hat{x}}+(\sin{l}\cos{b}) \bm{\hat{y}}+(\sin{b}) \bm{\hat{z}} \\
\bm{v'}_\mathrm{gal} =  \varpi\bm{v}_\odot + \varpi v_r \bm{\hat{r}} + (A_v \mu_l \cos{b}) \bm{\hat{l}}+ (A_v\mu_b) \bm{\hat{b}}
\end{array} 
\end{equation}

\noindent where $A_p=10^3$ mas$\cdot$pc, $A_v=4.74047$ yr \kms; $\lbrace \bm{\hat{x}},\bm{\hat{y}},\bm{\hat{z}} \rbrace$ are the unit vectors in the cartesian Galactocentric reference frame and $\lbrace \bm{\hat{r}},\bm{\hat{l}},\bm{\hat{b}} \rbrace$ are the unit vectors in a spherical heliocentric reference frame (see Appendix A in M11).

\emph{A significant advantage of the GC3 family of methods is that the computation of pole count maps works directly in observable space} (see M11 for a detailed description), instead of working with physical quantities like velocity, energy or angular momentum, 
for which errors are propagated in complicated ways because of the non-linearity of the transformations involved. 

Each of the variants in the GC3 method family\footnote{The Python toolkit PyMGC3 provides an implementation of the GC3 family of methods and is publicly available at \url{https://github.com/cmateu/PyMGC3}} associate stars to poles with different combinations of the criteria in Eq. \ref{e:rcriterion} and \ref{e:vcriterion}:

\begin{itemize}
\item GC3: 3D positional information only (Eq. \ref{e:rcriterion})
\item mGC3: 3D position and 3D velocity  (Eqs. \ref{e:rcriterion} \& \ref{e:vcriterion}) 
\item nGC3: 3D position and proper motion (Eqs. \ref{e:rcriterion} \& \ref{e:vcriterion}, without the $\varpi v_r \bm{\hat{r}}$~ term in Eq. \ref{e:rgal_vgal_prime})

\end{itemize}

As we have shown in Secs. \ref{s:gerrs} and \ref{s:gaia_can_see}, demanding radial velocities will severely restrict the volume we can probe with \Gaia. In what follows we will use the nGC3 method, which only requires proper motions, and we will also limit our samples to stars with proper motion relative errors less than 50\%, as we will discuss in more detail in Sec. \ref{s:pcm_a2}. 

Using proper motions alone and disregarding radial velocities has the advantage of allowing us to probe a much larger volume of the halo, as we have shown in Fig.~\ref{f:lRhel}. Also, for distant streams ignoring $v_r$ makes no difference; as the Sun-GC distance becomes negligible, the radial component of the velocity is approximately contained in the orbital plane by construction, so its contribution to the dot product in Eq. \ref{e:vcriterion} will tend to zero. On the other hand, ignoring $v_r$  will affect the contribution of contaminants to the PCMs. For planes roughly perpendicular to the Sun-GC direction ($\phi_{\rm pole}\sim0^\circ,180^\circ$), the line of sight is off the plane, so in these directions fewer fore/background contaminants will be filtered, and the PCM background level will be higher than for planes going through the Sun-GC line  ($\phi_{\rm{pole}}\sim90^\circ,270^\circ$).

\subsection{Signatures of Individual Streams in PCMs}\label{s:indiv_streams}

\begin{figure*}
\begin{center}
 \includegraphics[width=2.2\columnwidth]{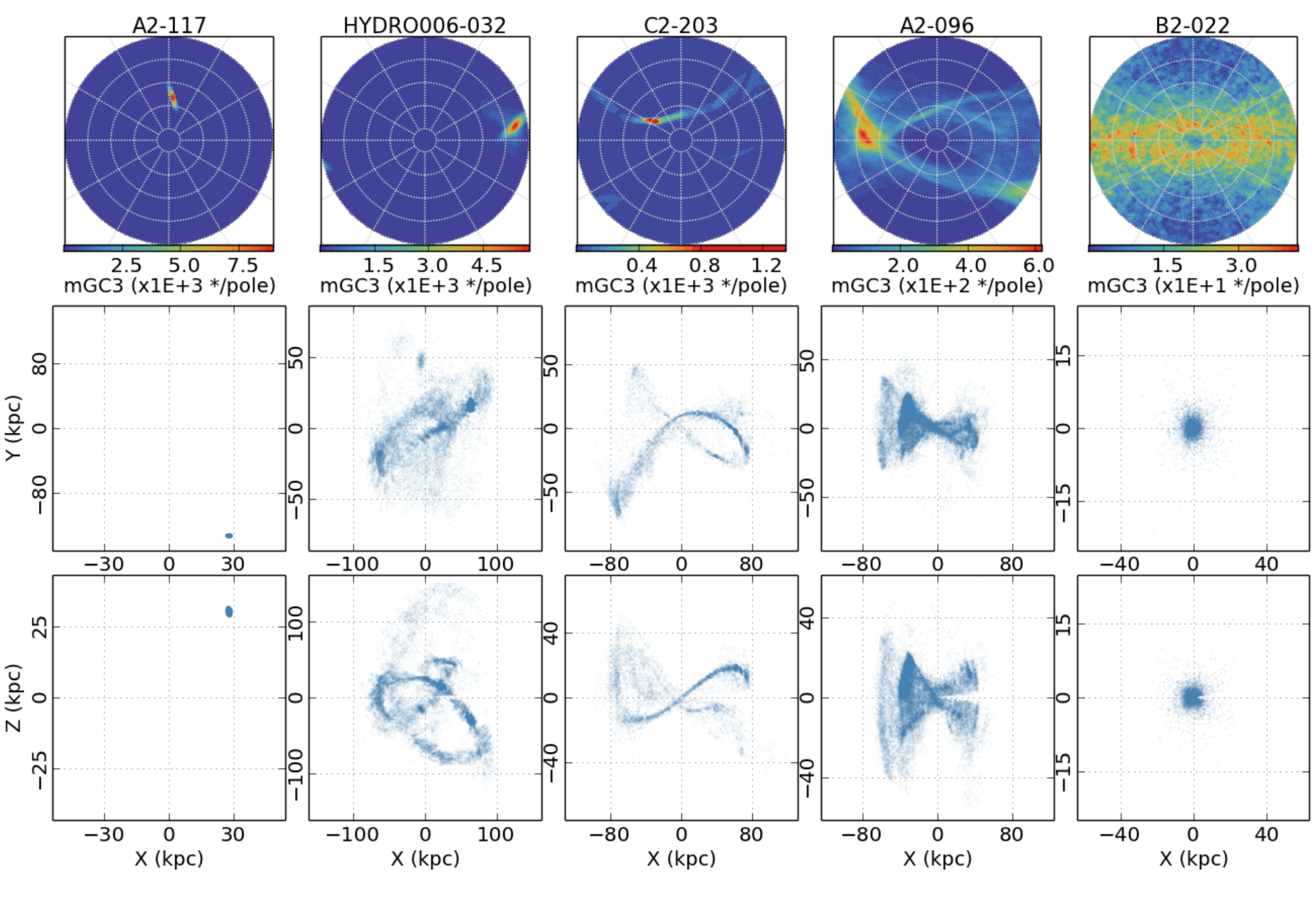} 
 \caption{PCMs and spatial distributions for different progenitors in the Aquarius and \ljmu~ K giants mock catalogues. The five progenitors shown go from completely bound to completely disrupted from left to right. The top, middle and bottom rows show respectively the \ngc~PCMs, with a colour scale proportional to the number of stars per pole; the face-on Y vs X projection and the edge-on Z vs X projection, where the colour shade is proportional to density. Only \Gaia~observable stars are shown in these plots. The Sun is located at $\mathrm{X}=-8.5$~kpc and $Z=0$ is the plane of the disc (in the Aquarius simulations the X and Y axes correspond to the major and intermediate axes of the potential). PCMs are shown in a north-polar azimuthal projection. }
\label{f:nongc_exs}
\end{center}
\end{figure*}

In this section we will illustrate how streams produced in cosmological simulations do in fact produce recognisable peaks in the PCMs. This is a crucial test, as in all previous applications of the \gc~methods we have used N-body simulations in a fixed axisymmetric potential, where substructure may be unrealistically enhanced against a smooth background. In cosmological simulations the haloes are gradually assembled through time and so the potential is neither fixed nor even necessarily axisymmetric, so it is not obvious that the GC3 methods can still be applied in this case (note that this is also the case for the real Milky Way). 

The morphology of PCMs for different progenitors is illustrated in Fig.~\ref{f:nongc_exs}. Five progenitors at different stages of disruption were chosen from the Aquarius and \ljmu~haloes. In the figure, each column corresponds to one progenitor; for each, the top row shows the corresponding \ngc~PCM, the middle and bottom rows two orthogonal projections: Y vs X and Z vs X respectively. The \ngc~PCMs are plotted in a north-polar azimuthal equatorial projection  showing the north pole at the centre; the concentric circles are parallels drawn at $20\degr$ intervals, and meridians are drawn at  $30\degr$ intervals in longitude with $\phi = 90\degr,180\degr$ in the right horizontal and top vertical axes at the centre of the plot. The PCMs were computed with a tolerance $\delta\theta=1\fdg5$ on a uniform grid with $1\degr$ spacing. 

The first column (A2-117) shows the \ngc~PCM for a completely bound progenitor. The signature is a very localised peak in pole counts around the orbital plane's pole, thanks to the use of kinematical information (proper motions in this case). Not all bound progenitors will necessarily produce such a well localised peak in an \ngc~or \mgc~PCM. The maxima in the pole counts will tend to stretch more and more along a great circle for more radial orbits, simply due to the geometric effect of the orbit collapsing into a line for the case of a perfectly radial orbit. The second column (HYDRO006-032) shows a tidal stream that has been largely disrupted but still produces a  strong main peak with a second, less prominent lobe. The third column (C2-200) shows a more complex morphology where there is u-shaped maximum. 
Here, two effects are in play: orbital precession causes the u-shape, and the low angular momentum of the orbit causes its stretching, as explained before.  In the fourth column example (A2-096) the spatial morphology is more shell-like and the corresponding signature in the PCM more intricate, although there is still a recognisable maximum. Finally, the last column (B2-022) shows a completely phase-mixed event that produces no discernible or significant maximum. These events will end up contributing to the PCM background.   

\section{Full Halo Pole Count Maps: A fiducial example}\label{s:full_PCMs}

In this section we will show in detail how the morphology of the PCMs is affected by the \Gaia~selection function and observational errors (Sec. \ref{s:pcm_a2} ), how the peaks are detected (Sec \ref{s:peak_detection}) as well as the effect of the choice of tracer (Sec. \ref{s:tracer}). For clarity, we consider a single fiducial halo in this section as an illustrative example of these different aspects of the method. In Sec. \ref{s:all_pcms} we will describe the results obtained for all the Aquarius and \ljmu~haloes.

\subsection{Selection function and errors}\label{s:pcm_a2}

\begin{figure*}
\begin{center}
 \includegraphics[width=2.2\columnwidth]{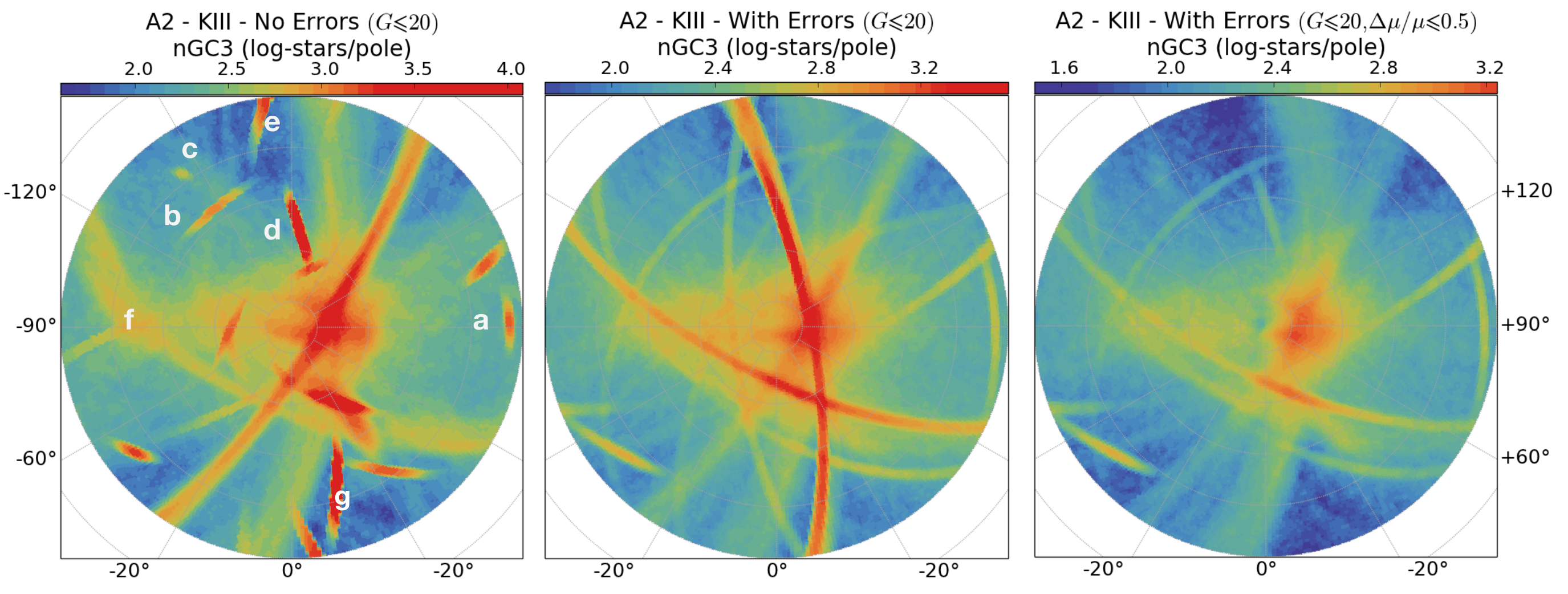} 
 \caption{\ngc~PCMs for \Gaia~observable stars in the Aquarius A2 Halo. \emph{Left}: without errors ($G\leqslant20$). \emph{Middle}: with errors ($G\leqslant20$). \emph{Right}: with errors, after proper motion error cut ($G\leqslant20,\Delta\mu/\mu\leqslant0.5$).  The three panels have the same colour scale limits. Some peaks have been labelled in the left panel for reference (see discussion in text). PCMs are shown in a north-polar azimuthal projection. Meridians and parallels (light grey) are drawn at uniform $30\degr$ and $20\degr$ intervals, their corresponding labels are shown where they intersect the vertical and horizontal axes respectively.}
\label{f:pcm_comp}
\end{center}
\end{figure*}

The \ngc~PCM for all \Gaia~observable stars ($G\leqslant20$) in the Aquarius A2 halo is shown in the left panel of Fig.~\ref{f:pcm_comp}. Some peaks have been labelled in this plot to facilitate discussion. The middle panel shows the effect of adding the \Gaia~selection function and observational errors. The right panel shows the PCM after a cut in the proper motion errors has been imposed ($\Delta\mu/\mu\leqslant0.5$). The colour scale is proportional to the logarithm of the star counts in each cell. The tolerance used to compute all \ngc~PCMs hereafter is $\Delta\theta=1\fdg5$. 

Several well-defined, localised peaks are very noticeable, as well as some other more extended features corresponding to streams that have undergone more significant phase-mixing. The two examples from the Aquarius A2 halo shown in Fig. \ref{f:nongc_exs} (1st and 4th columns) are easily recognisable here as peaks $\rm{d}$ and $\rm{f}$. 

When the observational errors are added, as illustrated in the middle panel of this figure, some peaks 
that were well-defined are now stretched along great circles to different degrees (e.g. $\rm{a}$, $\rm{b}$) and some appear a bit more fuzzy (e.g.  $\rm{c}$). The stretching is mostly due to the degradation of the kinematic information, particularly in cases where the progenitor is either mostly bound (as peak $\rm{d}$, corresponding to A2-117 in Fig. \ref{f:nongc_exs}) or when few stars in the tidal stream can be detected. The use of bad proper motion data causes a given peak to be stretched out along the great circle defined by the poles of all possible planes that go through the clump and the Galactic centre. Another way to look at this is to think of \ngc~PCMs as tending towards their \gc~counterparts as the precision of the kinematic data worsens. An effect due to distance errors is also present, but here it is minimized by our choice of tracers with reasonably small photometric distance errors (Sec. \ref{s:photdists}). 
This severe stretching of the peaks into great circles is problematic because it will increase the contamination in the features detected,  and will also make the detection of spurious peaks more likely at the intersection of great circles.
To mitigate these effects, we keep only those stars with proper motion errors less than 50\%. The resulting PCM is shown in the right panel of Fig.~\ref{f:pcm_comp}. Some features are inevitably lost because some progenitors do not give rise to streams that have stars sufficiently close or bright enough to have \Gaia~proper motions with errors smaller than our 50\% cut. That is the case of peak $\rm{e}$ in the left panel of Fig.~\ref{f:pcm_comp}, which has completely disappeared in the right panel. On the other hand, some other features like $\rm{d}$ and $\rm{g}$ would merge into one great circle if the cut were not imposed, so we believe this relatively relaxed cut offers a good compromise.

\subsection{Detecting peaks in PCMs}\label{s:peak_detection}

\begin{figure*}
\begin{center}
 \includegraphics[width=2.2\columnwidth]{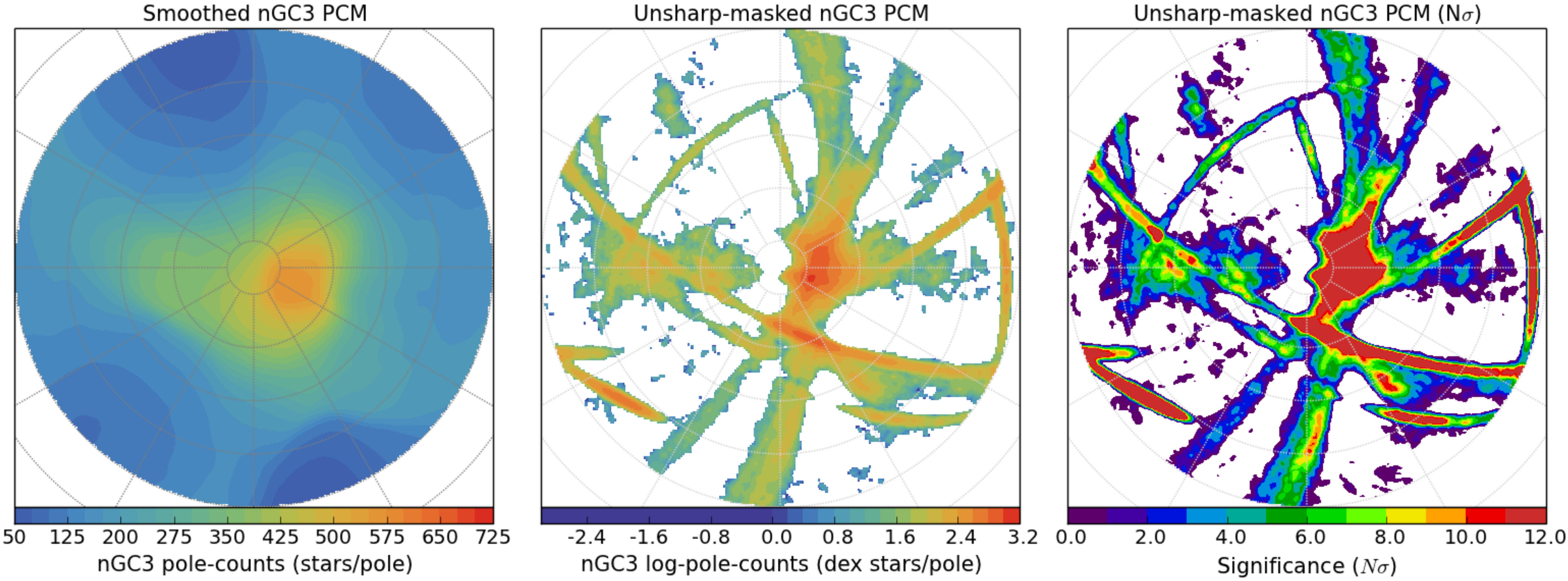} 
 \caption{The process of unsharp-masking illustrated for the \ngc~PCM of Aquarius Halo A2, with the proper motion error cuts (right panel Fig.~\ref{f:pcm_comp}). \emph{Left:} Smoothed (median-filtered) PCM. \emph{Middle:} Unsharp-masked PCM computed by subtracting the smoothed map from the original one. \emph{Right:} Unsharp-masked PCM in $N\sigma$ units, computed dividing the subtracted PCM by the square root of the smoothed PCM.}
\label{f:aqa2_unsharp}
\end{center}
\end{figure*}

We begin by first removing the contribution of the smooth background by \emph{unsharp-masking}, as in M11. A smoothed map is produced by applying a median filter to the PCM, assigning to each pixel the median counts computed in a neighbourhood of fixed size, selected to be $\sim20\degr-22\degr$ ($\sim15$ times the great circle tolerance $\delta\theta$), i.e. much larger than the typical size of the peaks one is interested in finding. The left panel in Fig.~\ref{f:aqa2_unsharp} shows the smoothed PCM for the Aquarius A2 Halo example in Fig. \ref{f:pcm_comp} (right panel). This smoothed map reflects the contribution of the well-mixed halo background stars to the PCM, with the effect of the selection function folded in. The middle panel of Fig. \ref{f:aqa2_unsharp} shows the unsharp-masked PCM, obtained by subtracting the smoothed map in the left panel. In this unsharp-masked PCM the peaks are clearly highlighted. The colour scale in this panel is proportional to the log-counts. To give a sense of the significance of the peak height with respect to the background, the right panel in Fig. \ref{f:aqa2_unsharp} shows the unsharp-masked PCM now in $N\sigma$ units. This is computed dividing pixel-by-pixel the unsharp-masked PCM (middle panel) by the square root of the smoothed PCM (left panel), which assumes the pole counts follow a Poisson distribution. 

Peaks are detected in the unsharp-masked PCM using the FellWalker\footnote{The FellWalker algorithm is part of the \href{http://starlink.eao.hawaii.edu/starlink}{Starlink Software Distribution}.} algorithm from \citet{Berry2015}. As explained in detail in this reference, Fellwalker uses a watershed algorithm that divides the pixels in an image into disjoint clumps, each of which contains one local maximum. This is done only for those pixels above some noise threshold, so background pixels below it are not assigned to any clump. This is a very efficient and general algorithm that allows detecting peaks without any particular shape, a crucial point since stream signatures in PCMs can significantly differ from simple Gaussian peaks as illustrated in Fig.~\ref{f:nongc_exs}.

Ideally one would want the peak detection algorithm to exploit the fact that peak signatures in PCMs tend to stretch along great circles (see Fig. \ref{f:nongc_exs}), particularly since very elongated features are frequently fragmented into multiple peaks by the detection algorithm. However, implementing this is out of the scope of the present paper, so we defer it for a future work. To reduce this excessive fragmentation we simply apply the position-only  \gc~method (see Eq.\ref{e:rcriterion} and Sec. \ref{s:nmgc3}), but this time on the $\phi-\theta$ coordinates of the FellWalker peak detections in the PCM themselves. This way we can merge peaks for which the majority of pixels lie on great circles within a tolerance of $\sim1^\circ$. This last step is done in an interactive way to ensure only the detections that lie along the most obvious great circles are merged into a single detection. The end result of the peak detection is shown in the top panel of Fig.~\ref{f:pcm_peaks_tracer} for the Aquarius A2 halo PCM, where each of the identified peaks is marked with a labelled circle. 

\subsection{The choice of tracers}\label{s:tracer}

\begin{figure}
\begin{center}
 \includegraphics[width=0.85\columnwidth]{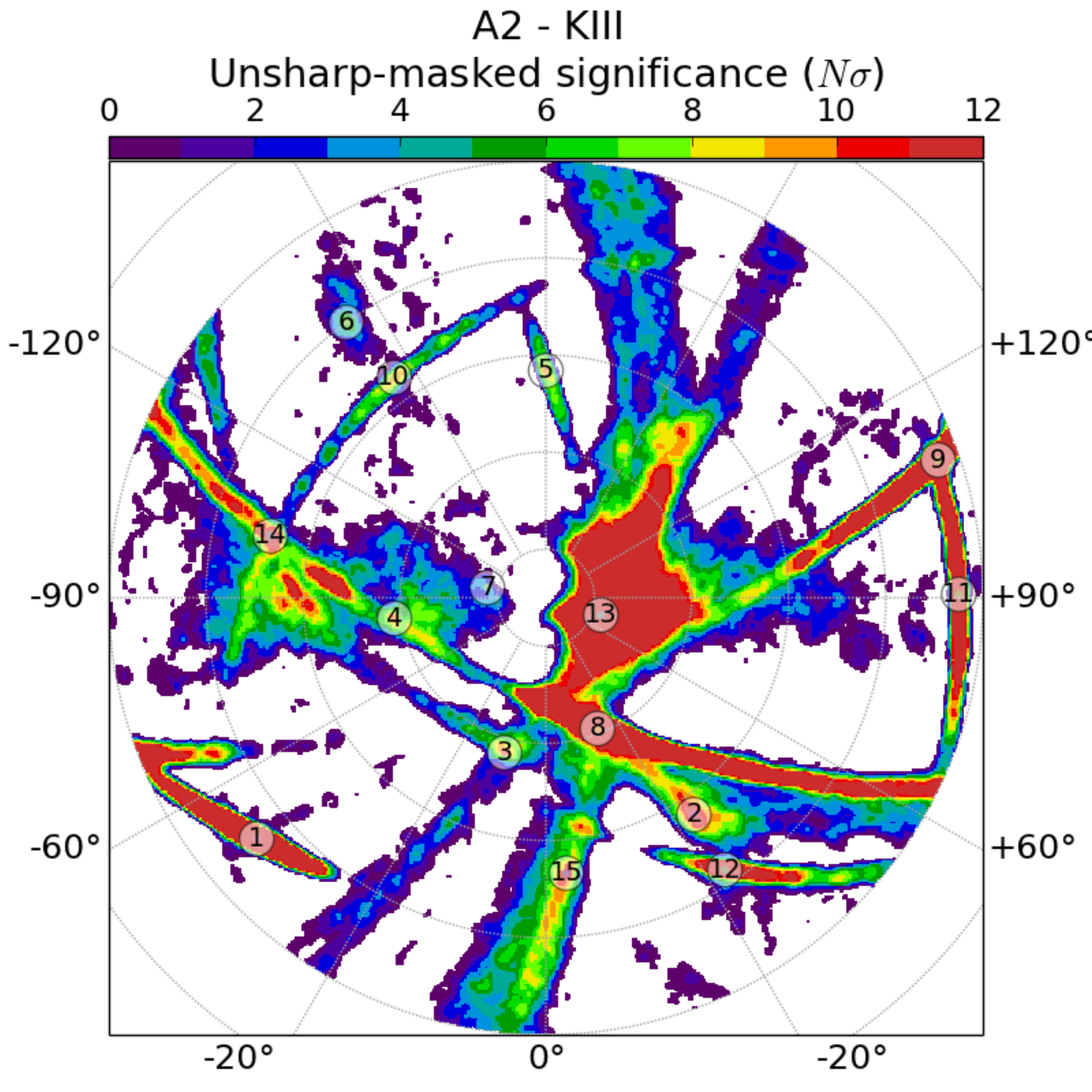}
 \includegraphics[width=0.85\columnwidth]{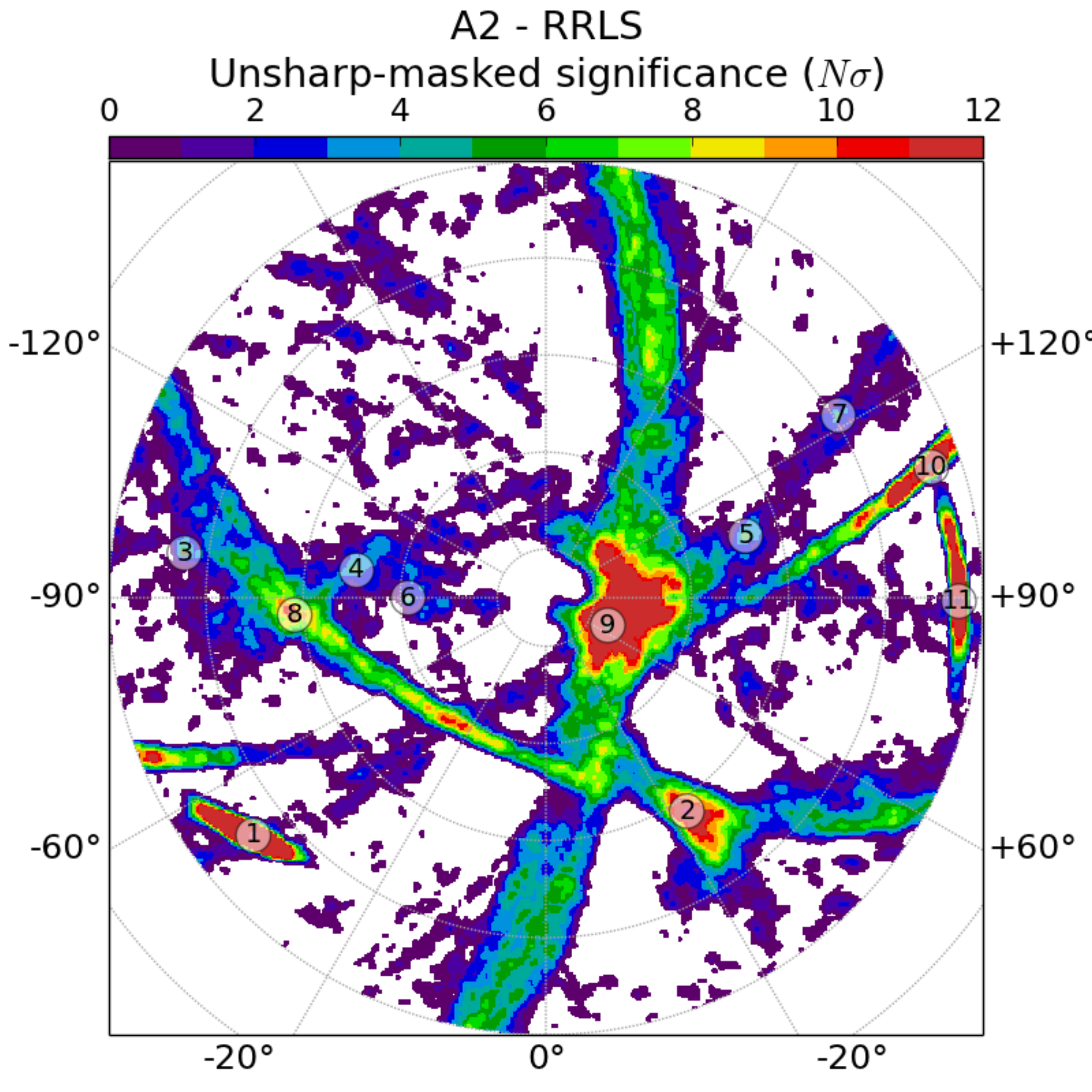}  
 \caption{Peak detections in the unsharp-masked \ngc~PCM for the Aquarius A2 halo. \emph{Top:} K giants. \emph{Bottom:} RRLS. Individual peaks are labeled.  The colour scale corresponds to the pixel's significance in $N\sigma$ units.}
\label{f:pcm_peaks_tracer}
\end{center}
\end{figure}

The choice of tracer will have different effects in the PCM. To illustrate this, the top and bottom panels of Fig.~\ref{f:pcm_peaks_tracer} show respectively the \ngc~PCM for the Aquarius A2 halo K giants and RRLS, observable with \Gaia~with errors and after the cut in relative proper motion error. The numbers of both tracer stars per halo, as observable by \Gaia, with and without the proper motion error cut, are summarised in Table \ref{t:NK_aq}.

RRLS are less numerous than K giants by factors ranging from $\sim$4 to 10 (see Table \ref{t:NK_aq}). Consequently, the RRL PCMs appear noisier than their K giant counterparts due to Poisson noise. RRLS are also fainter than K giants (the latter are giants brighter than the HB), so several peaks for the most distant structures that are observable with K giants are absent in the RRLS PCM; e.g. peaks 5, 6, and 12 in the top panel are not present in the bottom one. 
On the other hand, typical RRLS distance errors are smaller (see Sec. \ref{s:tracer_choice}) producing sharper features in the PCMs for the progenitors that do contain observable stars, e.g. compare peaks 1 and 11 in the top panel to peaks 1 and 11 in the bottom panel respectively. Producing more concentrated features could also translate into some peaks being detectable with RRLS and not with KIII stars, either because they are easier to resolve or because the contrast between the peak signal and the background noise is larger, as is the case for peak 5 which is just detectable in the bottom panel and not in the top one. 

The relative importance of these competing effects cannot be gauged a priori, as it will depend on the particular accretion history of each halo, as we will show in Sec. \ref{s:all_pcms}. Simply, each tracer offers its own advantages: K giants can probe a larger halo volume, whereas RRLSs can provide a more detailed view within the inner halo. 

\begin{table}
\caption{Number of K giants and RRLS observable by \Gaia~and with relative proper motion errors smaller than~$50\%$,   for the Aquarius haloes.}\label{t:NK_aq}
\centering
\tabcolsep=0.11cm
\begin{tabular}{lrrrr}\hline\hline
     & \multicolumn{2}{c}{KIII}  & \multicolumn{2}{c}{RRLS} \\
  \cmidrule(r{0.5em}l{0.5em}){2-3} \cmidrule(r{0.5em}l{0.5em}){4-5}
Halo & $G\leqslant20$ & $\Delta\mu/\mu\leqslant0.5$ & $G\leqslant20$ & $\Delta\mu/\mu\leqslant0.5$ \\
\hline
A2   &     458,626 &    283,419 &    98,870   &     64,126 \\    
B2   &     771,795 &    641,795 &   140,511   &    114,983 \\   
C2   &     511,452 &    282,512 &    75,506   &     45,176 \\    
D2   &   1,815,093 &  1,103,009 &   264,739   &    190,288 \\   
E2   &   1,586,759 &  1,338,395 &   155,834   &    132,549 \\   
\hline
\end{tabular}
\end{table}

\section{Full Halo Pole Count Maps: All Haloes}\label{s:all_pcms}

\begin{figure*}
\begin{center}
\includegraphics[width=0.52\columnwidth]{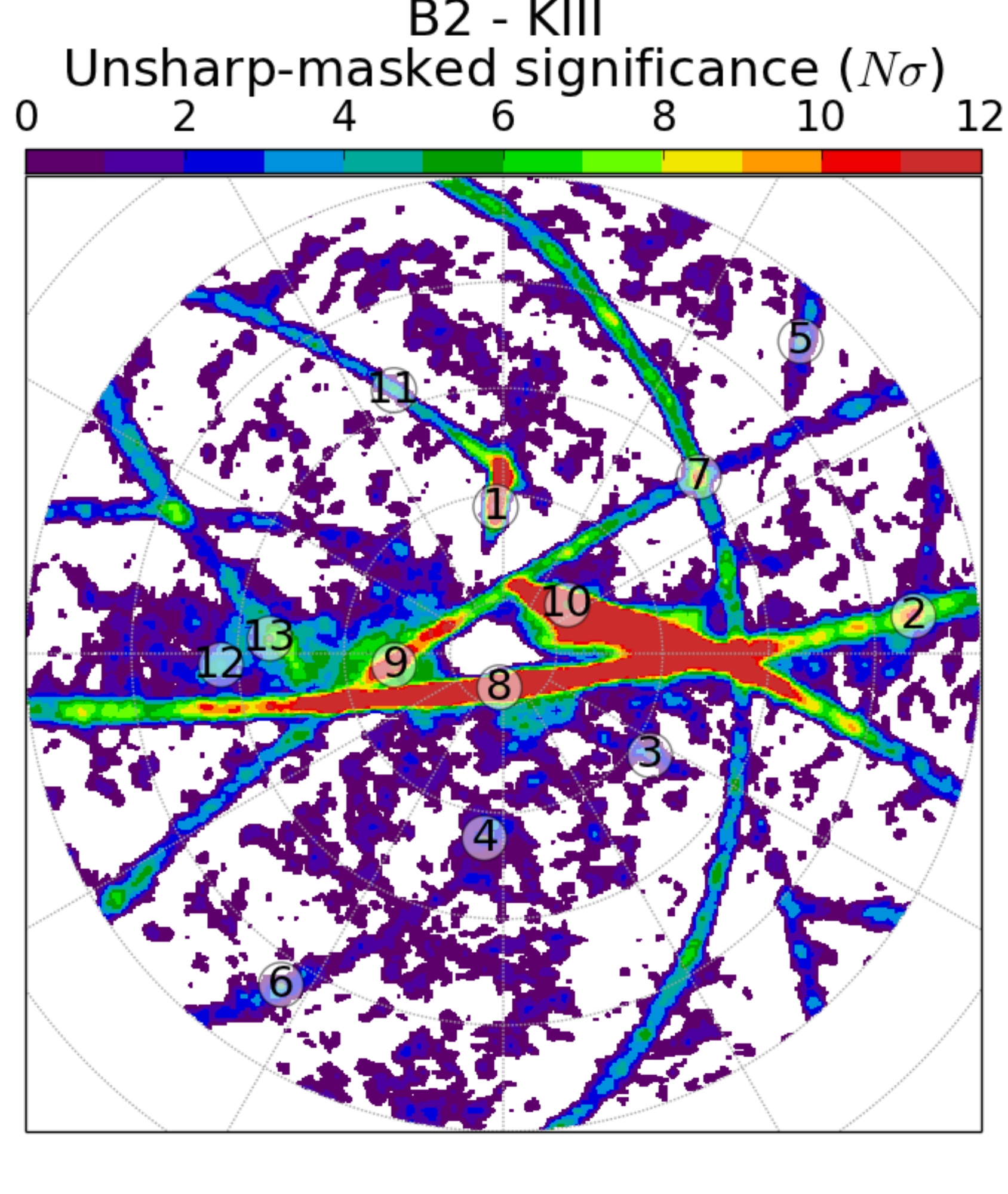} 
\includegraphics[width=0.52\columnwidth]{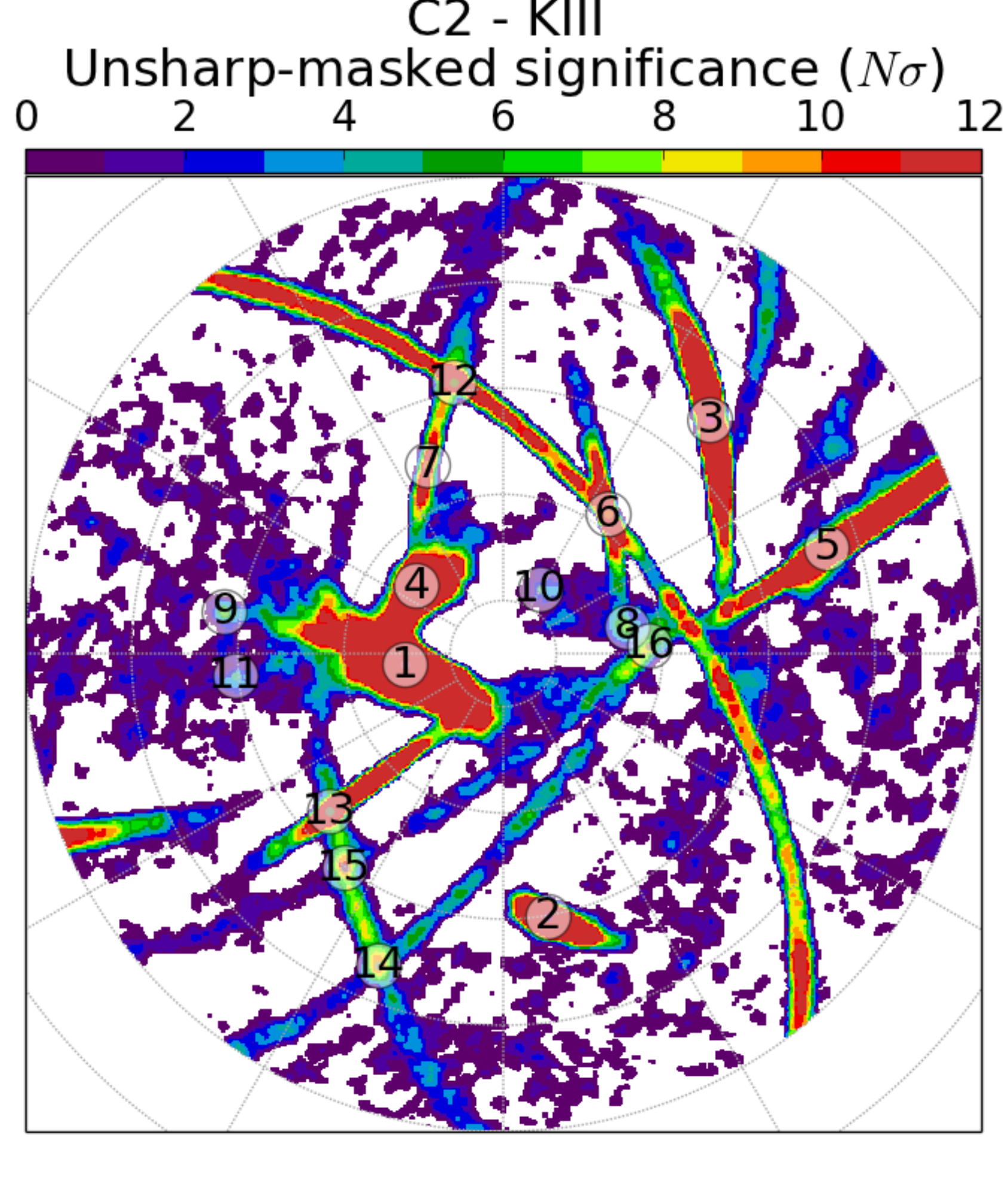} 
\includegraphics[width=0.52\columnwidth]{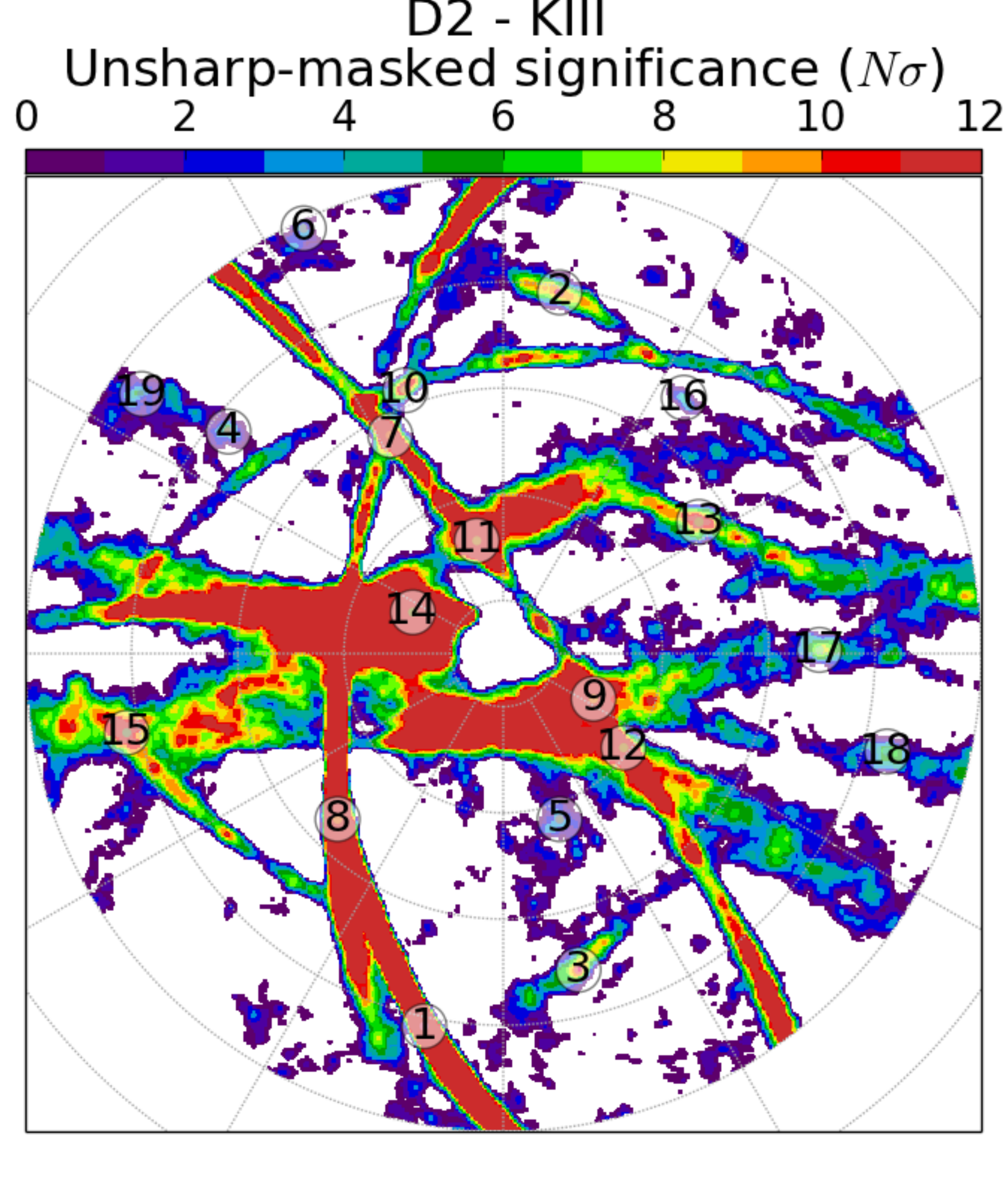} 
\includegraphics[width=0.52\columnwidth]{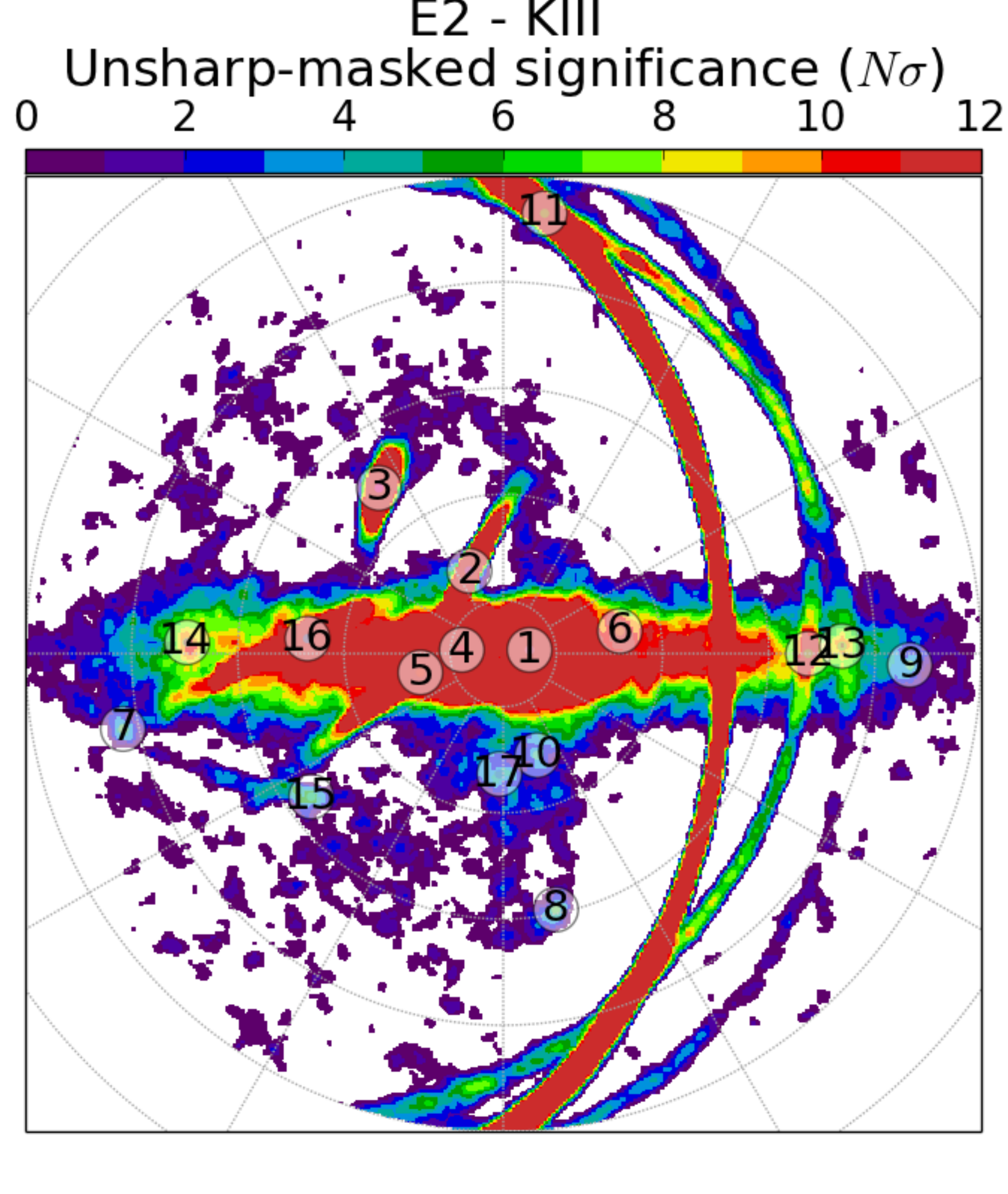} 
\includegraphics[width=0.52\columnwidth]{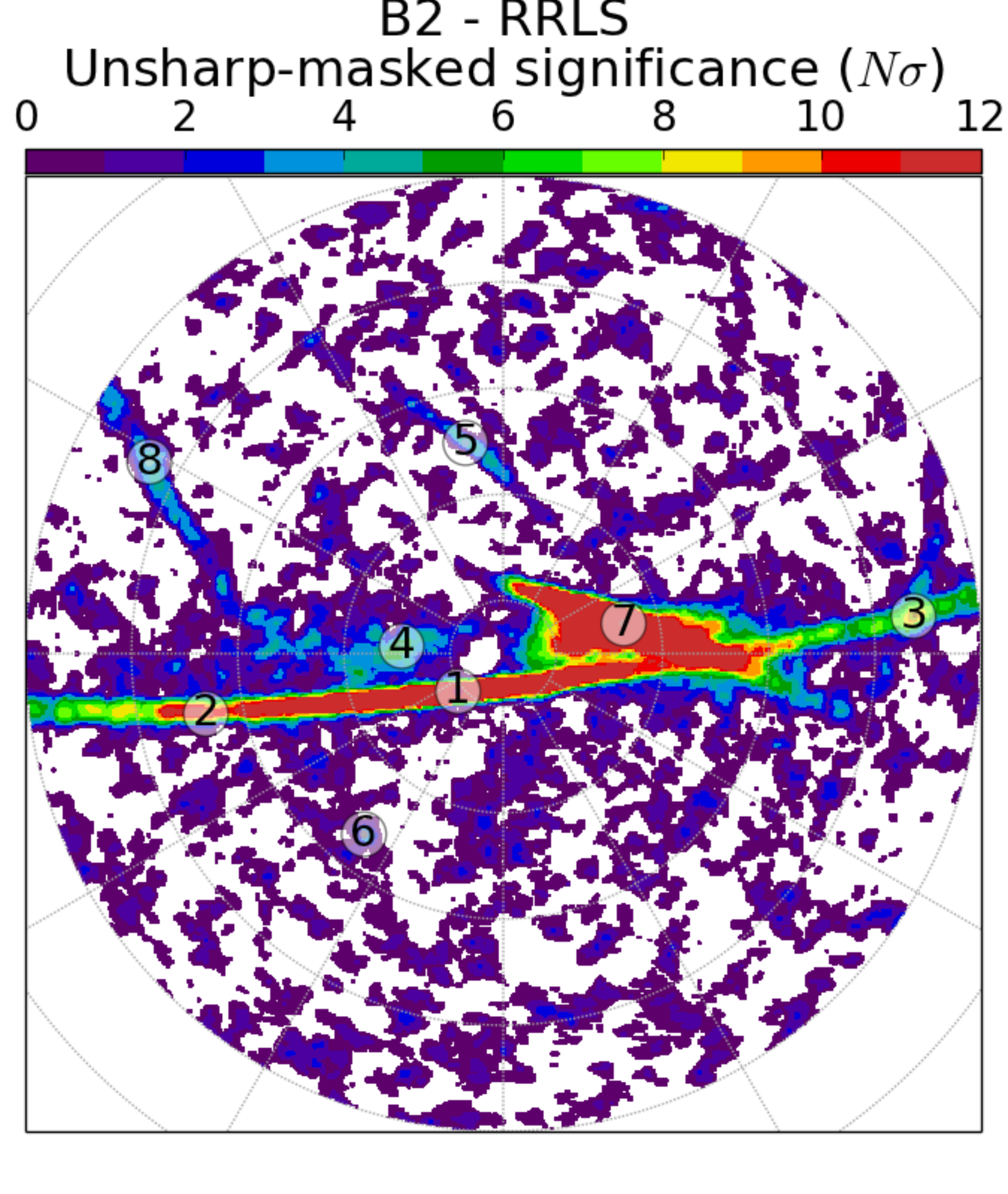} 
\includegraphics[width=0.52\columnwidth]{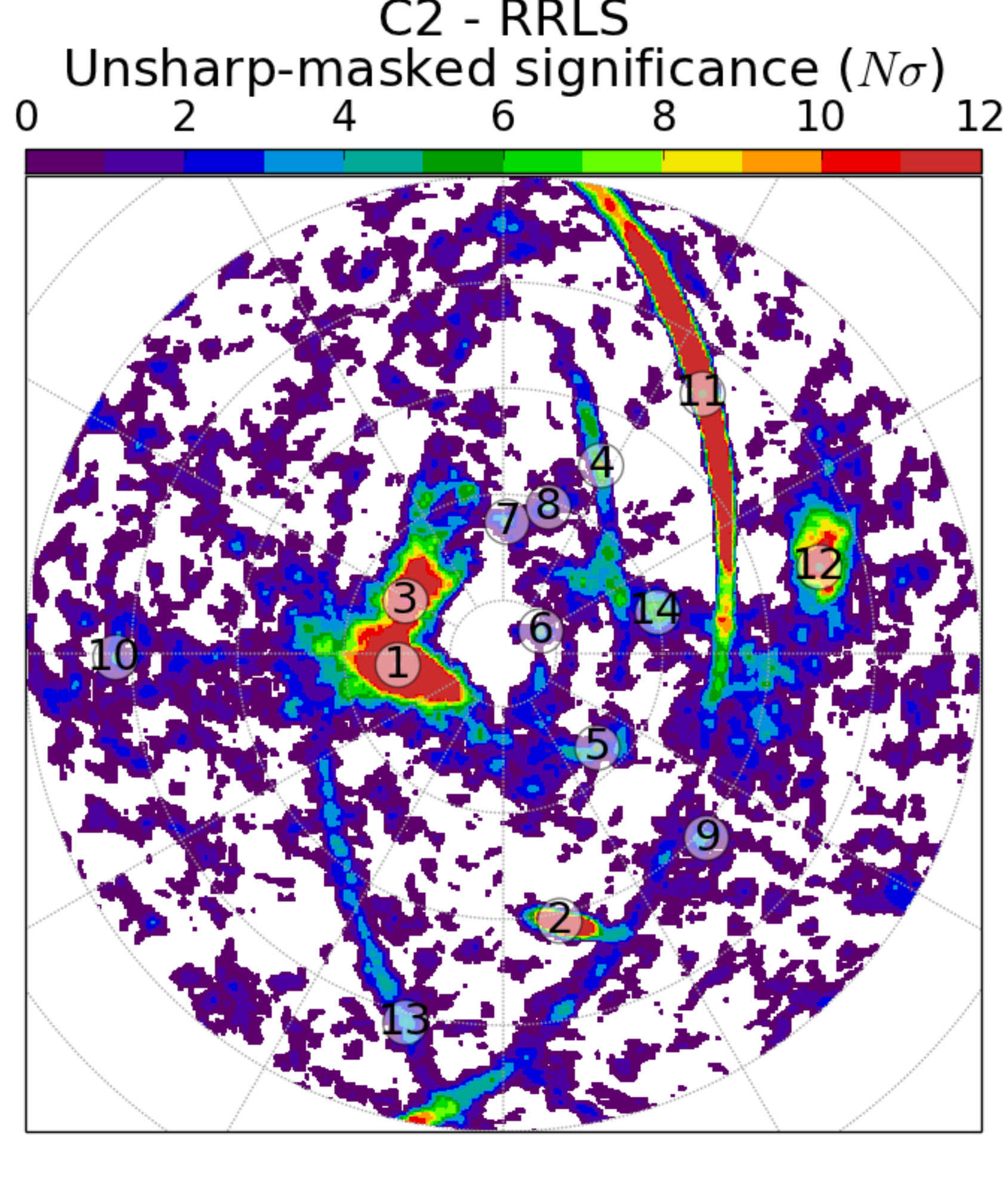} 
\includegraphics[width=0.52\columnwidth]{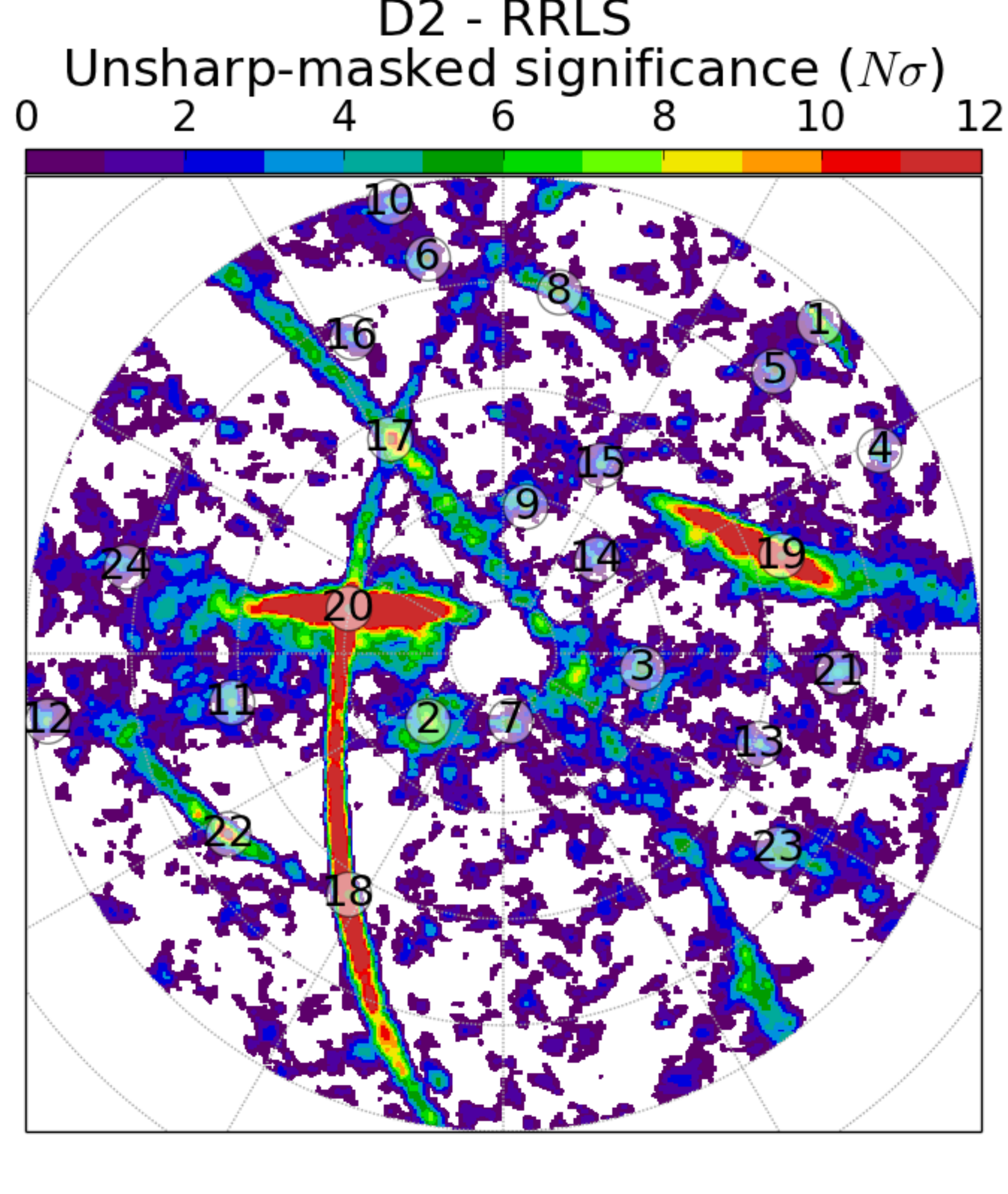} 
\includegraphics[width=0.52\columnwidth]{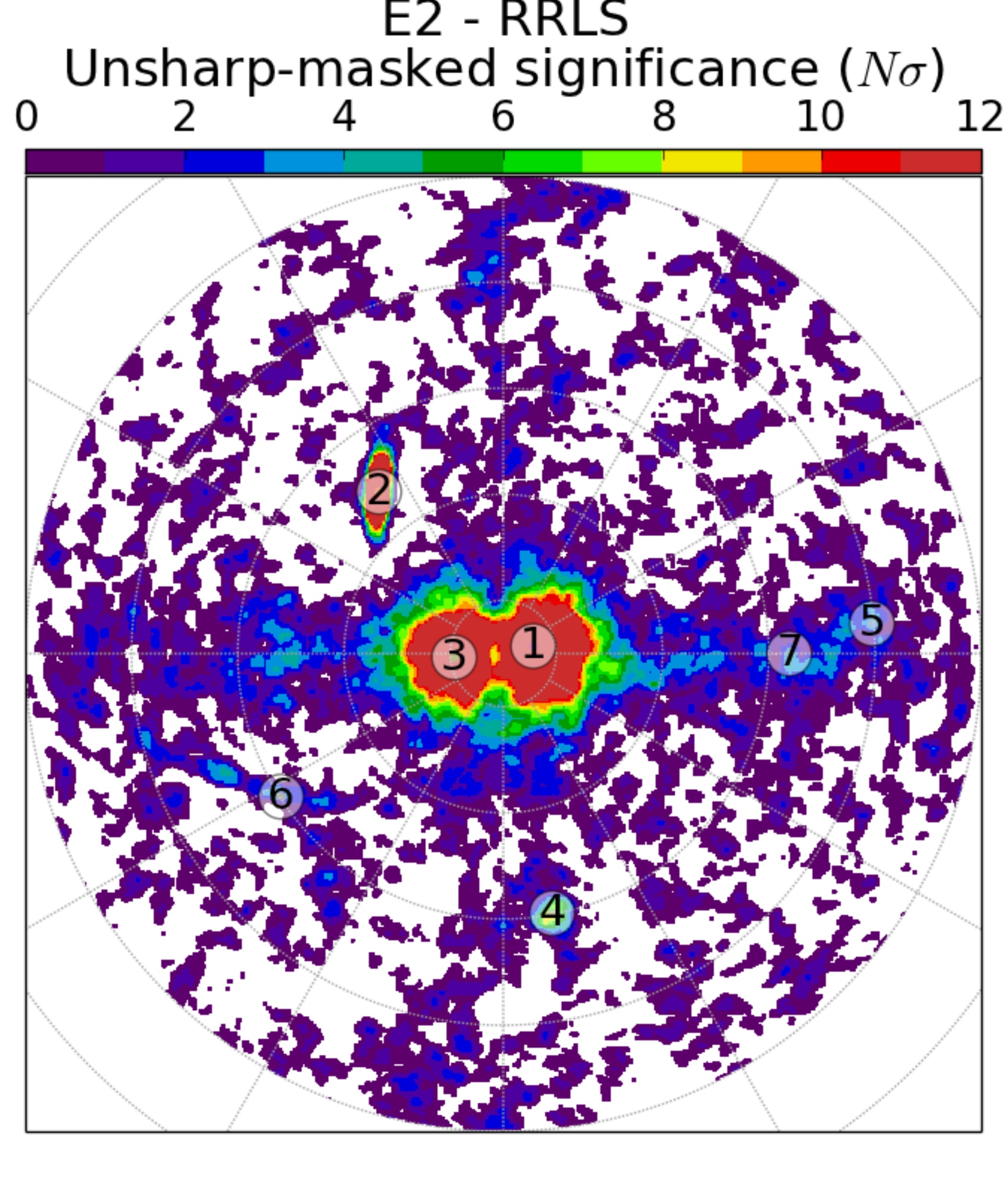} 
 \caption{Unsharp-masked \ngc~PCM for Aquarius Haloes B, C, D and E from left to right, for \Gaia~observable K giants (\emph{top}) and RRLS (\emph{bottom}), with errors, after proper motion error cut ($G\leqslant20,\Delta\mu/\mu\leqslant0.5$).  The colour scale corresponds to the pixel's significance in $N\sigma$ units. Labelled circles indicate the peaks detected using the procedure described in Sec.~\ref{s:peak_detection}.}
\label{f:pcm_aq_all}
\end{center}
\end{figure*}

So far we have used one halo, Aquarius A2, as a fiducial example to show in detail what the PCM of a cosmological halo looks like,  the effects of the \Gaia~selection function and errors, and the procedures we use to detect maxima in the PCMs. In what follows, we will discuss the PCMs of all Aquarius (Sec.~\ref{s:aqall_pcms}) and \ljmu~haloes (Sec.~\ref{s:hydro_pcms}). 

\subsection{Aquarius PCMs}\label{s:aqall_pcms}

Figure~\ref{f:pcm_aq_all} shows the \ngc~PCMs for the four remaining Aquarius haloes B2, C2, D2 and E2 from left to right. The top and bottom rows correspond to K giant stars and RRLS respectively. The halo-to-halo variation in the degree of substructure is evident, ranging from halo E2 with very little substructure to halo D2 with the most. 

The effect of the choice of tracer is also illustrated in Fig.~\ref{f:pcm_aq_all}. In all haloes, but more noticeably in haloes B2 and E2, the predominant effect is that more peaks are detected with K giants than with RRLS. The notable differences, in particular for E2, come from very strong great-circle peaks present in the K giant PCMs which are absent in the RRLS PCMs. These correspond to fully bound and distant progenitors that lie beyond the reach of \Gaia's RRLS. By contrast, in haloes C2 and D2 there are some examples of peaks detected with RRLS and not K giants, e.g. peaks 12 in halo C2 and 19 in halo D2. In this case, it is the better precision of RRLS distances that makes these peaks detectable with RRLS and not K giants. These examples show that each tracer has its own advantages and disadvantages, as we had discussed in Sec.~\ref{s:tracer}.

Since \ngc~and all great-circle methods are linear, it is possible to combine the data by simply adding the PCMs for different tracers. It would be advisable to do this after unsharp-masking, i.e. to combine the PCMs \emph{after the smooth background has been subtracted}. In cases like those presented here, where one tracer is much more numerous than the other, this would prevent the dilution of peaks detected only with the more scarce tracer by the overall background of the more numerous one. In what follows, however, we consider the data for different tracers separately in order to analyse which progenitors are recovered with each tracer.

\subsection{HYDRO-zoom PCMs}\label{s:hydro_pcms}

\begin{table}
\caption{Number of K giants observable by \Gaia~and with relative proper motion errors $\epsilon_\mu$ smaller than~$50\%$ in total and excluding stars with $|b|\leqslant10^\circ$ \& $R\leqslant20$ kpc, for the HYDRO-zoom haloes.}\label{t:NK_ljmu}
\centering
\tabcolsep=0.09cm
\begin{tabular}{lrrrrr}\hline\hline
     & \multicolumn{2}{c}{KIII (accreted)}  & \multicolumn{3}{c}{KIII (in situ)} \\
  \cmidrule(r{0.5em}l{0.5em}){2-3} \cmidrule(r{0.5em}l{0.5em}){4-6}
 \multirow{2}{*}{Halo}   & \multirow{2}{*}{$G\leqslant20$} & \multirow{2}{*}{$\Delta\mu/\mu\leqslant0.5$} & \multirow{2}{*}{$G\leqslant20$} & \multirow{2}{*}{$\Delta\mu/\mu\leqslant0.5$} & \multirow{1}{*}{$\Delta\mu/\mu\leqslant0.5$}\\
        &                          &                                               &                          & & + Excl. zone      \\

\hline
001 & 2,048,136 & 1,789,181 & 9,953,712 & 9,496,893 & 5,540,500 \\
004 & 2,476,549 & 2,053,819 & 7,595,529 & 7,059,015  & 3,642,743 \\
006 &   803,328 &   602,513   & 8,577,449 & 7,825,379 & 3,101,889 \\
008 & 1,170,131 &   549,300  & 7,438,652 & 7,241,107 & 4,418,790 \\
009 & 1,937,388 & 1,806,513 & 6,123,721 & 5,750,542 & 3,078,387 \\
\hline
\end{tabular}
\end{table}

\begin{figure*}
\begin{center}
\includegraphics[width=0.85\columnwidth]{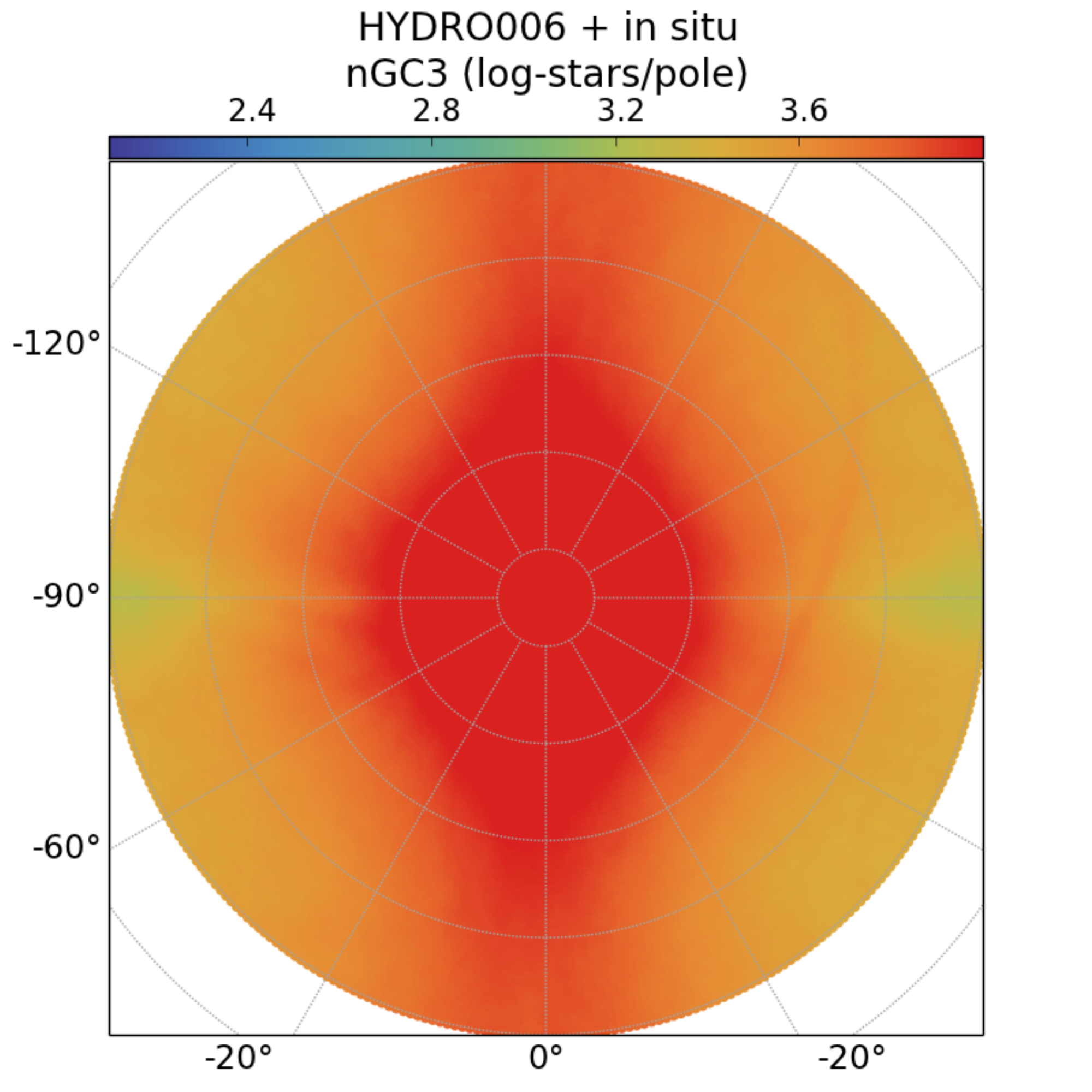} 
\includegraphics[width=0.85\columnwidth]{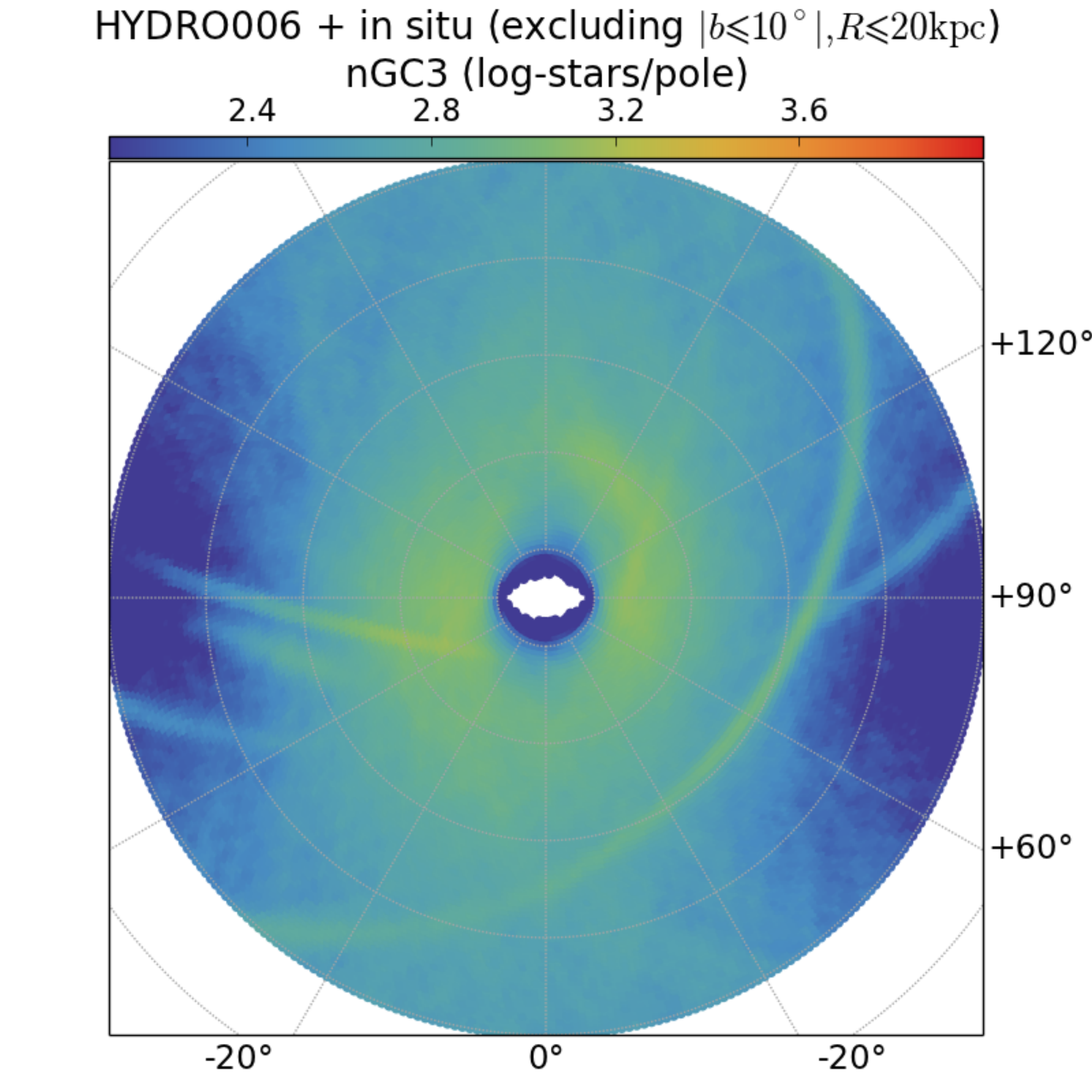} 
 \caption{PCMs for HYDRO-006 illustrating the effect of including the \insitu~component. \emph{Left:} full \insitu~background. \emph{Right}: excluding stars with $|b|\leqslant10\degr$ and $R_{\rm gal}\leqslant20$ kpc. The colour scale is the same for both plots and shows \ngc~pole counts. Note how the PCM pole counts are reduced by about an order of magnitude after including the disc avoidance zone.}
\label{f:ljmu6_insitu}
\end{center}
\end{figure*}

In the following, we will focus on the \Gaia~mock distribution of K giant stars in the \ljmu~haloes. Table \ref{t:NK_ljmu} summarises, for each halo, the number of K giants observable within the \Gaia~ magnitude limit and with an additional cut $\Delta\mu/\mu\leqslant0.5$, respectively, shown separately for the accreted (left) and \insitu~components (right). This shows that the number of \insitu~ K giant stars is quite large, more than double in most cases than the number of K giant accreted stars. This is to be expected since not only halo, but also the disc stars are included. 

Since the disc is the most important source of contamination, we introduce a cut to eliminate all low latitude ($|b|\leqslant10\degr$) stars inside a given galactocentric (cylindrical) radius ($R\leqslant20$ kpc). This way we avoid eliminating distant stars that may belong to streams beyond the disc radius. The cut is introduced in galactocentric radius, taking advantage of the fact that our assumed tracer provides reasonably precise distances (errors $<20$\%); if this were not the case, it would be preferable to define the cuts using a direct observable (see e.g. M11).  Fig.~\ref{f:ljmu6_insitu} shows the \ngc~PCM for halo 006 using all \insitu~stars (left) and including the disc avoidance zone (right), which clearly illustrates how the overall pole counts are reduced and several features are revealed in the PCM using this simple cut.

\begin{figure*}
\begin{center}
\includegraphics[width=0.58\columnwidth]{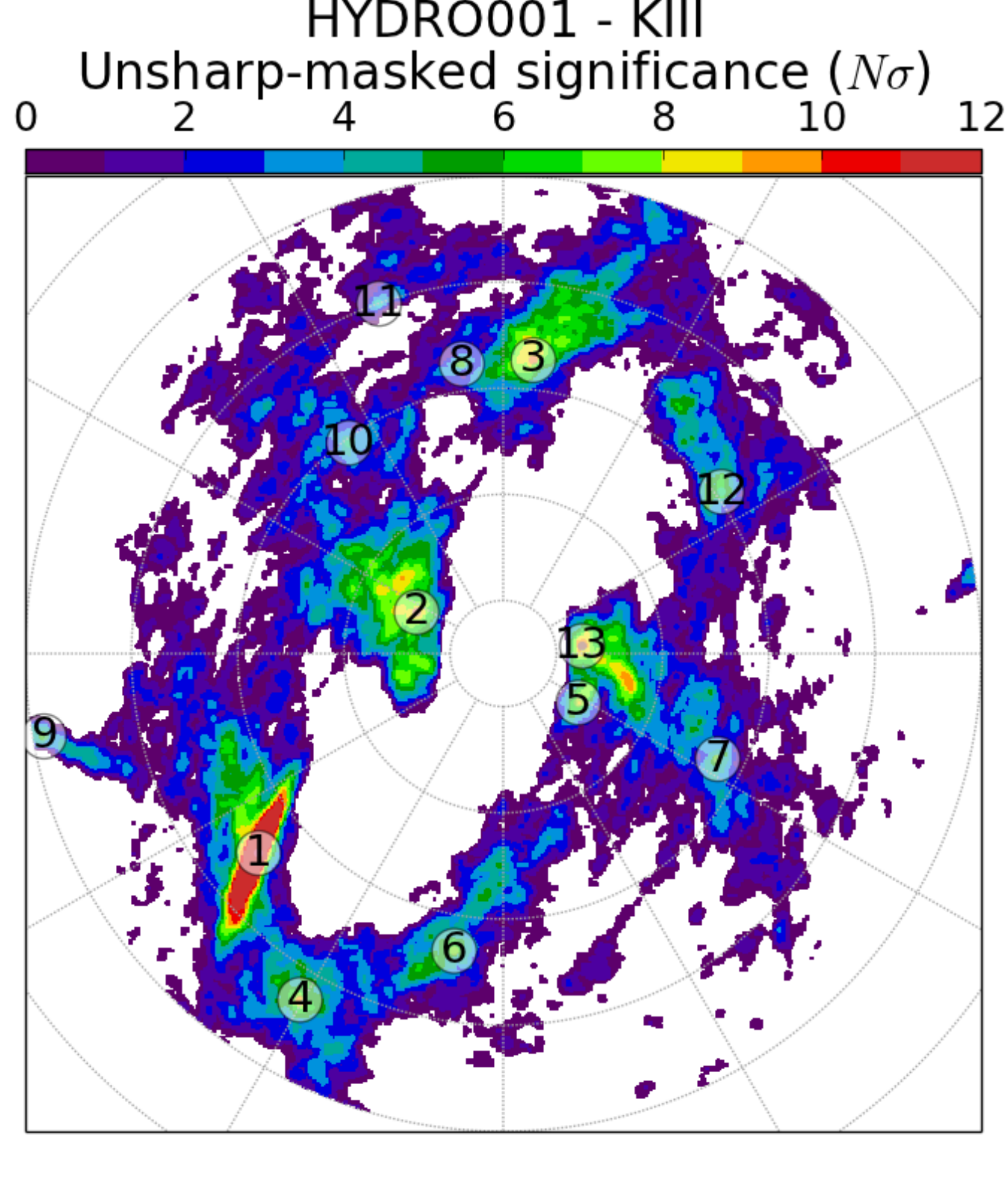}
\includegraphics[width=0.58\columnwidth]{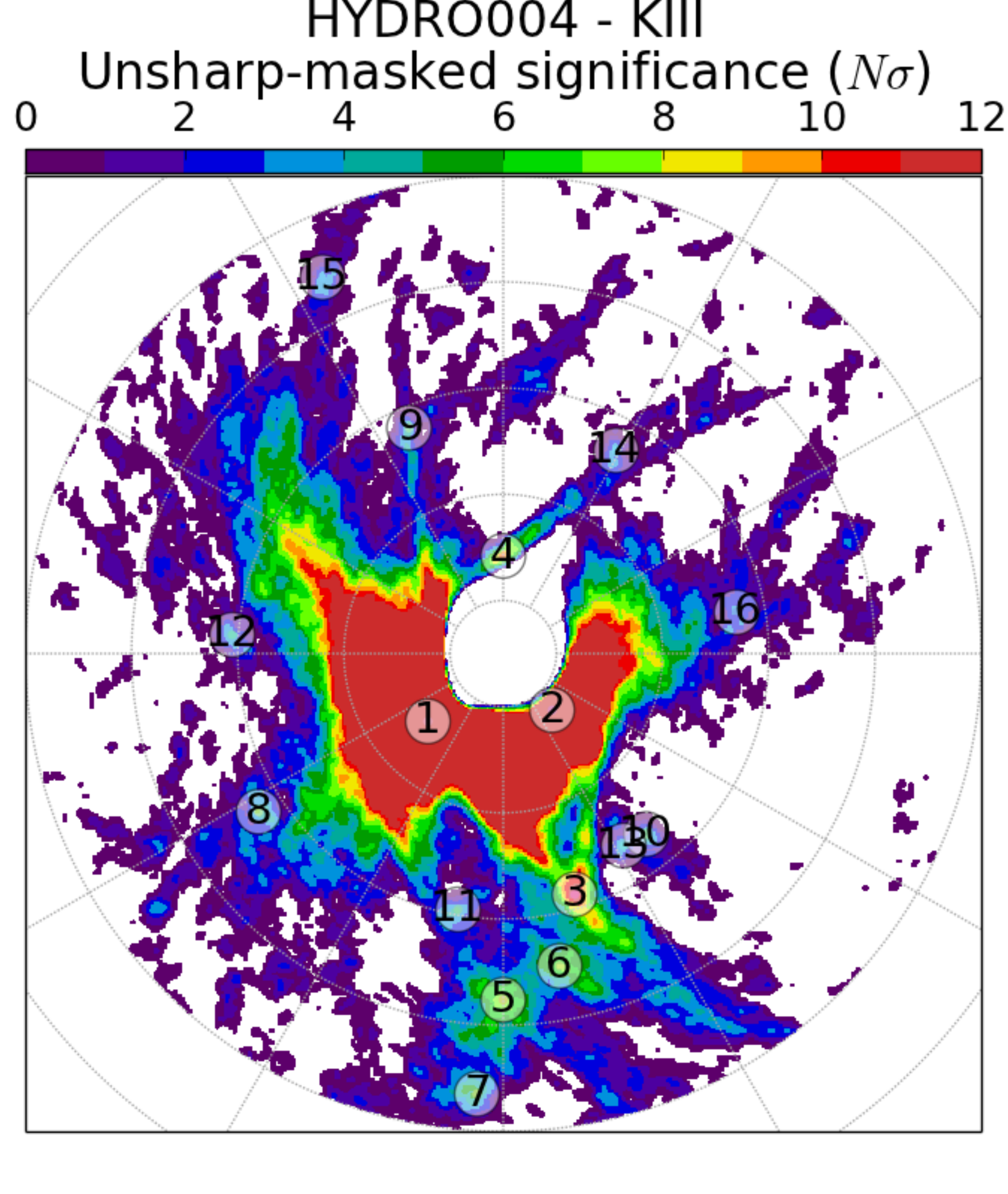} 
\includegraphics[width=0.58\columnwidth]{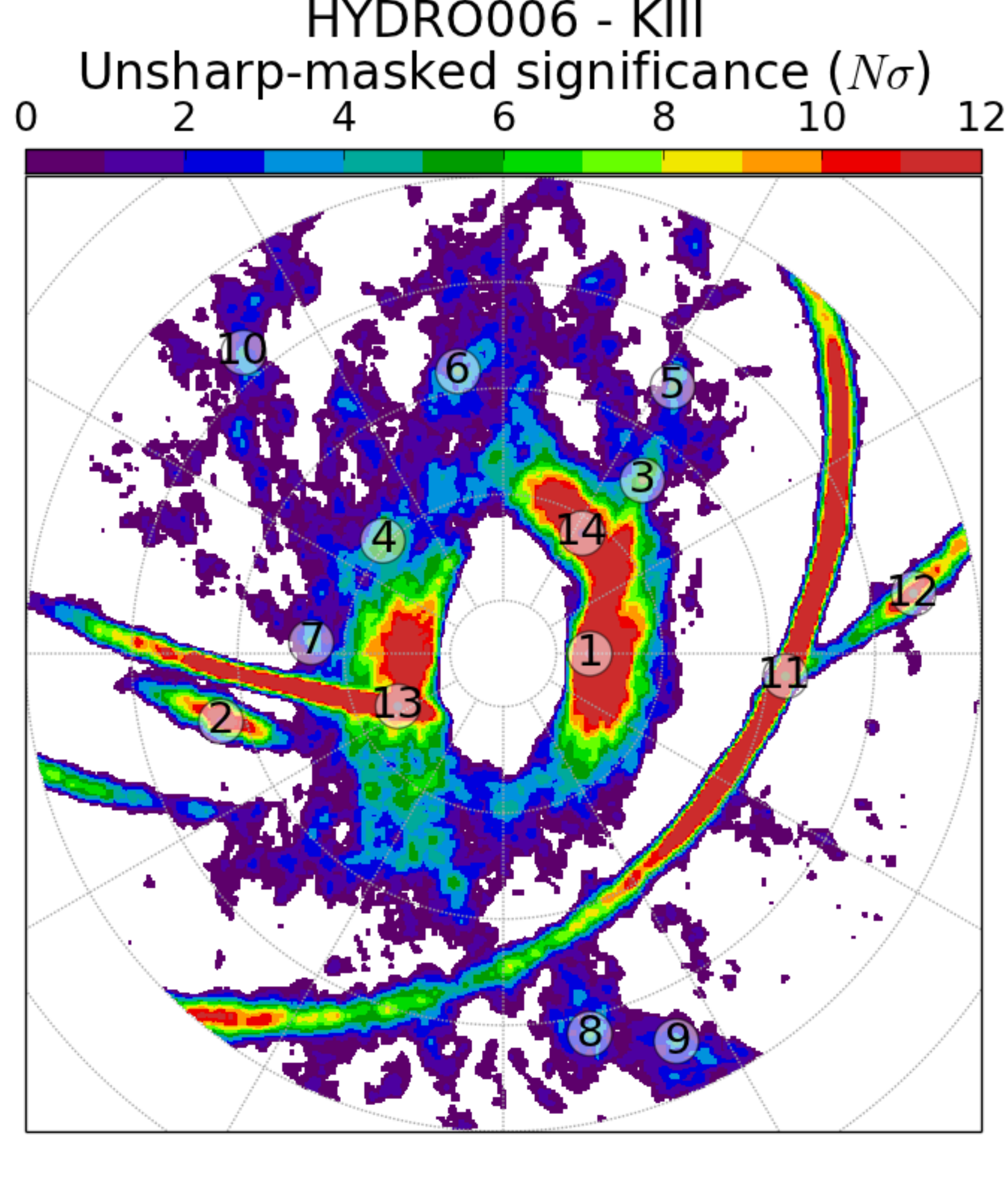} 
\includegraphics[width=0.58\columnwidth]{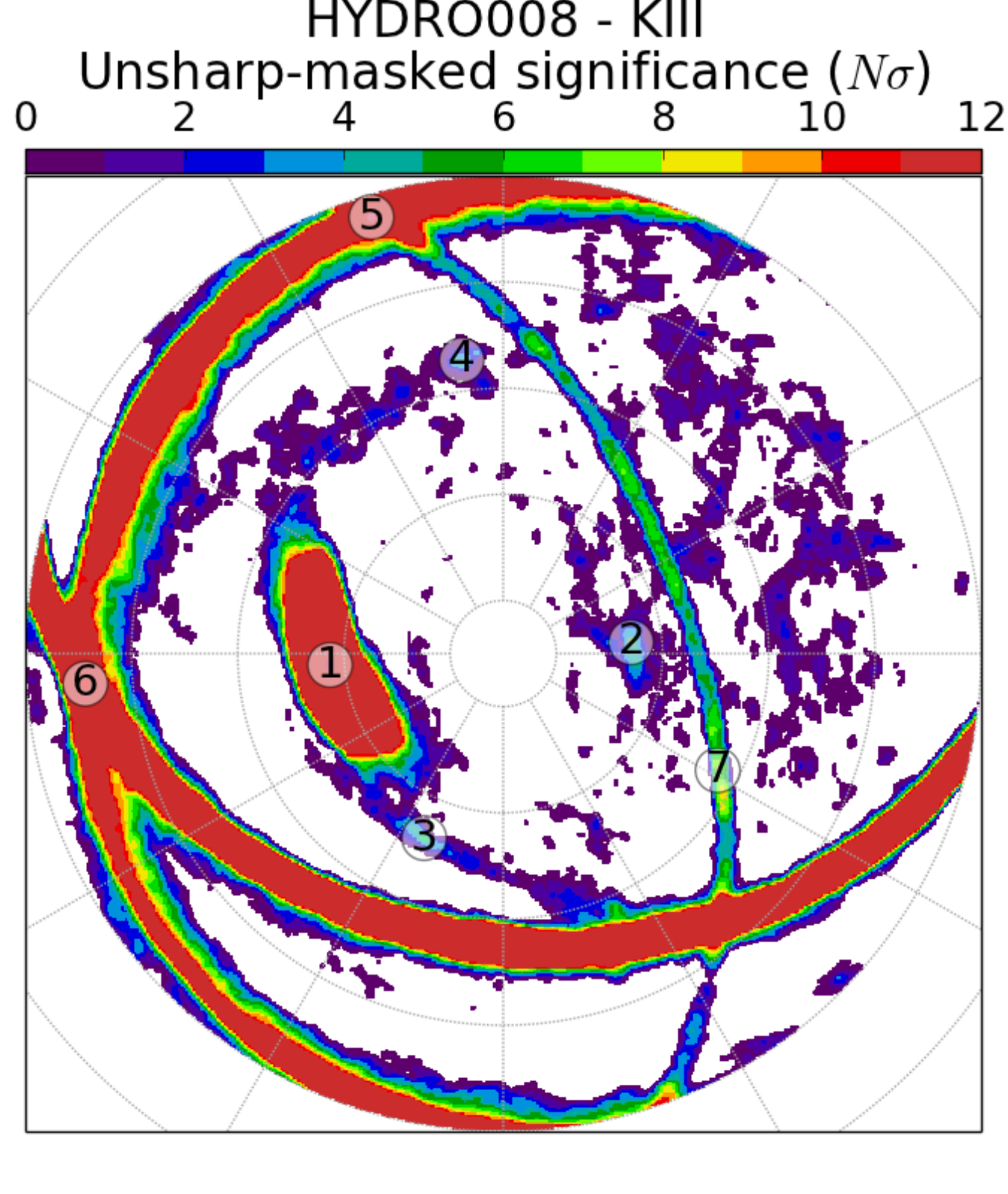} 
\includegraphics[width=0.58\columnwidth]{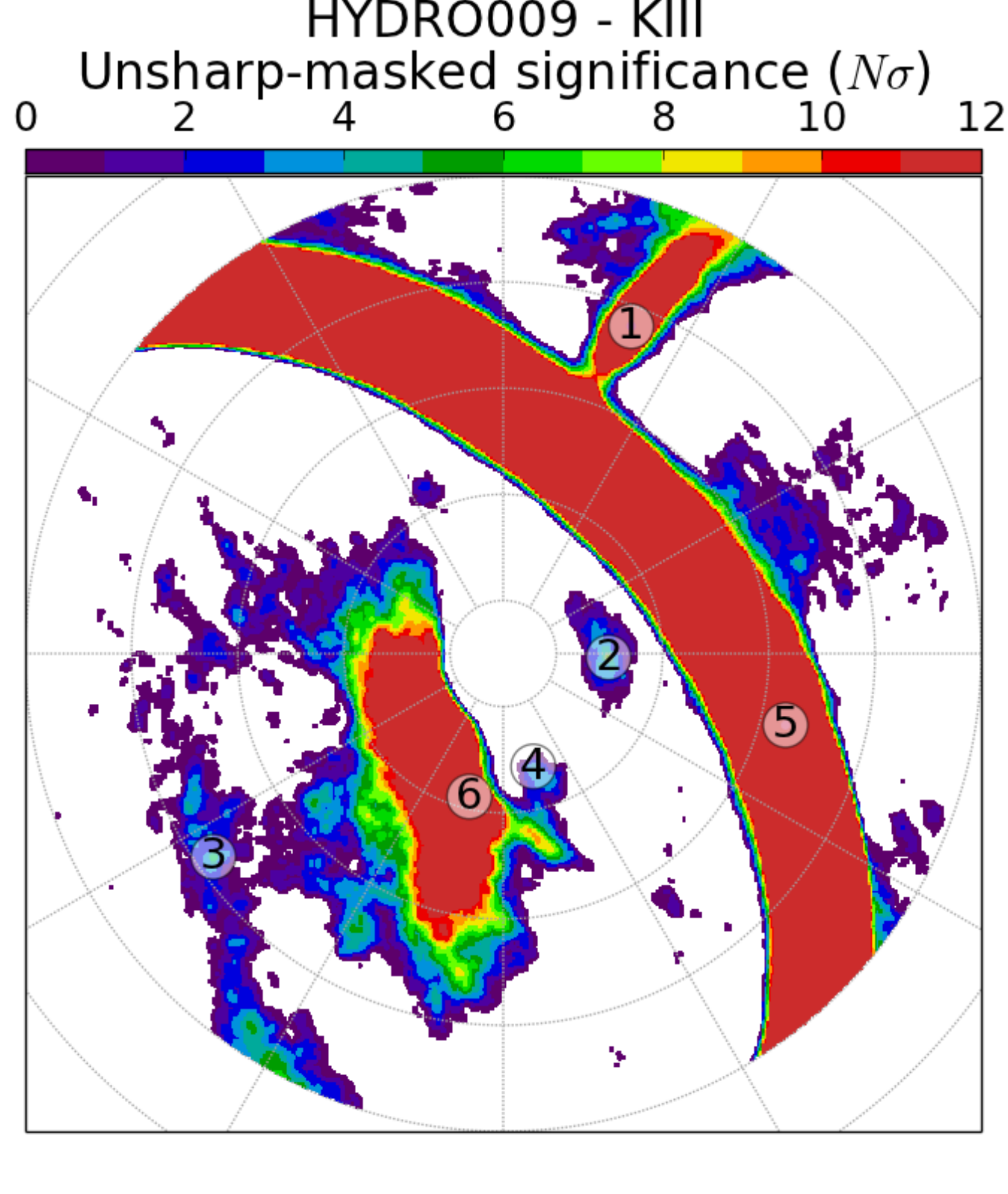}  
 \caption{Unsharp-masked \ngc~PCM for \ljmu~haloes 001 to 009, for \Gaia~observable K giants with errors, after proper motion error cut ($G\leqslant20,\Delta\mu/\mu\leqslant0.5$). The colour scale corresponds to the pixel's significance in $N\sigma$ units. Labelled circles indicate the peaks detected using the procedure described in Sec.~\ref{s:peak_detection}.}
\label{f:pcm_ljmu_all}
\end{center}
\end{figure*}

Figure~\ref{f:pcm_ljmu_all} shows the K giant \ngc~unsharp-masked PCMs for all gas dynamical \ljmu~haloes (001 to 009), with the in situ background and excluding stars with $|b|\leqslant10\degr$ and $R\leqslant20$ kpc. The peak detections were made following the procedure described in Sec.~\ref{s:peak_detection}, and are shown with labelled circles. As seen here, there seems to be less substructure overall in the \ljmu~PCMs than in the Aquarius haloes. Even so, there is a large range in the amount of substructure present in the different \ljmu~haloes. Haloes 008 and 009 exhibit little substructure and a few very luminous progenitors producing strong great-circle maxima in each case, whereas halo 006 shows a level of substructure similar to some of the Aquarius haloes. 

To check whether the lower mass resolution translates into fewer progenitors overall, Table \ref{t:Nstreams_mass} lists the  number of progenitors with masses higher than $10^6M_\odot$ (the mass limit of the \ljmus). The table shows, on average, 
there are \emph{more} progenitors per halo in the \ljmus~compared to Aquarius, so this is not the reason why less substructure is visible. In addition, the stellar mass of the accreted component in the \ljmu~haloes is, in general, higher than that of the Aquarius haloes, with three haloes (006,008,009) being as massive as the most massive Aquarius halo D2, and with haloes 001 and 004 being close to 2 and 4 times as massive. This causes the number of K giants in \ljmu~haloes to be generally higher than in the Aquarius haloes, as Tables~\ref{t:NK_aq}~and \ref{t:NK_ljmu} show, which translates into much higher backgrounds in the PCMs (this will be clearly illustrated in the next section by $N_{\rm BG}$ in Tables~\ref{t:rec_stats_aq} and \ref{t:rec_stats_ljmu}). Note that here we are comparing the number of stars in the \emph{accreted} halo, so this effect is present even without taking the \insitu~component into account.

 Up to this point it is not possible to tell whether this will prevent us from using the \ljmu~simulations for our predictions on the number of detectable streams. We will address this in the next section, as we analyse the progenitors recovered in the two types of simulations.
 
\begin{table}
\caption{Stellar mass and number of all progenitors $N_{\rm all}$ and of streams $N_{\rm streams}$ with stellar masses $>10^6{\rm M}_\odot$.}\label{t:Nstreams_mass}
\centering
\tabcolsep=0.09cm
\begin{tabular}{lrcc}\hline\hline
Halo & $M_{\rm stellar}$ & $N_{\rm all}$ & $N_{\rm streams}$   \\
        & (${\rm M}_\odot$) & ($>10^6{\rm M}_\odot$) & ($>10^6{\rm M}_\odot$) \\
\hline
A2   &   7.2e+08 & 41  &   19   \\
B2   &   6e+08   & 23  &   15   \\
C2   &   1.3e+09 & 36  &   20   \\
D2   &   2.3e+09 & 39  &   21   \\
E2   &   1.2e+09 & 28  &   14   \\
\hline
001  &   8.1e+09 & 52  &   32   \\
004  &   4e+09   & 33  &   21   \\
006  &   2.5e+09 & 37  &   34   \\
008  &   2e+09   & 24  &   17   \\
009  &   2.6e+09 & 32  &   28   \\
\hline
\end{tabular}
\end{table}

\section{Recovering Streams in Cosmological Simulations}\label{s:recovery_all}

\subsection{Which progenitors can we recover?}\label{s:recovery_example}

\begin{figure*}
\begin{center}
 \includegraphics[width=2.2\columnwidth]{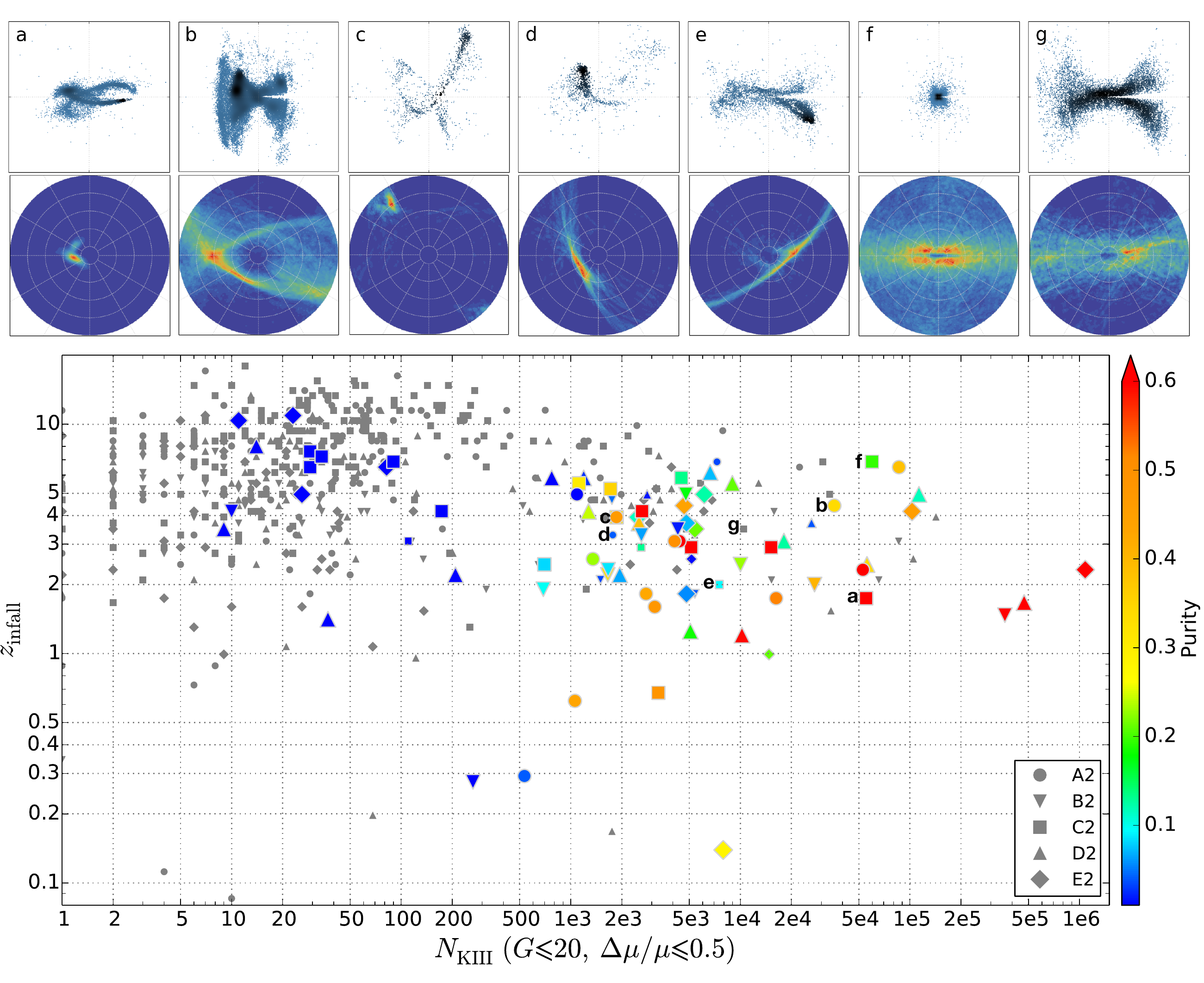} 
 \includegraphics[width=2.2\columnwidth]{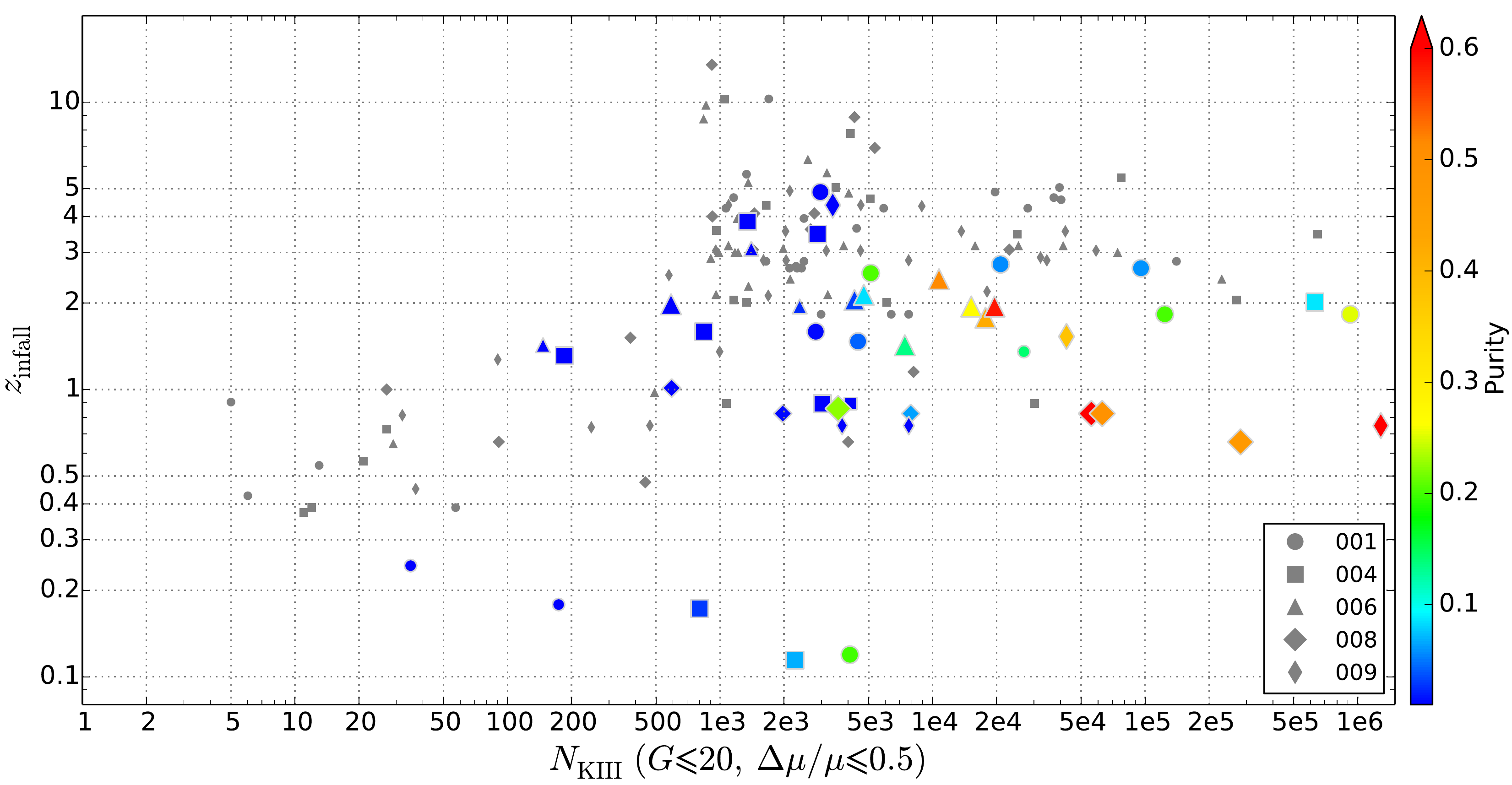} 
 \caption{\emph{Middle and top panels:} Infall redshift $z_{\rm{infall}}$ vs number of observable K giants $N_{\rm KIII}$ for each of the progenitors in the Aquarius haloes. Recovered progenitors are shown with filled symbols, primary and secondary detections are denoted by large and small symbols. The colour scale is proportional to the purity of the corresponding pole detection, with the upper end of the scale meaning a purity of 0.6 or higher. In the smaller sub-panels at the top, we show the spatial $Z$ vs $X$ distributions and the \ngc~PCMs, respectively. The sub-panels correspond to a few selected progenitors, labeled from ${\rm a}$ to ${\rm g}$ in the Aquarius A2 ($\rm{b,c,d}$) and C2 ($\rm{a,e,f,g}$) haloes, respectively. Darker colours in the spatial distribution plots correspond to higher density. \emph{Bottom panel} The Infall redshift $z_{\rm{infall}}$ vs number of observable K giants $N_{\rm KIII}$ for each of the progenitors in the \ljmu~haloes. Again, the colour scale is proportional to the purity of the corresponding pole detection.}
\label{f:z_N_a}
\end{center}
\end{figure*}

First, we need to decide when a progenitor is considered `recovered', as well as one or more quantities that will help us define the quality of the recovery. Two useful definitions are: the \emph{fraction of recovered stars} $f_{\rm rec}$, i.e. the fraction of progenitor stars in a given pole detection; and the \emph{purity}, defined as the number of progenitor stars within the detected peak, divided by the total number of stars within that detection. With this definitions $f_{\rm rec}=0.4$ means 40\% of the total (observable) progenitor stars are recovered in the pole detection and purity $=1$ means there are no contaminant stars from other progenitors (or the in situ halo). 

Since different progenitors can produce maxima that overlap in the PCM, any given pole detection can be associated to more than one progenitor and vice versa. We allow for multiple progenitors to be associated to any given pole, and hence considered as detected, provided a minimum fraction  $f_{\rm rec}>0.1$ of progenitor stars is recovered. For a given pole we will consider the progenitor recovered with the highest purity as the primary detection, and the remaining as secondary detections, so there will be as many primary detections as poles detected in the PCM. Although secondary detections are indistinguishable from primary ones with \ngc, we will consider them as valid detections as we expect follow up methods will be able to disentangle them, e.g. through radial velocities or colour-magnitude diagram analyses.

The distribution of recovered and unrecovered progenitors in the Aquarius and \ljmu~haloes is illustrated in Fig.~\ref{f:z_N_a}. The main plots (central and bottom panels) show the infall redshift $z_{\rm{infall}}$ versus $N_{\rm KIII}$, the number of \Gaia~observable K giants with proper motion errors $\Delta\mu/\mu\leqslant0.5$. We define the infall redshift as the time of the first simulation output at which a progenitor is identified as a subhalo of the main halo. Different symbols represent different haloes, as shown in the plot legend. Recovered progenitors are represented as filled colour symbols, large and small representing respectively primary and secondary detections and with a colour proportional to the purity. Unrecovered progenitors are shown with filled grey symbols. The labelled points indicate examples of recovered (${\rm a}$-${\rm f}$) and unrecovered (${\rm g}$) progenitors, for which the corresponding spatial distribution (X-Z plot) and \ngc~PCMs (with errors) are shown in the top row of the figure. 
Labelled points indicate progenitors selected to illustrate examples of detections, primary and secondary, and non-detections.

Fig.~\ref{f:z_N_a} shows that recovered streams exhibit a wide variety of morphologies in their spatial distributions as well as in their signatures in PCMs. 

Panels ${\rm a}$ to ${\rm c}$ show three progenitors recovered with high purity. Progenitor ${\rm a}$ has produced a bright and well defined stream, it has a very prominent bound core and tidal tails with several wraps that have undergone some precession, which has spawned a second lobe (light blue) in the PCM. Progenitor ${\rm c}$ is recovered with a similar purity as ${\rm a}$. Even though it was accreted as early as $z_{\rm{infall}}\sim4$, its tidal stream is quite cold and and produces a very well-defined peak in the PCM. Progenitor ${\rm b}$ was accreted even earlier than ${\rm c}$ and has produced a much more disrupted stream with a more complicated signature in the PCM, but that is still recovered with purity $>0.3$. 

Panels ${\rm d,e,f}$ show some intermediate cases that illustrate the effect of contamination and overlapping signatures in the PCM. Progenitor ${\rm d}$ has roughly as many visible stars as ${\rm c}$ and was accreted slightly earlier ($z_{\rm{infall}}\sim3$). The low purity ($<0.1$) of this detection is caused by its main peak overlapping in the PCM with the signature of the much brighter progenitor    ${\rm b}$. This can also be clearly seen in Fig.~\ref{f:pcm_peaks_tracer} (top), where pole detection 4, which corresponds to progenitor ${\rm d}$, is located in a PCM region with a higher than average background (see also Fig.~\ref{f:aqa2_unsharp}, left panel). Progenitor ${\rm e}$ is an example of a secondary detection. The signature it produces in the PCM is readily evident in Fig.~\ref{f:pcm_aq_all} (top row, second panel) and is detected as pole 14. This pole detection, however, is associated to the much brighter progenitor ${\rm f}$; a completely disrupted progenitor that produces a PCM signature, that although diffuse, dominates pole counts around progenitor ${\rm e}$'s peak. Hence, progenitor ${\rm f}$ also serves as a false positive example, as it is a spurious detection of a completely disrupted stream that we should not expect to recover with our method.

Panel ${\rm g}$ shows a progenitor that is not recovered. It was accreted at a relatively high redshift $z_{\rm{infall}}>3$ and is an example of a stream that has been completely phase-mixed, which produces no clear signature in the PCM and hence, as expected, is not detectable. 

The central panel of Fig.~\ref{f:z_N_a} shows that, in the Aquarius simulations, streams can be recovered up to infall redshifts as high as $\sim$5--6 and with relatively good purity ($>0.3$) for progenitors with more than about a thousand observable stars.  The lower panel shows the distribution of the recovered progenitors in the gas dynamical \ljmu~haloes. The \ljmu~simulations have a lower resolution than Aquarius, so in Fig.~\ref{f:z_N_a} the plots are more sparsely populated with only the most massive (and hence the most luminous) progenitors. This explains why there are only a few points corresponding to small numbers of observable stars, with a clear deficiency of objects below $N_{\rm KIII}\sim500$. This lack of objects is very evident, particularly at redshifts higher than $\sim$2 where there are no progenitors with $<700$ stars. 
This shows that \emph{close to $N_{\rm KIII}\sim$500--1000, our results might be hampered by the lower resolution of the \ljmu~simulations}. 

Overall, infall redshifts are lower for gas dynamical progenitors compared to Aquarius progenitors and, in particular, recovered progenitors are detected up to infall redshifts $\sim$2--3, i.e. accreted more recently than those recovered in the Aquarius simulations $\sim$5--6. However, it is not possible to tell from this plot whether the recovery of progenitors at lower redshifts in the gas dynamical simulations is simply due to the fact that, overall, there are fewer progenitors accreted at larger redshifts,  or whether at a particular redshift, progenitors are more easily disrupted due to the presence of a disc making its detection harder. Hence, we should look at the angular thickness of a stream, which \emph{will} have a direct effect on its detectability with our method and serves as a proxy for the dynamical age of a stream.

To estimate the angular thickness $\Delta\theta$, we rotate each stream so that its mid-plane coincides with the galactic equator and look at the distribution of stars in latitude. This can give us a sense of how thick the stream is in the direction perpendicular to its orbital plane. We fit a four component Gaussian mixture model to this distribution and compute $\Delta\theta$ as the sum in quadrature of the standard deviations of the two main Gaussian components, weighted by their amplitudes. We find that this gives a good representation of the angular thickness of the streams, as it balances the contribution of outliers, the actual tidal tails, and the bound core (where one exists).

\begin{figure}
\begin{center}
 \includegraphics[width=1.\columnwidth]{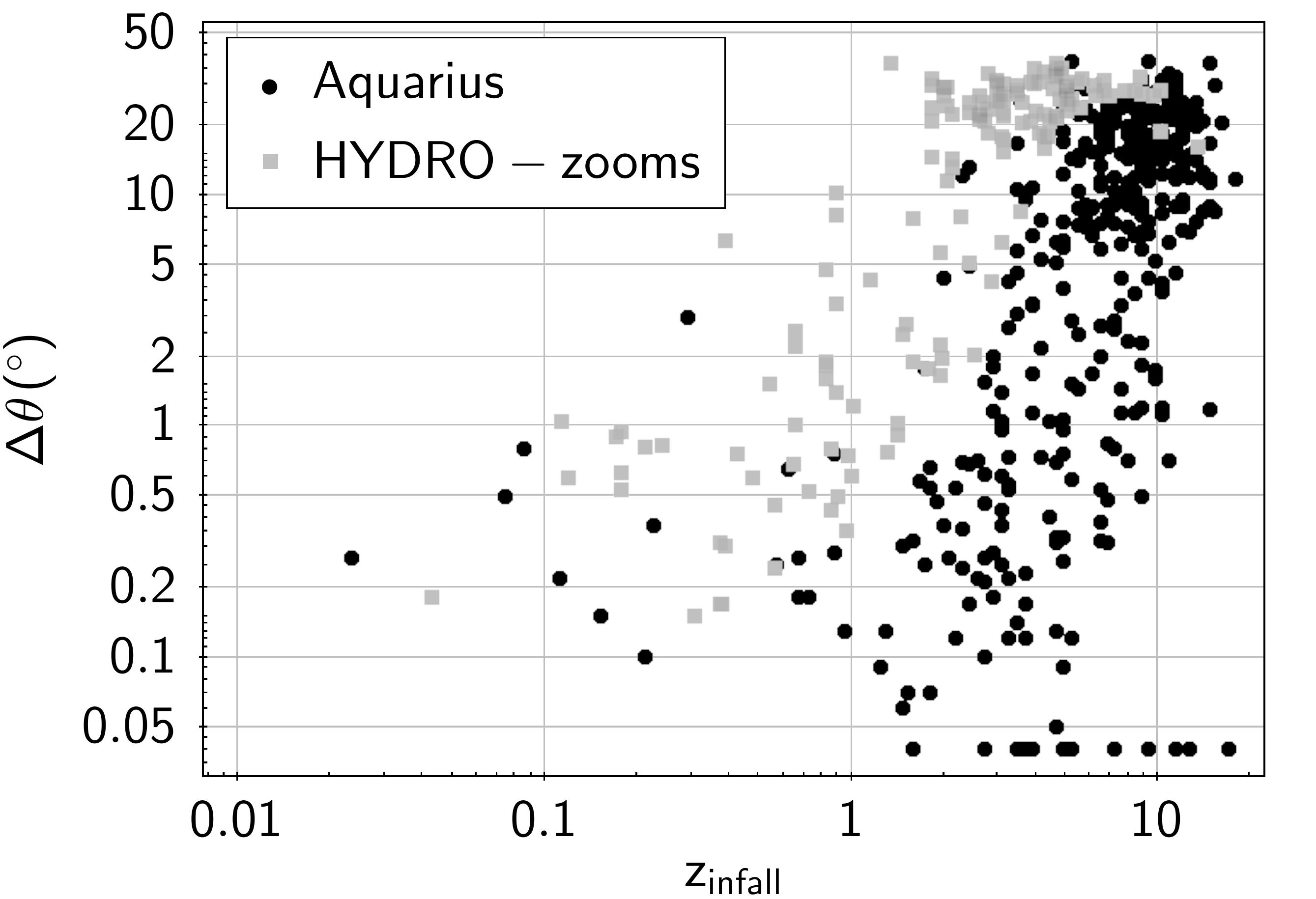} 
 \caption{Angular thickness $\Delta\theta$ vs infall redshift $z_{infall}$ for all progenitors in the Aquarius (black circles) and HYDRO-zoom (grey squares) simulations.}
\label{f:dtheta_infallz}
\end{center}
\end{figure}

Figure~\ref{f:dtheta_infallz} plots the angular thickness $\Delta\theta$ as a function of infall redshift $z_{\rm infall}$, for all progenitors in the Aquarius and HYDRO-zoom simulations. This plot clearly shows that, at a given infall redshift, progenitors in the \ljmu~gas dynamical simulations (grey) produce thicker streams than in the dark-matter only Aquarius simulations (black). Hence progenitors are more effectively disrupted, as we anticipated might be the case due to the effect of the disc, which explains why streams are detected up to lower redshifts in the \ljmus.  Also, for both simulations, the correlation of angular width and infall redshift is clear, albeit with large scatter, which confirms that the angular width is a suitable proxy for the infall redshift, and hence for dynamical age.

In the $z_{\rm infall}-N_{\rm KIII}$ plots of  Fig.~\ref{f:z_N_a}, additionally, detections and non-detections are not segregated, as is to be expected, since progenitors accreted at the same $z_{\rm{infall}}$ but on different orbits will be disrupted to different degrees (for example, ${\rm d}$ and ${\rm g}$). In order to look for a clear boundary that separates detections from non-detections, we will examine a plane of observables that have a direct effect on detectability with our method.

\subsection{The Detection Boundary}\label{s:det_boundary}

\begin{figure*}
\begin{center}
 \includegraphics[width=2.1\columnwidth]{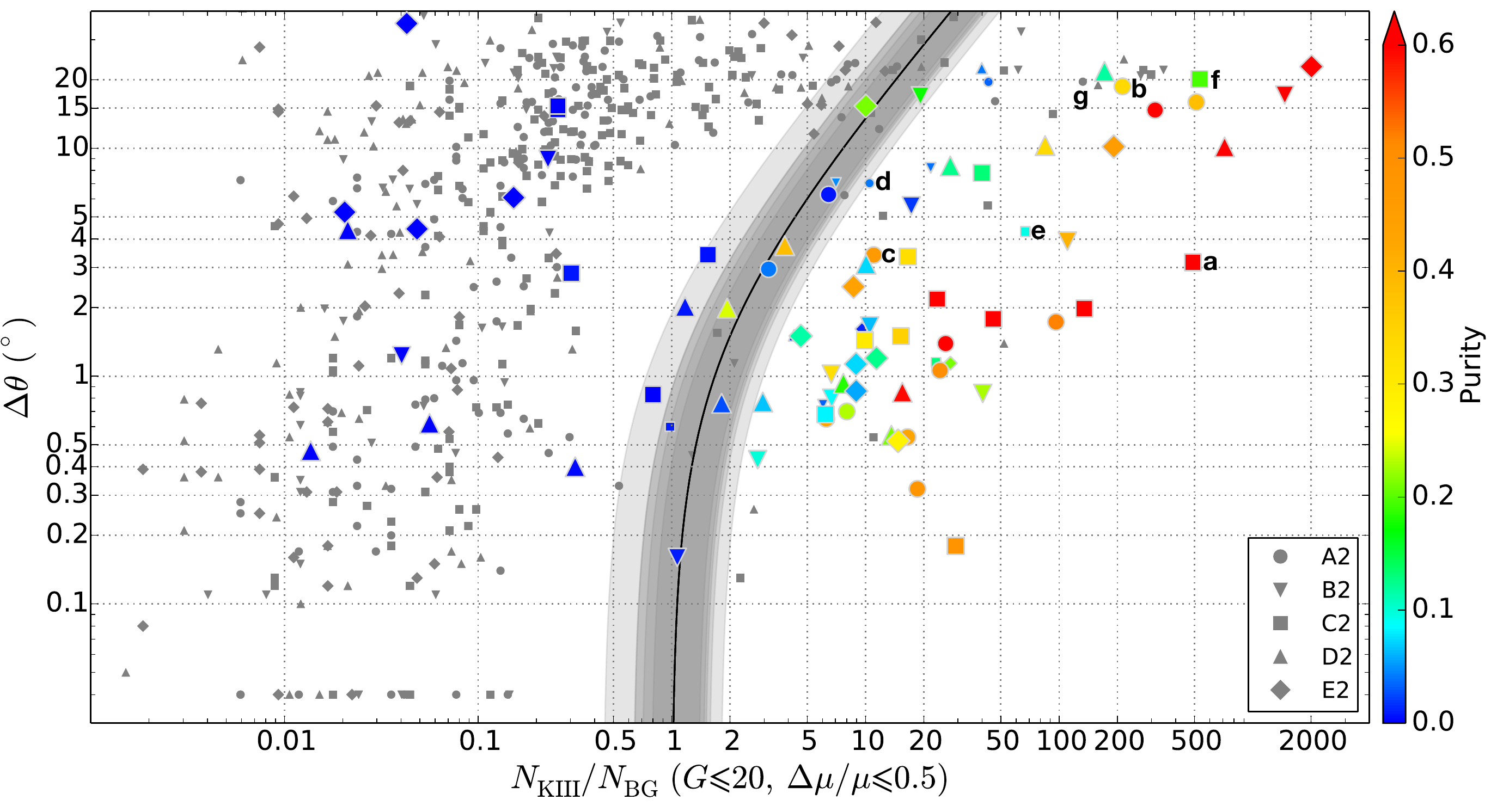} 
 \includegraphics[width=2.1\columnwidth]{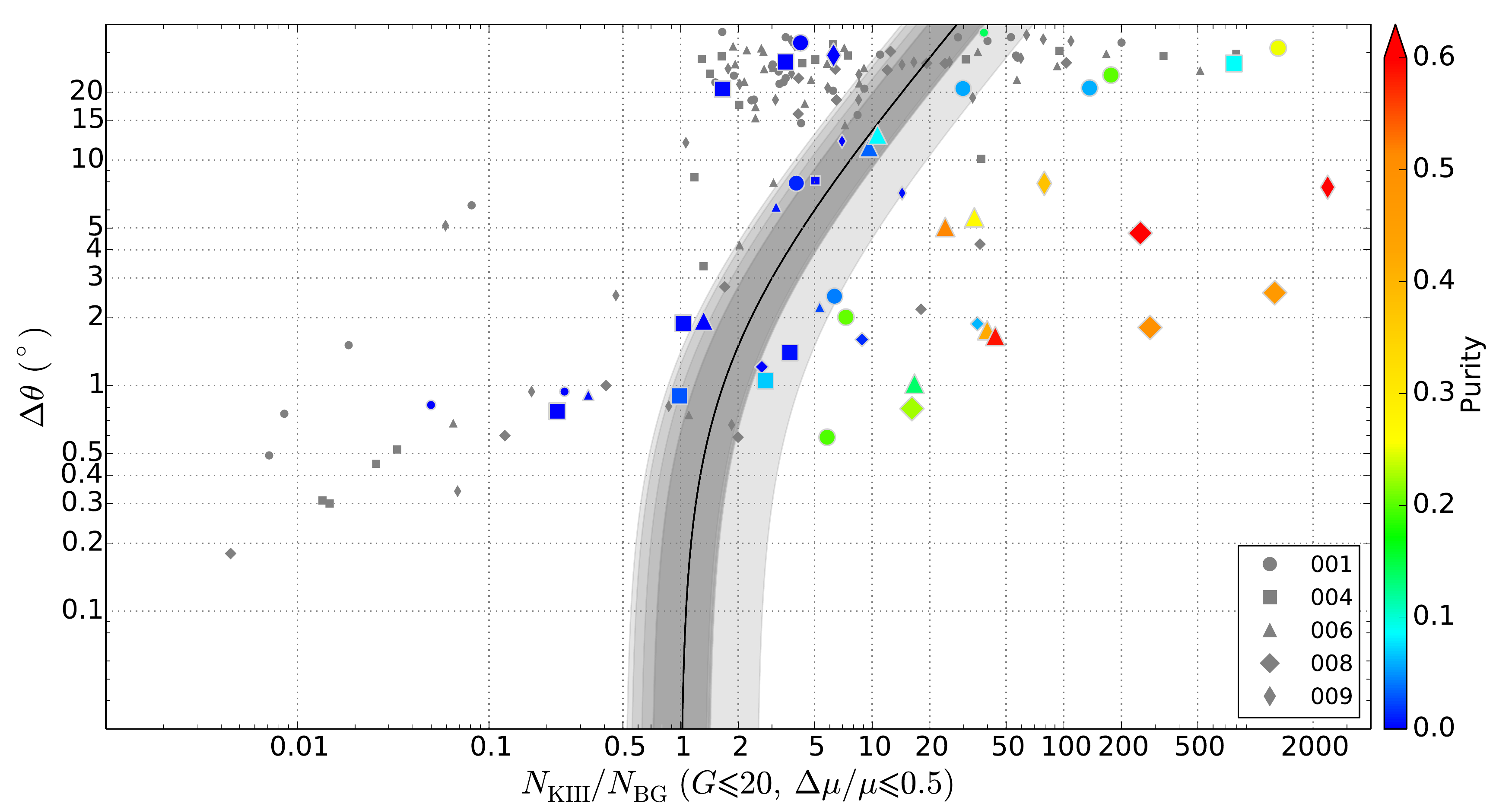} 
 \caption{Angular thickness $\Delta\theta$ vs number of observable K giants $N_{\rm KIII}$ for each of the progenitors of the Aquarius haloes. Large and small symbols denote progenitors recovered as primary and secondary detections, grey symbols indicate progenitors that are not recovered. The colour scale is proportional to purity, with the upper limit indicating a purity $=0.6$ or higher. The labelled points correspond to the same progenitors as in Fig.~\ref{f:z_N_a}. The black solid line and shaded regions in the right panel indicate respectively the median detection boundary and its edges computed from Eq.~\ref{e:boundary} using the 25th and 75th percentiles of the background counts.}
\label{f:dtheta_N_a}
\end{center}
\end{figure*}

The number of observable stars and the angular thickness of a tidal stream are two parameters that directly influence the detectability of a stream with \ngc~or any great-circle method in general. Obviously, the method is more efficient when more tracer stars are available and in cases when streams are dynamically colder.

Recovered and not recovered progenitors from the Aquarius and \ljmu~haloes shown in Fig.~\ref{f:z_N_a} are now shown respectively in the top and bottom panels of Fig.~\ref{f:dtheta_N_a} in the plane of angular thickness versus number of observable stars. The colour coding, symbols and labels are the same as in Fig.~\ref{f:z_N_a}. In this plot a clear segregation is evident as detected progenitors (coloured symbols) are fairly well separated from non-detections (grey symbols) and, in general, low purity detections tend to be those with fewer observable stars and larger angular thickness. 

In this plane we can estimate a priori where a detection boundary should lie based on how the great-circle methods work. Structures that are thinner than the assumed tolerance, and having more than some minimum number of stars above the background should be recovered, since all stars would fit inside a single great-circle cell. For wider structures to be detected, increasingly larger numbers of stars are needed to compensate for the fact that stars are dispersed into more than one great-circle cell, up to a certain angular width for which the method saturates.  Thus we propose the detection boundary can be expressed as

\begin{equation}\label{e:boundary}
{{\Delta\theta}\over{\delta\theta}} =
{{N_{\rm KIII}-N_{\rm{BG}}}\over{N_{\rm{BG}}}} =
{N_{\rm KIII}\over{N_{\rm{BG}}}}-1
\end{equation} 

In Eq.~\ref{e:boundary}, $\delta\theta$  is the tolerance used in producing the \ngc~PCMs (see Eqs. \ref{e:rcriterion}, \ref{e:vcriterion}) and $N_{\rm BG}$  the number of stars in the background, which can be estimated from  the PCM itself using the smoothed map computed during the unsharp-masking (e.g. Fig.~\ref{f:aqa2_unsharp}, left, for Aquarius A2). 
Therefore, \emph{the location of this boundary can be predicted without any free parameters, based on observables alone}. 
The solid black lines in Fig.~\ref{f:dtheta_N_a} represent the detection boundary given by Eq.~\ref{e:boundary}, taking $N_{\rm BG}$ to be the median of the counts in the smoothed PCM for each halo. The shaded regions were computed for each halo using the 25th and 75th percentiles of the respective smoothed PCM counts, so as to represent the uncertainty of this boundary due to the non-uniformity of the PCM smooth background. 

The eight progenitors labelled in Fig.~\ref{f:z_N_a} are also shown in Fig.~\ref{f:dtheta_N_a}. This clearly shows that all successfully recovered progenitors (${\rm a,b,c,d,e}$) are well inside the detection boundary. Note that progenitor ${\rm d}$ is a secondary detection, highly contaminated by the more luminous progenitor ${\rm b}$, and is located in the PCM (Fig.~\ref{f:pcm_peaks_tracer}) in a region with a relatively higher background. In Fig.~\ref{f:dtheta_N_a} it lies close to the grey bands marking the 75th percentile of the background counts, showing that it is almost a border-line detection. Progenitors ${\rm f}$ and ${\rm g}$, classified as a false positive and a non-detection respectively, are clearly seen here to be bright but very diffuse features with angular widths above $20\degr$. 

The two panels in Fig.~\ref{f:dtheta_N_a} show the detection boundary effectively separates detections from non-detections in both simulations, with most non-detections (grey points) and low purity detections (blue points) lying to the left of the boundary and good detections (green to red points) lying to its right. The qualitative behaviour of these plots is quite similar for both simulations. In both panels, most non-detections (grey symbols) lie above~$\Delta\theta\sim15\degr$, i.e. 10 times the great circle tolerance, so we take this to be the anticipate saturation limit of the method. Below this limit there are few non-detections, which shows that the overall recovery rate is quite good.

The statistics summarising the fractions and numbers of recovered progenitors are presented in Tables~\ref{t:rec_stats_aq}  and \ref{t:rec_stats_ljmu} for the Aquarius and \ljmu~haloes respectively. All statistics include both primary and secondary detections. In Table~\ref{t:rec_stats_aq}, for each tracer, the columns show: the median PCM background counts $N_{\rm BG}$; the overall fraction of streams and progenitors recovered $f_{\rm str}$, and in total, $f_{\rm all}$; the number of bound progenitors and streams inside the detection boundary in total $N^{\rm bnd}_T$, $N^{\rm str}_T$ and recovered $N^{\rm bnd}_T$ and $N^{\rm str}_T$. Here, we label as `streams' those progenitors with a fraction of bound stars $f_{\rm bound}\leqslant0.9$. This limit on $f_{\rm bound}$ is arbitrary but results are not very sensitive to the specific choice because, as noted by \citet{Cooper2010} for the Aquarius haloes, most progenitors are either completely bound ($f_{\rm bound}=1.0$) or almost completely unbound ($f_{\rm bound}<0.1$), which is also true for the \ljmus. Table~\ref{t:rec_stats_aq} also summarises the numbers of streams recovered with both RRLS \emph{and} K giants $N_{\rm RR\&K}$, those recovered with RRLS only $N_{\rm RRnotK}$ or K giants only $N_{\rm KnotRR}$, and when both tracers are combined $N_{\rm RR+K}$, which is simply the sum of the previous numbers ($N_{\rm RR+K}=N_{\rm RR\&K}+N_{\rm RRnotK}+N_{\rm KnotRR}$). 

Table~\ref{t:rec_stats_ljmu} presents the same summary statistics for the \ljmus. The results with the \insitu~background and the exclusion zone defined in Sec.~\ref{s:hydro_pcms} are presented in the right side of the table and, for comparison, results that would be obtained leaving out the \insitu~component are shown on the left side. The results obtained without the \insitu~component (left) are presented here to emphasize that the simple cuts used to filter out disc stars are very effective, since with them we recover as many progenitors (or more, in the case of halo 004) as in the case where we exclude the entire in situ background. Hence, \emph{the conclusions derived from the gas dynamical simulations are not driven by the presence of the in situ background}. For haloes 001, 006, 008 and 009, not only is the same number of progenitors recovered, but they are the same progenitors in both cases. For halo 004, as noted, two more progenitors are detected (one bound, one unbound) compared to the case when the \insitu~component is left out. In this particular case, this happens because many stars  from a massive and very heavily disrupted progenitor are removed by the exclusion zone, lowering the background enough for these peaks to be revealed.

For the Aquarius haloes, the overall fractions $f_{\rm str}$ and $f_{\rm all}$ in Table~\ref{t:rec_stats_aq} show that \emph{using K giants as tracers, a median 88\% of streams and 86\% of all progenitors (bound and unbound)  inside the boundary are recovered below the angular width limit of $15\degr$}. Out of these, a median 77\% are primary detections. \emph{When RRLS are used, the total recovery rate is only slightly lower but still very good, yielding a median of $88$\% for streams and $80$\% for all progenitors}, with 75\% of progenitors recovered as primary detections. The difference between the two tracers therefore does not lie in the relative efficiency, which is the same, but in the total number of progenitors that can be observed. 

For the \ljmu~haloes, using K giants, the median recovery fraction for streams is 75\%, for all progenitors it is 67\%, out of which a median 67\% are primary detections. Note that these recovery fractions are only slightly lower than obtained with Aquarius with the same tracer. \emph{This confirms that, within the boundary, the detectability is not significantly affected by the fact that progenitors are more easily disrupted in these simulations.}  

The detection boundary for the \ljmus~lies at $N_{\rm KIII}\sim N_{\rm BG}$ for small $\Delta\theta$, which Table~\ref{t:rec_stats_ljmu} shows is around 200--800. This overlaps with the limit of $N_{\rm KIII}\sim$500--1000, found in the previous section, where we estimate \ljmu~results might be hampered by a lack of progenitors due to the lower resolution of the simulations.  Therefore, in what follows, \emph{we will only use results from the Aquarius simulations in our analysis of the number of streams expected within the detection boundary}.

The Aquarius simulation results show that \emph{a total of 3--8 and 3--10 streams would be recovered successfully with \Gaia+\ngc~when observed with K giants and with RRLS respectively}. Note that, in the case of the Milky Way, \emph{this implies that Gaia could potentially double the number of known dwarf galaxy streams in the halo}. Since the detection limit of our method is well above the typical stellar mass associated with halos at the resolution limit of Aquarius (as shown by Fig.~\ref{f:dtheta_N_a}), we do not expect these results would be significantly different in a simulation with even higher resolution. The \ljmu{} results imply that \emph{a total of 2--7 streams would be recovered successfully with \Gaia+\ngc~with K giants}. 

A total of 2--6 or 0--1 bound progenitors would be recovered with K giants or RRLS respectively.  The difference between results with different tracers is more notable in the number of recovered bound progenitors than in the number of tidal streams. This is due to the combination of two factors: that K giants are observable up to distances twice as large as RRLS and that partially or completely unbound structures such as streams tend to spread stars out over larger ranges of (heliocentric) distance, making it more likely for these structures to have observable RRLS, in comparison to a bound progenitor, for which all stars lie at approximately the same distance. This is an interesting result as it shows that \emph{approximately the same number of streams can be recovered with RRLS as with K giants, even though RRLS probe a substantially smaller volume.}

In addition to the number of streams recovered being similar, there is the question of whether the two tracers recover the same streams or not. The last four columns of Table~\ref{t:rec_stats_aq} provide this information. $N_{\rm RRnotK}$ shows that there can be up to 5 streams recovered with RRLS that are not recovered with K giants. This demonstrates that \emph{there is something to be gained by using both tracers, instead of just the brighter one}. \emph{When results from RRLS and K giants are combined, 4--13 streams are recovered successfully ($N_{\rm RR+K}$), which implies a median gain of 2 extra streams compared to results obtained with K giants alone}.

The streams we are considering are tidal streams produced by dwarf galaxies, which is why the search tolerance has been tuned to the relatively large value of $\delta\theta=1\fdg5$ (see M11). With a lower tolerance, \ngc~could also identify the much narrower globular cluster streams, of which many are known in the Milky Way. For example, the Pal 5 tidal stream has a full width at half maximum (FWHM) of $0\fdg 3$ \citep{Odenkirchen2003} and the GD-1, Cocytos, Acheron and Lethe streams, all thought to have been produced by disrupted globular clusters, have FWHM of $0\fdg5$ \citep{Grillmair2006}, $0\fdg7$, $0\fdg9$ and $0\fdg4$ respectively \citep{Grillmair2009}. The resolution of the Aquarius (and \ljmu) simulations is not sufficient, however, to simulate globular clusters analogues. We therefore leave the exploration of the detectability of globular cluster streams for a future work.

Evidently, the estimates presented here are made under the assumption that the simulations we have used are representative of the Milky Way. Although this is not exactly the case (see Secs. \ref{s:aquarius} and \ref{s:gas_dynamical}), they still provide a useful estimate of the number of streams we can expect to detect with \Gaia+\ngc~and the selected tracers. 

The works of \citet{Sharma2011} and \citet{Elahi2013} have also explored the performance of other stream-finding algorithms, EnLink and S-tracker, VELOCIraptor, ROCKSTAR and HOT-6D respectively; and used the same definition of purity, among other statistics, to quantify their results. Our findings regarding the purities of the streams recovered by \ngc~(see Fig.~\ref{f:dtheta_N_a}) are competitive with their results: we find median purities of 0.44 and 0.55 for K giants and RRLS respectively; \citet{Sharma2011} obtain purities of $\sim0.66-0.70$ for their 2MASS M-giant and RRLS LSST synthetic samples; and \citet{Elahi2013} find purities of $\sim0.85$ for tidally disrupted subhaloes and $0.40$ for \emph{completely} disrupted subhaloes. This is remarkable, as our simulations include the effects of the \Gaia~selection function and observational errors, while the \citeauthor{Elahi2013} simulations are error-free and although \citeauthor{Sharma2011} simulate their M giant sample with similar distance errors (18\%) as our K giants, for the RRLS they assume the much deeper LSST selection function ($m_r=24.5$) and assume slightly better distance errors (5\%).

\begin{table*}
 \caption{Statistics of recovered progenitors in the Aquarius Haloes (primary + secondary detections combined). The columns are: the median PCM background counts $N_{\rm BG}$; the overall fraction of all progenitors and streams recovered $f_{\rm all}$ and $f_{\rm str}$ respectively; the number of progenitors inside the detection boundary ($\Delta\theta\leqslant15\degr$) in total and recovered respectively for bound progenitors $N^{\rm bnd}_T$, $N^{\rm bnd}_{\rm rec}$ and for streams $N^{\rm str}_T$, $N^{\rm str}_{\rm rec}$; and the numbers of recovered streams detected in common with RRLS and K giants $N_{\rm RR\&K}$, only RRLS $N_{\rm KnotRR}$ , only K giants $N_{\rm KnotRR}$ and combined $N_{\rm RR+K}$.}\label{t:rec_stats_aq}
\centering
\tabcolsep=0.11cm
\begin{tabular}{lrrrrrrrrrrrrrrrrrr}\hline\hline
& \multicolumn{7}{c}{RRLS}  & \multicolumn{7}{c}{KIII} & \multicolumn{4}{c}{RRLS+KIII ($N^{\rm str}_{\rm rec}$)} \\
 \cmidrule(r{0.5em}l{0.5em}){2-8} \cmidrule(r{0.5em}l{0.5em}){9-15} \cmidrule(r{0.5em}l{0.5em}){16-19}
Halo & $N_{\rm BG}$  & $f_{\rm all}$ & $f_{\rm str}$ &  $N^{\rm bnd}_{\rm T}$ & $N^{\rm bnd}_{\rm rec}$ &  $N_{\rm T}^{\rm str}$ & $N_{\rm rec}^{\rm str}$ &
$N_{\rm BG}$  &  $f_{\rm all}$ & $f_{\rm str}$ &  $N^{\rm bnd}_{\rm T}$ & $N^{\rm bnd}_{\rm rec}$ & $N_{\rm T}^{\rm str}$ & $N_{\rm rec}^{\rm str}$  &
$N_{\rm RR\&K}$ & $N_{\rm RRnotK}$ & $N_{\rm KnotRR}$ & $N_{\rm RR+K}$  \\
\hline 
A2  & 45   & 0.71 & 0.67&   1  & 1  & 6  & 4 & 168  & 0.86  & 0.75   & 6 & 6  & 8 & 6 &  1 &3 &5 &9  \\
B2  & 55   & 1.00 & 1.00&   0  & 0  & 7  & 7 & 251  & 0.91  & 1.00   & 3 & 2  & 8 & 8  &  7 &0 &1 &8  \\
C2  & 22  & 0.83 & 0.83 &   0  & 0  &12 &10 & 114 & 0.67  & 0.67   & 6 & 4  & 12   & 8   &  5 &5 &3 &13 \\
D2  & 111 & 0.78 & 0.88 &   1 & 0   & 8 &  7 & 624 & 0.85  & 0.88   & 5 & 4  &  8 & 7   &  5 &2 &2 &9   \\
E2  &  56  & 0.80 & 1.00 &   2  & 1   & 3 &  3 & 537 & 1.00  & 1.00   & 6 &  6 &  3 & 3&  2 &1 &1 &4   \\
\hline
\end{tabular} 
\end{table*}

\begin{table*}
 \caption{Statistics of recovered progenitors in the \ljmu~haloes, with K giant stars (primary + secondary detections combined). The columns are: the median PCM background counts $N_{\rm BG}$; the overall fraction of all progenitors and streams recovered $f_{\rm all}$ and $f_{\rm str}$ respectively;  the number of progenitors inside the detection boundary ($\Delta\theta\leqslant15\degr$) in total and recovered respectively 
for bound progenitors $N^{\rm bnd}_T$, $N^{\rm bnd}_{\rm rec}$ and for streams $N^{\rm str}_T$, $N^{\rm str}_{\rm rec}$.}\label{t:rec_stats_ljmu}
\centering
\tabcolsep=0.11cm
\begin{tabular}{lrrrrrrrrrrrrrr}\hline\hline
& \multicolumn{7}{c}{No \insitu~component}  & \multicolumn{7}{c}{With \insitu~Excl. $|b|\leqslant10^\circ$ \& $R\leqslant20$ kpc} \\
 \cmidrule(r{0.5em}l{0.5em}){2-8} \cmidrule(r{0.5em}l{0.5em}){9-15}
Halo & $N_{\rm BG}$  & $f_{\rm all}$ & $f_{\rm str}$ &  $N^{\rm bnd}_{\rm T}$ & $N^{\rm bnd}_{\rm rec}$ &  $N_{\rm T}^{\rm str}$ & $N_{\rm rec}^{\rm str}$ &
$N_{\rm BG}$  &  $f_{\rm all}$ & $f_{\rm str}$ &  $N^{\rm bnd}_{\rm T}$ & $N^{\rm bnd}_{\rm rec}$ & $N_{\rm T}^{\rm str}$ & $N_{\rm rec}^{\rm str}$  \\
\hline
001 & 1441 &       0.67 & 1.00     &  1& 0   & 2 & 2  & 703 & 1.00 & 1.00 &  1  & 1 & 2 & 2 \\
004 & 1673 &      1.00  & 1.00     &  0& 0    & 1 & 1 &  815 & 0.67 & 0.50&   1  & 1  & 2 & 1\\
006 &   445&        1.00 & 1.00     &  1& 1   & 7 & 7 &  446 & 1.00 & 1.00 &   1  & 1  & 7 & 7\\
008 &   138&       0.70  & 1.00     &  4& 1   & 6 & 6 &  223 & 0.60 & 0.67 &  2 & 3 & 6 & 4\\
009 &   626&       0.50  & 0.50     &  0& 0   & 4 & 2 &  540 & 0.50 & 0.75 &   0  & 0 & 4 & 3 \\ 
\hline
\end{tabular} 
\end{table*}

\subsection{The Progenitor Stellar Masses}\label{s:mass}

\begin{figure*}
\begin{center}
\includegraphics[width=2.2\columnwidth]{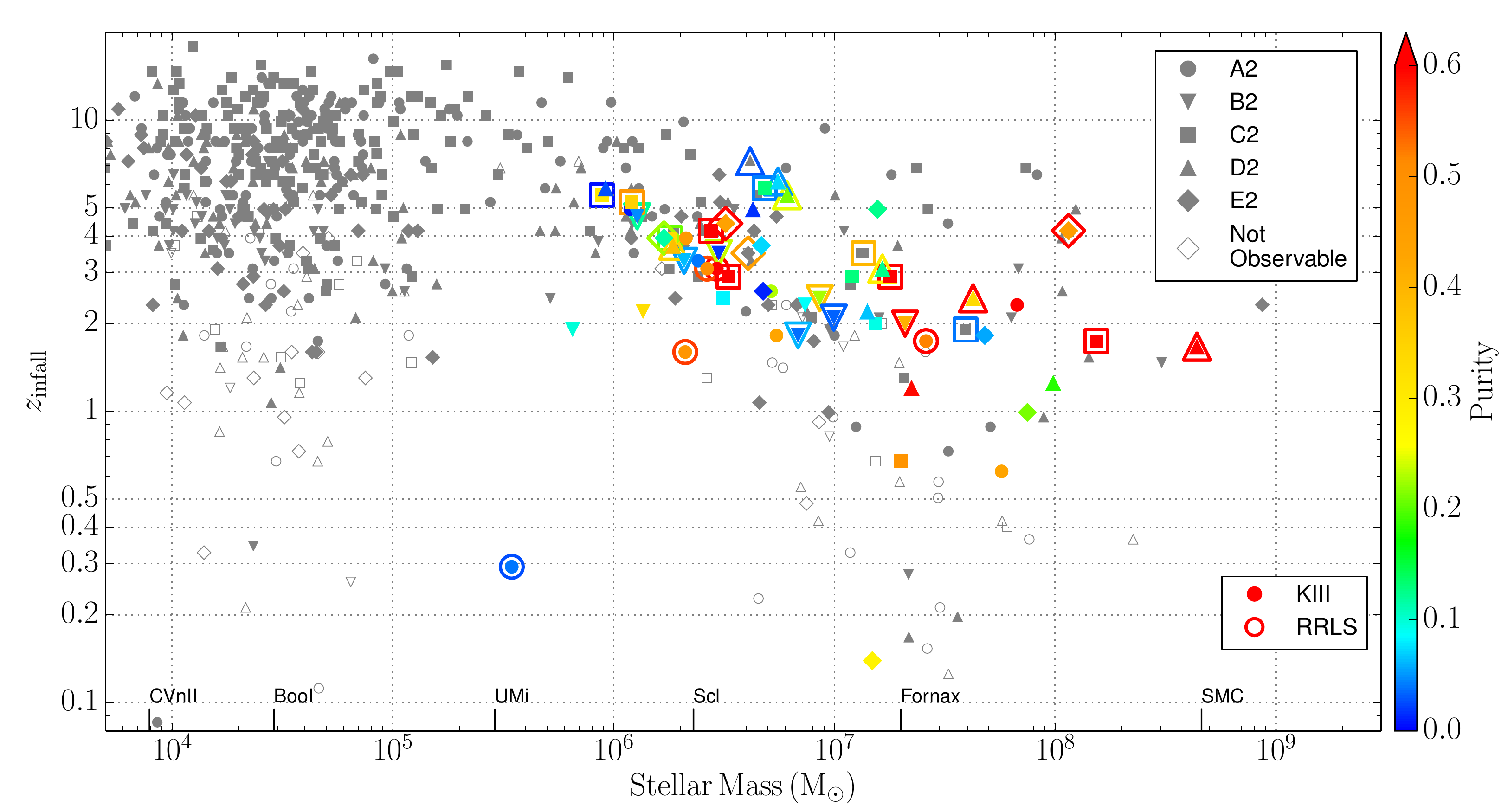} 
 \caption{Infall redshift vs Stellar Mass for progenitors in the Aquarius haloes. Progenitors successfully recovered (with $\Delta\theta\leqslant15\degr$) with K giants and RRLS are shown respectively with filled and empty coloured symbols, with a colour proportional to the purity of the detection. Small grey symbols denote progenitors that are: observable ($G\leqslant20$ and $\Delta\mu/\mu\leqslant0.5$)~but not recovered (grey filled), or not observable at all~(grey empty). The stellar mass of the SMC and a few classical (Fornax,Scl,UMi) and ultra-faint (BooI,CVnII) dwarf spheroidal Milky Way satellites from \citet{McConnachie2012} are shown in the bottom axis for reference. 
 The stellar mass scale divided by $3$ is roughly equivalent to an $L_V$ luminosity scale, for galaxies dominated by an old metal-poor population ($>10$ Gyr, $\FeH=-1.5$). }
\label{f:z_Mass}
\end{center}
\end{figure*}

A plot of the infall redshift versus total stellar mass for all progenitors in the Aquarius haloes is shown in  Fig.~\ref{f:z_Mass}. All progenitors with at least one observable star ($G\leqslant20$ and $\Delta\mu/\mu\leqslant0.5$) are plotted with  grey filled symbols, progenitors with no observable stars are plotted as open grey symbols. Recovered progenitors inside the detection boundary, below the angular width threshold $\Delta\theta=1.5\degr$, are shown with coloured symbols, the colour scale being proportional to the purity. Spurious detections (left of the detection boundaries in Fig.~\ref{f:dtheta_N_a}) are not shown.  Filled and empty coloured symbols denote progenitors recovered with K giants and RRLS, respectively. 

Fig.~\ref{f:z_Mass} shows that recovered progenitors have masses down to a few times $10^6$ $\Msun$, similar to or lower than that of the Sculptor dwarf spheroidal galaxy \citep{McConnachie2012}, and even below $10^6\Msun$ in a handful of cases. This mass limit is of the order of the least massive `classical' dwarf spheroidal satellites of the Milky Way ($\sim3\times10^5\Msun$, Ursa Minor and Draco). Interestingly, \emph{progenitors can be recovered down to this mass limit and in the same mass range with both tracers}, as evidenced in the plot by the fact that there are empty and filled coloured symbols spanning the same mass range and overlapped in most cases.  
In fact, the only clear difference between detection with both tracers, in terms of their distribution in this plane, is that progenitors accreted relatively recently $z_{\rm{infall}}\lesssim1$ are detected only with K giants. Note also that the majority of progenitors that are not observable also lie in this redshift range, this is precisely because, having been accreted only recently by the main halo, most of these progenitors are almost completely bound and very distant so only their brightest stars are observable.  

As we have seen in Sec. \ref{s:det_boundary} and Table~\ref{t:rec_stats_aq}, the same number of \emph{streams} (unbound progenitors) in total are detectable with RRLS as with K giants, but the number of \emph{bound progenitors} is smaller because RRLS are intrinsically fainter and thus probe a smaller volume. Fig.~\ref{f:z_Mass} shows that \emph{progenitors recovered with RRLS are not limited to the most massive/luminous ones, but span the same mass range as  those recovered with K giants}. 

\subsection{The Progenitor Distance Distribution}\label{s:distance}

\begin{figure*}
\begin{center}
\includegraphics[width=1.5\columnwidth]{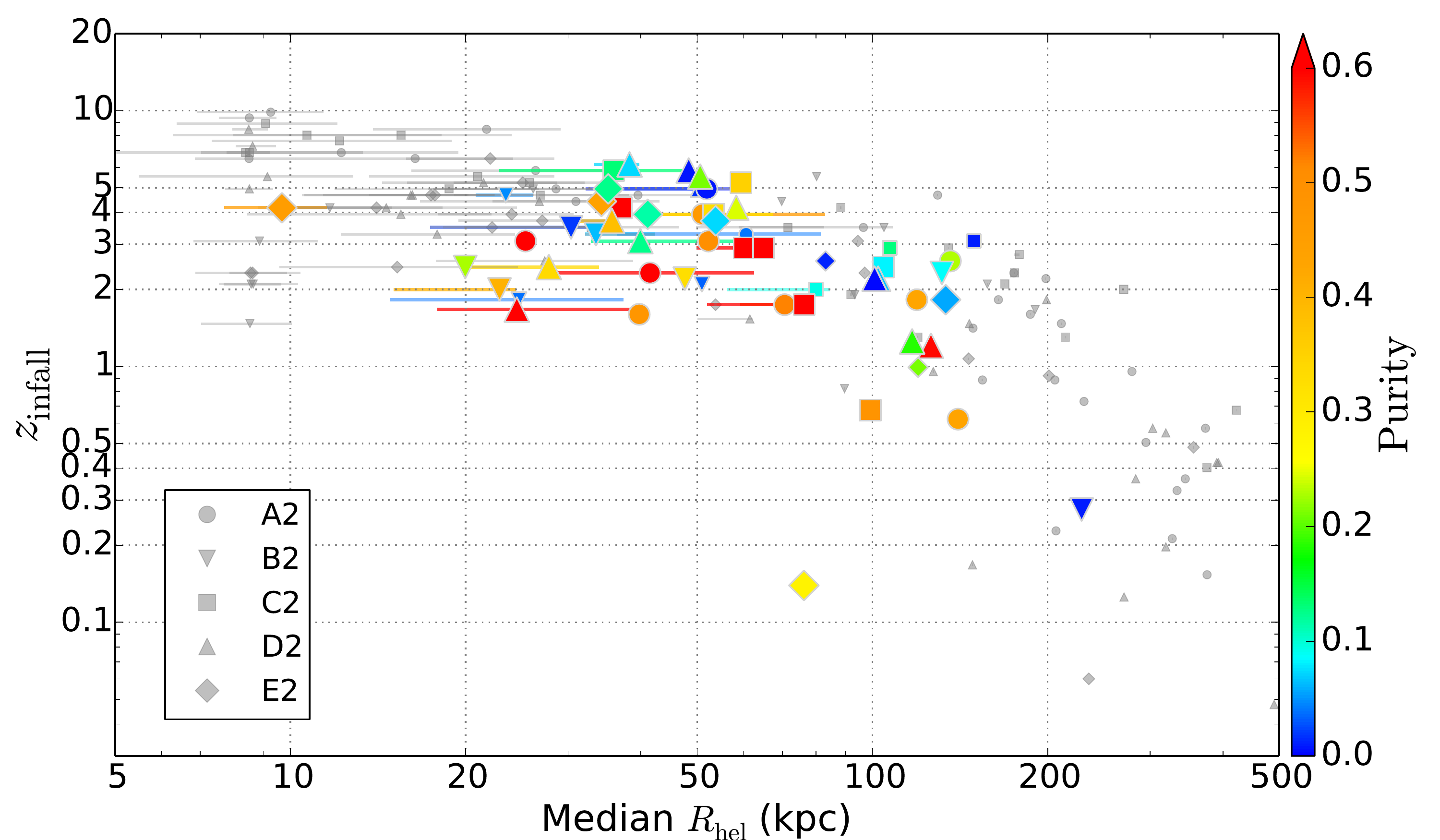} 
\includegraphics[width=1.5\columnwidth]{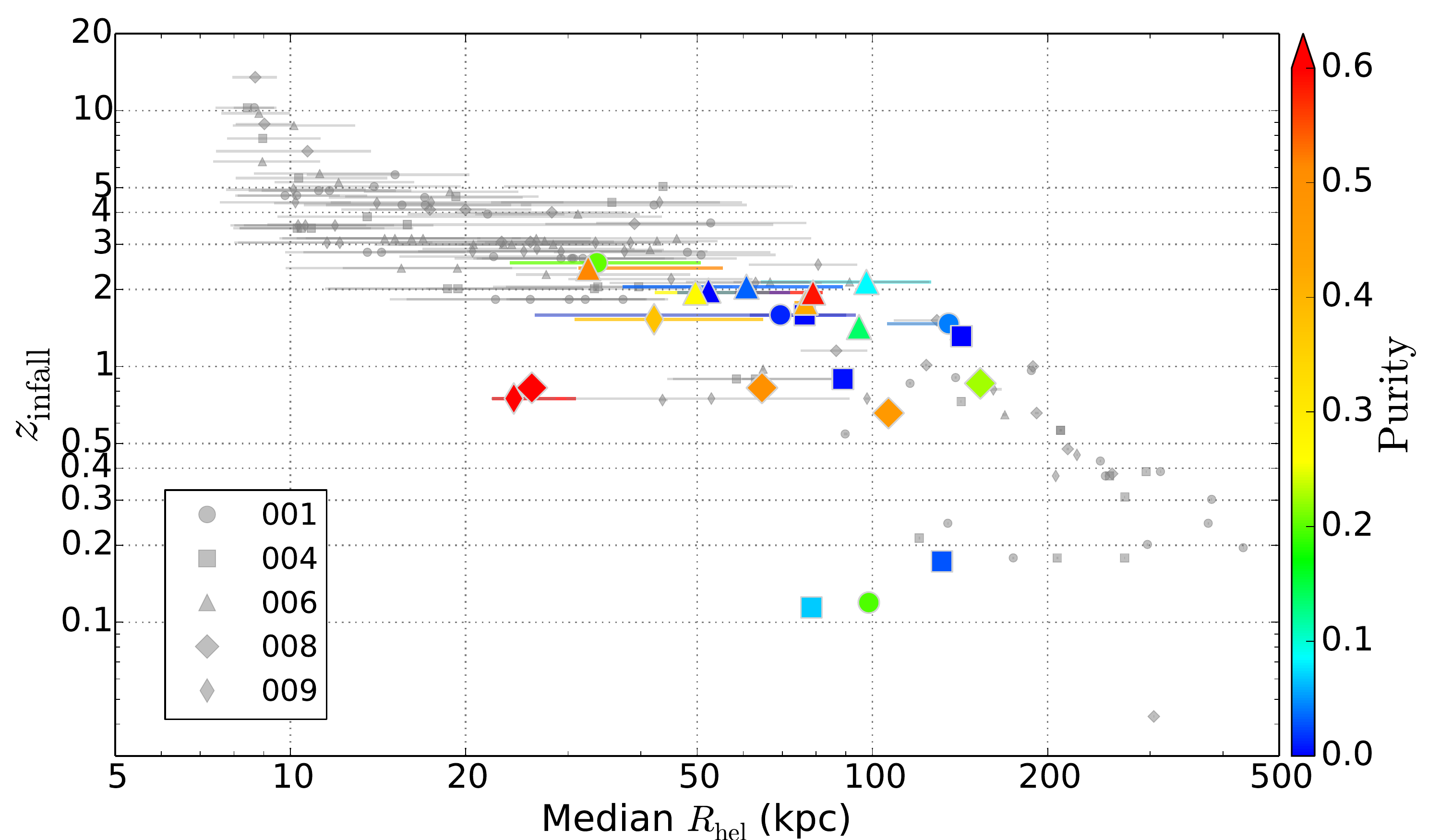} 
 \caption{Infall redshift vs median heliocentric distance for progenitors in the Aquarius (left) and \ljmu~(right) haloes. Only progenitors more massive than $10^6\Msun$ are shown. As in the previous figure, progenitors successfully recovered (with $\Delta\theta\leqslant15\degr$) with K giants and RRLS are shown respectively with filled and empty coloured symbols, with a colour proportional to the purity of the detection. Small grey symbols denote progenitors that are observable but not recovered (grey filled). The error bars denote the inter quartile range of the heliocentric distance distribution of the stars in each progenitor.}
\label{f:z_dist}
\end{center}
\end{figure*}

The heliocentric distance distribution of progenitors is illustrated in Fig.~\ref{f:z_dist}. The panels show the infall redshift as a function of the median heliocentric distance for progenitors more massive than $10^6\Msun$ in the Aquarius (left) and \ljmu~(right) haloes. As in the previous figures, filled and open symbols coloured symbols denote progenitors recovered respectively with KIII and RRLS and grey filled and open symbols progenitors that are not recovered or not observable respectively. Error bars depict the inter-quartile range of the heliocentric distance distribution of stars in each progenitor, so points with no visible error bars correspond mostly to bound progenitors and, less commonly, to very distant streams in almost perfectly circular orbits. 

The two panels in Fig.~\ref{f:z_dist} show that progenitors overall are recovered in the $\sim$20 to 130 kpc distance range. At the lower distance end, the plot shows detections down to $\sim$20--30 kpc in both types of simulations, the dark-matter-only Aquarius and the gas-dynamical \ljmus. As noted in M11 and \citet{Smith2016}, great circle methods are expected to work best at intermediate to large distances ($>$20--30 kpc), where tidal streams are less affected by the presence of the disc and by phase mixing given the longer dynamical time scales. This is reinforced by the $z_\mathrm{infall}-R_\mathrm{hel}$ trend observed in the plot, that shows streams in the inner Galaxy ($<$20kpc) were accreted at higher $z_\mathrm{infall}$, which translates into larger stream widths $\Delta\theta$ (Fig.~\ref{f:dtheta_infallz}) making detection more difficult. At the higher distance end, progenitors are much more recent infallers and there are no streams (i.e. unbound progenitors) beyond 100 kpc in the Aquarius simulations, and only a couple in the \ljmus.

The \ngc~method is most efficient at detecting streams in the distance range from $\sim$30 to $\sim$90 kpc. As 
the left panel shows, most progenitors are unbound ones and the majority are recovered (84\% and 80\%  with KIII and RRLS respectively). Therefore, based on the Aquarius simulation results, \emph{in this distance range we expect a search for streams using either KIII or RRLS to be fairly complete ($>$80\%) for progenitors more massive than $10^6\Msun$}. As discussed in Sec.~\ref{s:det_boundary}, we cannot draw any conclusions about the completeness of the search with the \ljmu~simulations due to limitations caused by mass resolution.

The distribution of open and filled symbols in the left panel also shows no distance bias in the stream detections made with KIII compared to RRLS up to a median distance of $\sim$90 kpc. This is an interesting result as it is rather counter-intuitive since the maximum distance up to which RRLS will be observable with \Gaia~is $\sim$50 kpc (see Sec. \ref{s:gaia_can_see} and Figs.~\ref{f:Rhel_Mv_gaia} and \ref{f:lRhel}). However, in streams, stars are scattered across a range of distances making them still detectable at larger median distances. Combined with results from Secs.~\ref{s:det_boundary} and \ref{s:mass}, this means that \emph{RRLS probe the same effective volume ($\sim$20--90 kpc) and mass range ($\gtrsim10^6\Msun$), with a similar completeness as KIII stars when it comes to streams}.

\section{Gaia Mission Lifetime Extensions}\label{s:gaia_extensions}

In what follows we will evaluate the effect a possible extension of the \Gaia~mission will have on the number of progenitors we expect to detect with \Gaia+\ngc.

An extension of the \Gaia~mission lifetime will translate into improved proper motion and parallax precisions. The survey completeness magnitude will remain the same, $G=20$, as this is set by the available antenna time to download data for the stars detected on-board up to the set magnitude limit. Since we have assumed throughout this work that photometric distances will be used for our tracers and we have so far neglected radial velocities, in what follows we will only consider the effect of the increase in the proper motion precision. 

We will consider the following three scenarios:
\begin{itemize}
\item a two-year extension of \Gaia, increasing the total mission time to 7 yr 

\item a five-year extension of \Gaia, increasing the total mission time to 10 yr

\item a \Gaia~twin mission launched in 20 yrs time, increasing the overall mission baseline to 25 yr
\end{itemize}

The first two are realistic short-term scenarios, depending on the satellite's fuel budget and instrument performance at the end of the nominal mission lifetime of 5 yr in 2019 (A. G. A. Brown, private communication). In these two scenarios, we assume the proper motion errors will decrease by the expected factor of $(t_{\rm nom}/\tm)^{3/2}$, where $\tm$ is the new mission lifetime and $t_{\rm nom}=5$ yr is the nominal mission duration\footnote{The factor of $t^{3/2}$ comes from a factor of $t^{1/2}$ due to photon noise reduction and a factor $t$ due to the longer time baseline (A.G.A. Brown and J. de Bruijne, private communication).}.

The third scenario is a medium term possibility. In this case we assume the proper motion errors will decrease by a more conservative factor of $(t_{\rm nom}/\tm)$, as even though there would be a much longer baseline of 25 yr, there will be a gap in the data for the $\sim20$ yr in between the two missions (A.G.A. Brown and J. de Bruijne, private communication).

\begin{figure}
\begin{center}
 \includegraphics[width=1.05\columnwidth]{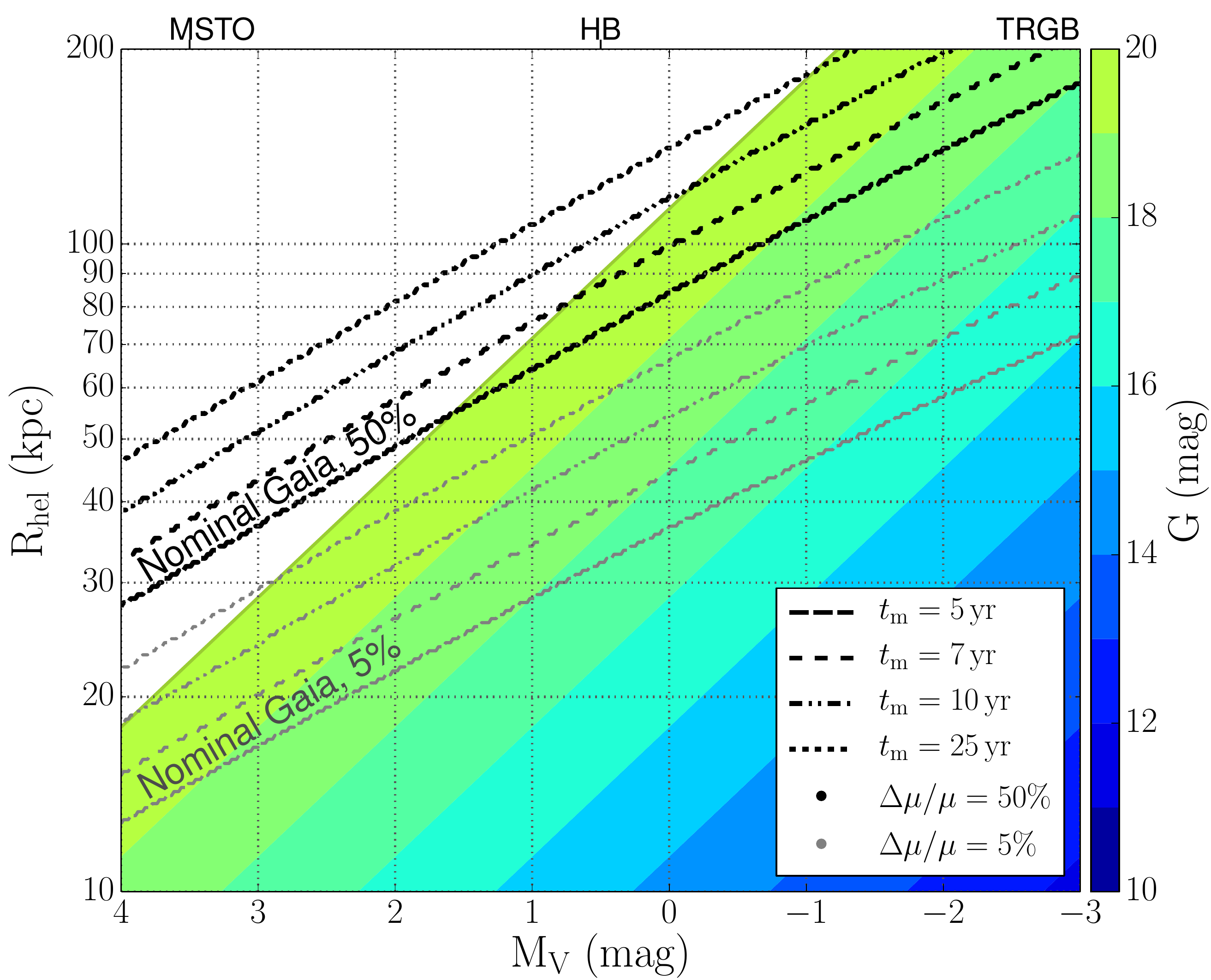} 
 \caption{Proper motion relative error horizons for the different \Gaia~mission lifetime scenarios, in the heliocentric distance $R_{\rm hel}$ versus absolute magnitude $M_V$ plane. The colour scale is proportional to the apparent $G$ magnitude and goes up to the \Gaia~magnitude limit ($G\leqslant20$), assuming a fixed $V-I=1$ colour and $A_V=0$. Grey and black lines respectively show horizons for proper motion relative errors of 5 and 50\%. The nominal mission lifetime $\tm=5$ yr horizon is indicated with the long-dashed line (also shown in Fig.~\ref{f:Rhel_Mv_gaia} with black long-dashes); the three possible extension scenarios of $\tm=7$ yr, $\tm=10$ yr  and $\tm=25$ yr are indicated respectively with short-dashed, dash-dotted and dotted lines. The absolute magnitudes of MSTO, HB and TRGB stars are shown for reference on the top axis.}
\label{f:Rhel_Mv_gaia_future}
\end{center}
\end{figure}

Fig.~\ref{f:Rhel_Mv_gaia_future} shows a new version of Fig.~\ref{f:Rhel_Mv_gaia} illustrating the proper motion error prescriptions expected for each of the three scenarios. The scaling factors for the proper motion errors are $0.60$, $0.35$ and $0.20$ for the $\tm=7$ yr, $\tm=10$ yr and $\tm=25$ yr scenarios respectively.

For each scenario we produce \Gaia~mock catalogues of the Aquarius simulations, rescaling the proper motion error prescriptions by the appropriate factors. As we have already shown, the \ngc~has a clear detection boundary, we simply need to compute the full \ngc~PCM for each halo to estimate the median pole-counts in the smoothed PCM (see Sec.\ref{s:peak_detection} and, using Eq.~\ref{e:boundary}, count how many progenitors lie inside the detectability boundary below the $\Delta\theta=15\degr$ threshold we have used so far. This gives us $N_{\rm T}$, the total number of detectable progenitors. We also compute $f_{\rm 70}$, the fraction of progenitors which have more than 70\% of their stars observable by \Gaia~with proper motion precision better than 50\%. These results are summarised in Table~\ref{t:future_gaia} for RRLS and KIII, starting with the nominal mission lifetime $\tm=5$ yr for comparison (leftmost columns), followed by the three scenarios considered.
 
Table~\ref{t:future_gaia} shows that, in general, the number of detectable progenitors $N_{\rm T}$ increases by only 1 in each scenario with respect to the previous one, so that \emph{for any halo an overall increase of 2-3 progenitors at most is expected in the last scenario} of a second \Gaia~mission in 25 yrs time, with respect to the nominal 5 yr mission lifetime. 
This seems like a relatively small gain, but $f_{\rm 70}$ shows why this should be the case:  in all the Aquarius haloes, \emph{more than half the progenitors will have over 70\% of their tracer stars meeting the proper motion criterion, even for the nominal 5 yr mission lifetime}. This fraction increases up to $\sim70$\% for the $\tm=7$ yr mission lifetime scenario and up to $\sim80$ to $95$\%, depending on the halo, for the $\tm=25$ yr scenario of a second \Gaia~mission. 

The last column of Table~\ref{t:future_gaia} gives $N_\infty$, the number of detectable progenitors expected in the limiting case considering \Gaia~observable stars but without any observational errors. This gives us an estimate of how many more progenitors we could expect to get in an ideal case, with a fixed $G=20$ \Gaia~completeness limit. This shows, that a maximum number of 15 to 26 progenitors, depending on the halo, could be detectable with KIII stars and \Gaia+\ngc~in the ideal error-free case. So, with K giants, there would still room for improvement since in the $\tm=25$ yr scenario the number of detectable progenitors could be increased by 20 to 50\% depending on the halo. 

On the other hand \emph{for RRLS it is clear that, even for the nominal mission time, the number of detectable progenitors is quite close to $N_\infty$}; and it reaches this limit, for all five Aquarius haloes, in the $\tm=25$ yr scenarios. In other words, \emph{all progenitors that could be detected with RRLS in an ideal \Gaia~error-free case, are indeed detectable as they do lie inside the method's detection boundary}. 
The fact that this happens for RRLS and not K giants is most likely due to the notably smaller distance errors RRLS have in comparison to K giants; which even though the \gc~methods have been implemented so as to minimise the effect of distance errors,  are still expected to have an impact (see M11). This result highlights the importance of obtaining radial velocities for RRLSs identified during the nominal mission time, which will be necessary for removing contaminants and for detailed modelling of the stream candidates found.

Although there is an increase in the number of detectable progenitors, the expected improvement in the future \Gaia~scenarios explored is relatively modest, considering these numbers could improve by up to a further $\sim50$\% for K giants.
In the next section we comment on possible strategies to improve upon these results.

\begin{table}
 \caption{Statistics of the progenitor recovery for the different \Gaia~lifetime scenarios. $N_{\rm T}$ is the total number of detectable progenitors and $f_{\rm 70}$ the fraction of progenitors with more than 70\% of stars observable by \Gaia.}\label{t:future_gaia}
\centering
\tabcolsep=0.11cm
\begin{tabular}{cccccccccc}
\hline\hline
\multicolumn{10}{c}{KIII}\\
\hline 
& \multicolumn{2}{c}{$\tm=5$ yr} & 
\multicolumn{2}{c}{$\tm=7$ yr} & 
\multicolumn{2}{c}{$\tm=10$ yr}  &
\multicolumn{2}{c}{$\tm=25$ yr} & 
\\
 \cmidrule(r{0.4em}l{0.4em}){2-3} \cmidrule(r{0.4em}l{0.4em}){4-5} \cmidrule(r{0.4em}l{0.4em}){6-7} \cmidrule(r{0.4em}l{0.4em}){8-9}
Halo & $N_{\rm T}$ & $f_{\rm 70}$ &  $N_{\rm T}$ & $f_{\rm 70}$ &  $N_{\rm T}$ & $f_{\rm 70}$ &  $N_{\rm T}$ & $f_{\rm 70}$ & $N_\infty$ \\
\hline
A2 & 14 & 0.66 & 15 &  0.75 &  16 &  0.77 &  17 &  0.84 & 26 \\
B2 & 11 & 0.75 & 13 &  0.75 &  14 &  0.82 &  14 &  0.90 & 16 \\
C2 & 18 & 0.66 & 18 &  0.71 &  18 &  0.82 &  19 &  0.86 & 26 \\
D2 & 13 & 0.51 & 14 &  0.63 &  14 &  0.69 &  14 &  0.78 & 18 \\
E2 &   9 & 0.51 & 10 &  0.66 &  11 &  0.75 &  12 &  0.81 & 15 \\
\hline
\multicolumn{10}{c}{RRLS}\\
\hline
A2 & 7   & 0.70 &  9  & 0.77 &   9 &  0.86 &   9 &  0.93  &  9\\
B2 & 8   & 0.70 &  8  & 0.83 &   9 &  0.87 &  11 &  0.93 & 11\\
C2 & 12 & 0.59 & 12 & 0.67 & 13 &  0.85 &  13 &  0.94 & 13\\
D2 & 10 & 0.60 & 11 & 0.76 & 11 &  0.86 &  11 &  0.95 & 11\\
E2 &  5  & 0.67 &  6  & 0.86 &   6 &  0.95 &   7 &  0.95  &  7\\
\hline
\end{tabular}
\end{table}

\section{Pushing the detection boundary: recommendations and improvements}\label{s:recommendations}

Throughout this work we have analysed the performance of the \ngc~method detecting 
tidal streams and satellites in cosmological simulations and we have discussed some possible
recommendations and improvements for the time this can be applied to real data.
Our suggestions and recommendations can be summarised as follows:

\begin{itemize}
\item Radial velocities will help reduce foreground/background contamination, in as much as they can be obtained for large samples of stars, and they are also necessary to disentangle different streams that share an orbital plane. 
Spectroscopic surveys planned and ongoing like LAMOST, WEAVE, 4MOST and DESI \citep{Liu2014,Dalton2012,deJong2015,Eisenstein2015} will make an important contribution in this respect, providing radial velocities for K giants spanning large portions of the volume probed by \Gaia. 

\item \Gaia~radial velocities could also be incorporated, when available, by combining the \mgc~PCM for stars with full 6D information with \ngc~PCMs for the remaining stars.

\item Great circle cell count methods are linear, so PCMs from different tracers could be combined by simple addition. As we discuss in Sec. \ref{s:tracer}, for a given survey, it would be optimal to add PCMs from different tracers \emph{after} unsharp-masking.

\item The use of simple cuts can effectively reduce background contamination minimising its effect in progenitor detectability, as we have shown in Sec. \ref{s:det_boundary}. Thus, it would prove useful to analyse the use of other cuts that can help reduce the background even further. 

\item Further improvements of this method can be made by combining it with the chemical abundance information. For example, knowing that many intermediate and metal-rich stars in the halo belong to tidal debris from massive satellite galaxies or from those accreted more recently \citep{Gilbert2009}, the number of detections can be maximised by targeting this metallicity range preferentially. A broad classification as metal-poor, intermediate or metal-rich will be feasible with \Gaia~BP/RP spectro-photometry, which can be used to produce separate PCMs in each metallicity bin. We intend to test in the future how the combination of chemical abundances and GC3 methods can improve on the recovery of substructure in the halo.

\end{itemize}

Other benefits will come from improvements in the peak detection algorithm and the pole-counting strategy such as: (i)  using a deblending algorithm in the peak detection and incorporating the fact that peaks in PCMs stretch along great circles arcs \citep{Torii2005} (ii) weighing the contribution of stars to poles proportionally to the observational errors, (iii) assigning pole-membership probabilities to each star and (iv) using the full sphere in pole-space to differentiate structures with different sense of rotation.

\section{Conclusions}\label{s:concl}

Tidal streams are widely recognised for their usefulness in the inference of the Galactic accretion history, one of the key science drivers for the Gaia mission \citep{deBruijne2012}. However, any such inference demands a thorough understanding of the selection biases that may affect tidal stream detection methods. 

Motivated by this, and the prospects that the Gaia mission opens up for all-sky homogeneous stream surveying, we have explored the detectability of tidal streams in Gaia mock catalogues using \ngc, a great-circle cell counts method that uses positional information and proper motions \citep{Abedi2014}. We have built mock catalogues for two standard candle tracers: K giants and RRLSs, reproducing the Gaia selection function and observational errors, and assuming photometric distance errors of 20 and 7\% respectively for each tracer. These mock catalogues were made from a set of 5 haloes from the Aquarius N-body simulations and 5 haloes from the \ljmu~gas dynamical simulations.
The diversity of orbits and progenitors in these allows us to characterise the \ngc~method's completeness and detection limits in a realistic setting. We have also explored how the \insitu~stellar halo background in \ljmu~gas dynamical simulations affects the detection of streams, and the improvements in proper motion errors expected for three possible extensions of the \Gaia~mission.

We summarise our results as follows:

\begin{enumerate}

\item The \ngc~method is able to identify realistic tidal streams produced in cosmological N-body and gas dynamical simulations, even when contamination from a smooth halo background is included.

\item The method has a \emph{well defined parameter-free detection boundary} in the plane of angular width vs. ratio of observable to PCM background stars, defined in Equation~\ref{e:boundary}.

\item Progenitors are recovered up to infall redshifts as large as $z_{\rm{infall}}\sim3$ based on results with the gas dynamical simulations, in which progenitors are more prone to disruption.

\item A total of 9 to 12 progenitors, bound and unbound, are expected to be detectable with \Gaia+\ngc~using KIII stars as tracers; and 4 to 10 using RRLS. These correspond respectively to a median 86\% and 80\% of all progenitors inside the detection boundary, below our selected threshold of $\Delta\theta=15\degr$. \label{i:res_i}

\item A total of 3 to 8 \emph{streams} would be recovered successfully with Gaia+nGC3 when observed with K giants and 3 to 10 with RRLS. Depending on the specific merger history of the Milky Way this means that \Gaia~ has the potential to almost double the number of known tidal streams in the halo. Also, approximately the same number of streams can be recovered with RRLS as with K giants, even though RRLS probe a substantially smaller volume. 

\item When results from RRLS and K giants are combined, 4 to 13 streams are recovered successfully ($N_{\rm RR+K}$), which implies a median gain of 2 extra streams when compared to results obtained with K giants alone. 

\item The stellar masses and luminosities of recovered progenitors go down to $\sim10^6\Msun$ and $\sim4\times10^5\Lsun$ respectively, i.e. similar to the classical dwarf spheroidal MW satellites.\label{i:res_f}

\item Our forecasts in items \ref{i:res_i}-\ref{i:res_f} are based on results from the Aquarius simulations alone, since HYDRO-zoom results may be hampered due to their lower mass resolution.

\item Progenitors are recovered down to the same stellar mass limit and the same infall redshift range with either tracer, RRLS or K giants. 

\item Recovered progenitors span the heliocentric distance range from $20$ to $130$ kpc, with the best completeness ($>$80\%) achieved in the range from $\sim$30 to $\sim$90 kpc.

\item For streams (i.e. partially unbound progenitors), RRLS probe \emph{the same effective volume} ($\sim$20--90 kpc) and mass range ($\gtrsim10^6\Msun$), with a similar completeness, as KIII stars. For bound progenitors KIII stars probe a larger volume, reaching out to $\sim$130 kpc.

\item We analysed the detectability of progenitors also for gas dynamical simulations which naturally include the \insitu~background. Although, as expected, the contamination from this additional background hinders the detections, we find that using a simple cut to exclude the disc stars ($|b|\leqslant10\degr$ \& $R\leqslant20$ kpc) one can recover as many progenitors (or more in one case) as in the case when  the in situ component is not taken into account. 

\item We analysed how the detectability of progenitors would be improved by the smaller proper motion errors resulting from an extension of the \Gaia~mission lifetime. The three scenarios considered were a two-year extension, a five-year extension and a second \Gaia~mission launched in 20 yr. In these scenarios, proper motion errors would be reduced by factors of 0.6, 0.35 and 0.2 respectively. Increases of about one, two and three progenitors respectively are expected in each scenario with respect to the results found for the nominal mission lifetime. 

\end{enumerate}

Finally, the K giant and RRLS \Gaia~mock catalogues produced for both the Aquarius and \ljmu~simulations are publicly available at \href{https://cmateu.github.io/Cecilia_Mateu_WebPage/Gaia_Halo_Mocks.html}{this URL}. These catalogues include, for each star, all the position and velocity information in heliocentric spherical and galactocentric cartesian coordinate systems, with and without simulated \Gaia~errors, including the pole ID indicating  to which pole detection  (if any) it is associated in the \ngc~PCM. Each pole detection catalogue thus represents a realistic set of stream detections in which streams may overlap, stars will be missing and there will be contamination from the smooth background and foreground. These catalogues will be a useful benchmark for further studies on the inference of the Galactic accretion history and gravitational potential.

\section*{Acknowledgments}

We thank the anonymous referee for a careful reading of our manuscript and for suggestions that helped improve its clarity. CM and LA acknowledge support from DGAPA/UNAM grant IG100115. CM acknowledges the support of the post-doctoral fellowship of DGAPA-UNAM, Mexico; and the European Commission's Framework Programme 7, through the Marie Curie 
International Research Staff Exchange Scheme LACEGAL (PIRSES-GA- 2010-269264), 
and is grateful for the hospitality of LJMU and the Institute for Computational Cosmology (ICC), Durham University, where part of this research was carried out. LA also acknowledges the hospitality of the Institute for Computational Cosmology (ICC), Durham University. APC and WW are supported by the COFUND Junior Research Fellowship scheme under EU grant 267209 and acknowledge support from STFC (ST/L00075X/1). Gaia error simulations were carried out using ATAI, a high performance cluster, at IA-UNAM. The Aquarius Project was carried out at the Leibniz Computing Centre and Computing Centre of the Max-Planck-Society in Garching, Germany; on the ICC COSMA system; and on the STELLA supercomputer of the LOFAR experiment at the University of Groningen. The mock catalogues were generated with the DiRAC Data Centric system at Durham University, operated by the ICC on behalf of the STFC DiRAC HPC Facility (www.dirac.ac.uk). This equipment was funded by BIS National E-infrastructure capital grant ST/K00042X/1, STFC capital grants ST/H008519/1 and ST/K00087X/1, STFC DiRAC Operations grant ST/K003267/1 and Durham University. DiRAC is part of the National E- Infrastructure. The authors thank the organisers of the 2nd Gaia Challenge Workshop, where this project was initiated. The 2nd Gaia Challenge Workshop (October 2014, Max Planck Institute for Astronomy, Heidelberg) was supported by Collaborative Research Center SFB881 ``The Milky Way System". CM acknowledges the use of TOPCAT \citep{Taylor2005} through out the course of this investigation.

\bibliographystyle{mnras}

\label{lastpage}

\end{document}